\documentclass{elsarticle}
\usepackage{lineno}
\usepackage{hyperref}
\usepackage{amssymb,amsmath}
\usepackage{algorithm}
\usepackage{enumitem}
\usepackage{array}
\usepackage{xcolor}
\biboptions{authoryear}

\begin{document}

\begin{frontmatter}

\title{Fast Computation of Robust Subspace Estimators}
%\tnotetext[mytitlenote]{Fully documented templates are available in the elsarticle package on \href{http://www.ctan.org/tex-archive/macros/latex/contrib/elsarticle}{CTAN}.}

%% Group authors per affiliation:
%\author{Holger~Cevallos Valdiviezo\fnref{myfootnote}}
%\address{Escuela Superior Polit\'{e}cnica del Litoral, Facultad de Ciencias Naturales y Matem\'{a}ticas, Guayaquil, Ecuador\\
%Ghent University, Department of Applied Mathematics, Computer Science and Statistics,  Gent, Belgium}
%\fntext[myfootnote]{Since 1880.}
%
%\author{Stefan~Van Aelst}
%\address{KU Leuven, Department of Mathematics, Section of Statistics, Leuven, Belgium}

%% or include affiliations in footnotes:
\author[mymainaddress,mysecondaryaddress]{Holger~Cevallos-Valdiviezo}
\ead{holgceva@espol.edu.ec}

\author[myaddress]{Stefan~Van Aelst\corref{mycorrespondingauthor}}
\cortext[mycorrespondingauthor]{Corresponding author: KU Leuven, Department of Mathematics, Celestijnenlaan 200B, 3001 Leuven, Belgium. Phone: +3216372383, fax: +3216327998.
\ead{Stefan.VanAelst@kuleuven.be}}

\address[mymainaddress]{ESPOL Polytechnic University, Escuela Superior Politécnica del Litoral, ESPOL, (Facultad de Ciencias Naturales y Matemáticas, FCNM), Campus Gustavo Galindo Km. 30.5 Vía Perimetral, P.O. Box 09-01-5863, Guayaquil, Ecuador}
\address[mysecondaryaddress]{Ghent University, Department of Applied Mathematics, Computer Science and Statistics, Krijgslaan 281 S9, 9000 Gent, Belgium}
\address[myaddress]{KU Leuven, Department of Mathematics, Celestijnenlaan 200B, 3001 Leuven, Belgium}

\begin{abstract}
Dimension reduction is often an important step in the analysis of high-dimensional data.
PCA is a popular technique to find the best low-dimensional approximation of high-dimensional data. However, classical PCA is very sensitive to atypical data. 
Robust methods to estimate the low-dimensional subspace that best approximates the regular data have been proposed.
However, for high-dimensional data these algorithms become computationally expensive. Alternative algorithms for the robust subspace estimators are proposed that are better suited to compute the solution for high-dimensional problems. The main ingredients of the new algorithms are twofold. First, the  principal directions of the subspace are estimated directly by
iterating the first order solutions corresponding to the estimators. Second, to reduce the computation time even further five robust deterministic values are proposed to initialize the algorithms instead of using random starting values. 
It is shown that the new algorithms yield robust solutions and the computation time is largely reduced, especially for high-dimensional data. 
\end{abstract}


\end{frontmatter}

%\linenumbers

\section{Introduction}

Principal component analysis (PCA) is a popular exploratory tool for multivariate data.
In particular, PCA is extremely useful to find a low-dimensional representation of high-dimensional data that yields the best possible approximation to the original data. 
Classical PCA minimizes the squared euclidean distances between the original observations and their orthogonal projections onto the lower dimensional subspace. However, classical PCA is very sensitive to atypical data due to the use of quadratic loss. Therefore, several approaches to robustify PCA have been proposed. 

The earliest and easiest approach to robust PCA consists of taking the eigenvectors and eigenvalues of a robust scatter estimator instead of the standard sample covariance matrix~\citep[see e.g.][]{Campbell1980, Devlin1981, Naga1990, Croux2000, Salibian-Barrera2006a}. However, this approach cannot be used for high-dimensional data because calculating high-dimensional robust scatter matrices is computationally complex or even infeasible if the sample size is small compared to the dimension. Moreover, while the efficiency of robust scatter estimators increases with dimension  this comes at the expense of a loss of robustness. Therefore, \citet{Locantore1999} introduced spherical PCA which uses the covariance matrix of the data projected onto the unit sphere and is fast to compute. 

A second approach to robust PCA sequentially looks for univariate directions that maximize a robust estimator of scale and are orthogonal to each other. This robust projection pursuit (PP) approach has been studied by e.g. \citet{Li1985} and \citet{CrouxChristopheRuizGazen, Croux2005}. 

Instead of looking for one direction at a time as in PP, one can robustly estimate a lower-dimensional subspace directly~\citep[see e.g.][]{Liu2003,Croux2003}. ROBPCA \citep{Hubert2005} seeks for a lower-dimensional subspace via a multiple-step procedure. Briefly, the first step aims to identify a subset of $n/2\leq h<n$ observations based on the Stahel-Donoho outlyingness. Classical PCA is then applied on this $h-$subset and the optimal dimension of the subspace is determined. Next, the data points are projected onto the lower-dimensional subspace. Finally, the reweighted MCD estimator is computed on the projected data to obtain the estimates for the principal directions of the subspace. 

Principal Component Pursuit method (PCP) aims to decompose the data matrix into a low-rank component and a gross outlier component~\citep{Candes2011}. See e.g.  \citet{chiang2016} and~\citet{Rahmani2016} for related work.
However, PCP may fail to detect outliers in the orthogonal complement of the subspace~\citep[cfr.][]{Maronna2015,She2016}. %Therefore, \citet{She2016} and \citet{Brahma2017} modified PCP in order to target such outliers as well. 
Therefore, \citet{She2016} introduced a robust orthogonal complement
(ROCPCA) approach to deal with orthogonal complement outliers. The authors model the projected data onto the orthogonal complement subspace with a mean term, a sparse outlier matrix and a noise term. The sparse outlier matrix identifies orthogonal outliers. More recently, \citet{Brahma2017} combined ideas from PCP and ROCPCA to construct a reinforced robust PCP (RRPCP) method that considers outliers in both the observation space and in the orthogonal complement subspace. RRPCP decomposes the data matrix into a low-rank component, a sparse outlier matrix to represent outliers in the observation space, a noise term, a mean term and another sparse outlier matrix to represent outliers in the orthogonal complement space. The two latter components are represented in the orthogonal complement space and then transformed back to the observation space. However, according to their model formulation ROCPCA and RRPCP have to estimate the directions of the orthogonal complement space, which becomes computationally intensive for high-dimensional data.  

\citet{Maronna2005} proposed to robustly estimate the best lower-dimensional approximation by minimizing either an M-scale or a least trimmed squares (LTS) scale of the Euclidean distances corresponding to the observations. He also characterized the solutions by the orthogonal complement directions and showed that these directions correspond to the eigenvectors associated with the smallest eigenvalues of a weighted covariance matrix. Based on this characterization \citet{Maronna2005} proposed an iterative algorithm to compute the robust subspace estimators. 
The robustness of these subspace estimators has been widely investigated empirically. See e.g.   
\citet{Maronna2005,Serneels2008} and \citet{Tharrault2008} for the M-scale estimator, and \citet{Maronna2005,Engelen2005} and~\citet{robusttrimmedsub} for the LTS-scale estimator. Moreover, \citet{robusttrimmedsub} also provides a thorough theoretical study of the properties of the estimator based on the LTS scale.  

However, in case of a low-dimensional approximation for high-dimensional data Maronna's orthogonal complement algorithm requires to decompose a high dimensional covariance matrix and needs a large number of its eigenvectors to characterize the solution. This makes computing the subspace estimators time consuming or even infeasible in high dimensions.  Therefore, we propose an algorithm for the robust subspace estimators of \citet{Maronna2005} that directly calculates principal directions of the low-dimensional subspace. 

The main ingredients of our new algorithm are twofold. First, we use the first order conditions corresponding to the estimator to update the principal directions of the subspace iteratively. 
This approach only requires low-dimensional vector and matrix operations rather than manipulating high-dimensional covariance matrices. Second, instead of using random starting values, similarly to~\citet{Hubert2012} we propose five robust deterministic values to initialize the algorithm. 
These starting values yield robust fits that are usually close to the sought after robust solution, so that convergence occurs quickly. 

The remainder of the manuscript is organized as follows. Section \ref{sectionMVS} reviews the robust subspace estimators based on an M-scale or the LTS-scale.
Our definition is equivalent to the definition in~\citet{Maronna2005} but characterizes the solution by the principal directions of the subspace which better serves our needs for the development of the new algorithm in Section \ref{Sectionalgorithm}. In Section~\ref{globalmin} we introduce the robust deterministic values to initialize the algorithm. 
In Section \ref{sec:simulation} we compare the solutions calculated with the new algorithm to those obtained with the original algorithm by means of an extensive simulation study. We also include other robust subspace estimators in the simulations to compare their robustness properties. Since the deterministic starting values are not orthogonally equivariant, in Section \ref{equivariance} we empirically investigate the effect of orthogonal transformations of the data on our algorithm with deterministic starting values. We compare all methods in terms of their computation time in Section \ref{simcomptime}, while in Section~\ref{highdimsim} we extend the simulation study in Section \ref{sec:simulation} to higher-dimensions.
Section \ref{Examplech1} contains a real data illustration and Section~\ref{sec:conclusions} presents our final conclusions. 

\section{Robust subspace estimators} \label{sectionMVS}
Consider a data matrix $\mathbf{X}=(\mathbf{x}_1\dots\mathbf{x}_n)^{\mathrm{T}} \in \mathbb{R}^{n\times p}$ which contains the measurements of $p$ variables for $n$ observations.
The goal is to approximate the $n$ observations $\mathbf{x}_i$ by points $\widehat{\mathbf{x}}_{i}$ that lie in a $q$-dimensional subspace. That is, 
$\widehat{\mathbf{x}}_{i} \equiv 
\widehat{\mathbf{x}}_{i}( \mathbf{B}_{q},\mathbf{A}_{q}, \mathbf{m})= \mathbf{m} + \mathbf{B}_{q} \mathbf{a}_{i}$
for some $ \mathbf{m} \in \mathbb{R}^{p}$, $\mathbf{A}_{q}=(\mathbf{a}_{1},\dots,\mathbf{a}_{n})^\mathrm{T} \in \mathbb{R}^{n \times q}$ and orthogonal matrix $\mathbf{B}_{q} \in \mathbb{R}^{p \times q}$, i.e.\ $\mathbf{B}_{q}^{\mathrm{T}}\mathbf{B}_{q} = \mathbf{I}_{q}$.
Let $\mathbf{b}_{1},\dots,\mathbf{b}_{p}$ denote the rows of $\mathbf{B}_{q}$. 
The Euclidean distance between $\mathbf{x}_{i}$ and its approximation $\widehat{\mathbf{x}}_{i}$ is denoted by $d_{i}(\mathbf{B}_{q},\mathbf{A}_{q}, \mathbf{m})= d_i={\lVert \mathbf{r}_{i} \rVert}$, where $\mathbf{r}_{i} = \mathbf{x}_{i} -  \widehat{\mathbf{x}}_{i}$.
 
\citet{Maronna2005} proposed to robustly estimate the optimal subspace by
minimizing a robust scale estimator of the Euclidean distances $d_{i}(\mathbf{B}_{q},\mathbf{A}_{q}, \mathbf{m})$. 
Note that if the nonrobust standard deviation is used, then the estimator minimizes the sum of squared Euclidean distances and thus the classical PCA solution is retrieved. Maronna's estimators are thus robust extensions of classical PCA dimension reduction.   

Although the Euclidean distance between each observation $\mathbf{x}_{i}$ and its projection $\widehat{\mathbf{x}}_{i}$ onto the $q$-dimensional subspace is measured in the $p-q$ dimensional orthogonal subspace in \citet{Maronna2005}, this is equivalent to our current formulation in the $p$-dimensional space.

\subsubsection*{Subspace S-estimator}

The subspace S-estimator $(\widehat{\mathbf{B}}_{\mathrm{S}},\widehat{\mathbf{A}}_{\mathrm{S}}, \widehat{\mathbf{m}}_{\mathrm{S}})$  is obtained by minimizing an M-scale of the Euclidean distances $d_{i}(\mathbf{B}_{q},\mathbf{A}_{q}, \mathbf{m})$. That is, $(\widehat{\mathbf{B}}_{\mathrm{S}},\widehat{\mathbf{A}}_{\mathrm{S}}, \widehat{\mathbf{m}}_{\mathrm{S}})$ is a solution  of 
\begin{align}\label{definitionMVS}
\min_{\mathbf{B}_{q},\mathbf{A}_{q}, \mathbf{m}}
\widehat{\sigma}_{\mathrm{M}}(\mathbf{d}(\mathbf{B}_{q},\mathbf{A}_{q}, \mathbf{m})),
\end{align}
over all $ \mathbf{m} \in \mathbb{R}^{p}$, $\mathbf{A}_{q}=(\mathbf{a}_{1},\dots,\mathbf{a}_{n})^{\mathrm{T}} \in \mathbb{R}^{n \times q}$ and orthogonal matrices $\mathbf{B}_{q} \in \mathbb{R}^{p \times q}$. For any $\mathbf{d}(\mathbf{B}_{q},\mathbf{A}_{q}, \mathbf{m})
=(d_{1},d_{2},\ldots,d_{n})$, the corresponding M-scale $\widehat{\sigma}_{\mathrm{M}}(\mathbf{d}(\mathbf{B}_{q},\mathbf{A}_{q}, \mathbf{m}))$ is defined as the solution in $s$ of 
\begin{align}\label{definitionMscale}
\frac{1}{n}  \sum_{i=1}^{n} \rho \left( \frac{d_i}{s} \right) = b,
\end{align}
where $\rho$ is an even function that is differentiable and non-decreasing on the positive real line with $\rho(0) = 0$  (see e.g. \citealp{Maronna2006}). 

%Let $\mathbf{r}_{i}=(r_{i1},\dots,r_{ip})$ with $r_{ij}=x_{ij} - m_{j} - \mathbf{a}_{i}^{\mathrm{T}}\mathbf{b}_{j}$, then 
Similarly as in ~\citet{BoenteGracielaSalibian-Barrera2015}  
first order conditions for the subspace S-estimator can be obtained by implicitly differentiating the M-scale in (\ref{definitionMscale}). 
The solutions of these first order conditions can be written as
%\begin{align}
%\sum_{j=1}^p  \, (x_{ij} - m_{j}) \, \mathbf{b}_{j} &= \left(\sum_{j=1}^p \, \mathbf{b}_{j} \, \mathbf{b}_{j}^{\mathrm{T}} \right)  \mathbf{a}_{i} \, ,  1\leq i \leq n \, ,    
%\label{esteqMVS1}
%\\
%\sum_{i=1}^n w_{i} \, (x_{ij} - m_{j}) \, \mathbf{a}_{i} &= \left(\sum_{i=1}^n w_{i} \, \mathbf{a}_{i} \, \mathbf{a}_{i}^{\mathrm{T}} \right)  \mathbf{b}_{j} \, ,  1\leq j \leq p, \label{esteqMVS2}  \\ 
% \sum_{i=1}^n w_{i} \, (x_{ij} - \mathbf{a}_{i}^{\mathrm{T}} \mathbf{b}_{j})  &= \sum_{i=1}^n w_{i} \, m_{j}  \, , 1\leq j \leq p, \label{esteqMVS3} 
%\end{align}
\begin{align}
 \mathbf{a}_{i} &= \left(\sum_{j=1}^p \, \mathbf{b}_{j} \, \mathbf{b}_{j}^{\mathrm{T}} \right)^{-1}  \left(\sum_{j=1}^p  \, (x_{ij} - m_{j}) \, \mathbf{b}_{j}\right), \label{esteqMVS1}\\ \nonumber \\
 \mathbf{b}_{j} &= \left(\sum_{i=1}^n w_{i} \, \mathbf{a}_{i} \, \mathbf{a}_{i}^{\mathrm{T}} \right)^{-1} \left(\sum_{i=1}^n w_{i} \, (x_{ij} - m_{j}) \, \mathbf{a}_{i}\right), \label{esteqMVS2} 
\end{align}
 and
\begin{equation}
   m_{j} = \frac{\sum_{i=1}^n w_{i} \, (x_{ij} - \mathbf{a}_{i}^{\mathrm{T}} \mathbf{b}_{j})}{ \sum_{i=1}^n w_{i}}, \label{esteqMVS3} 
\end{equation}
with weights 
\begin{align}\label{weights}
w_{i} \equiv w(d_{i}) = \rho^\prime \left( \frac{d_{i}}{\widehat{\sigma}_{\mathrm{M}}} \right)\frac{\widehat{\sigma}_{\mathrm{M}}}{d_{i}},
\end{align}
for the rows of $\mathbf{A}_{q}$, $i=1,\ldots,n$, and the rows of $\mathbf{B}_{q}$ and components of $\mathbf{m}$, $j=1,\ldots,p$, respectively. 
Expressions~\eqref{esteqMVS1}-\eqref{weights} suggest an iterative reweighted least squares procedure.
Initial values $\mathbf{B}_{q}^{(0)}$, $\mathbf{A}_{q}^{(0)}$ and $\mathbf{m}^{(0)}$ can be used to obtain starting weights $w_{i}^{(1)}$ from expression (\ref{weights}). 
Based on $\mathbf{B}_{q}^{(0)}$ and $\mathbf{m}^{(0)}$ one can obtain updated scores $\mathbf{A}_{q}^{(1)}$ from~(\ref{esteqMVS1}). Next, based on $\mathbf{A}_{q}^{(1)}$, $\mathbf{m}^{(0)}$ and weights $w_{i}^{(1)}$ one can obtain $\mathbf{B}_{q}^{(1)}$ from~(\ref{esteqMVS2}). Finally, based on $\mathbf{A}_{q}^{(1)}$, $\mathbf{B}_{q}^{(1)}$ and weights $w_{i}^{(1)}$ one can obtain $\mathbf{m}^{(1)}$ from~(\ref{esteqMVS3}). This is the first iteration. 
%The second iteration starts by updating the weights to $w_{i}^{(2)}$ based on $\mathbf{B}_{q}^{(1)}$, $\mathbf{A}_{q}^{(1)}$ and $\mathbf{m}^{(1)}$ and continues in a similar fashion to obtain updates $\mathbf{A}_{q}^{(2)}$, $\mathbf{B}_{q}^{(2)}$ and $\mathbf{m}^{(2)}$. 
This procedure is repeated until we obtain a stable solution. Note that the resulting estimate of the matrix $\mathbf{B}_{q}$ is not necessarily orthogonal. Therefore, we orthogonalize this estimate at the end of the procedure and update the scores accordingly. Using the orthogonality of $\mathbf{B}_{q}$,  it can be seen from (\ref{esteqMVS1}) that $\mathbf{a}_{i}=\mathbf{B}_{q}^{\mathrm{T}}(\mathbf{x}_{i}-\mathbf{m})$.
Hence, once $\mathbf{B}_{q}$ and $\mathbf{m}$ are known, the corresponding scores $\mathbf{a}_{i}$ of the observations are easily obtained. 
By combining this result  with (\ref{esteqMVS3}) we can also obtain that $\mathbf{m}= \sum_{i=1}^n w_{i} \, \mathbf{x}_{i}/ (\sum_{i=1}^n w_{i} )$.
Note that if we put $w_{i}=1$ for all observations, then the procedure reduces to an  alternating least squares algorithm whose solution yields the classical PCA estimates~\citep[see also][]{RubenGabriel1979}. 
Finally, by combining these expressions and using that $\widehat{\mathbf{B}}_{\mathrm{S}}^{\mathrm{T}}\widehat{\mathbf{B}}_{\mathrm{S}}=\mathbf{I}_{q}$
it can also be derived that the subspace S-estimators ($\widehat{\mathbf{B}}_{\mathrm{S}}$, 
$\widehat{\mathbf{m}}_{\mathrm{S}}$) satisfy the equation
\begin{align}\label{preweightedcov}
 \sum_{i=1}^{n} w_{i} (\mathbf{x}_{i} - \mathbf{m})(\mathbf{x}_{i} - \mathbf{m})^{\mathrm{T}}\mathbf{B}_{q} = \mathbf{B}_{q}\, \Lambda.
\end{align}
where $\Lambda=\mathbf{B}^{\mathrm{T}} \sum_{i=1}^{n} w_{i} (\mathbf{x}_{i} - \mathbf{m})(\mathbf{x}_{i} -  \mathbf{m})^{\mathrm{T}} \mathbf{B}$.
From (\ref{preweightedcov}) and (\ref{definitionMVS}) it follows that the columns of $\widehat{\mathbf{B}}_{\mathrm{S}}$ correspond to the first $q$ eigenvectors of the weighted covariance matrix 
\begin{align}\label{weightedcov}
\mathbf{C}(\widehat{\mathbf{m}}_{\mathrm{S}}, \widehat{\mathbf{B}}_{\mathrm{S}}) = \frac{1}{n}\sum_{i=1}^{n} w_{i}(\mathbf{x}_{i} - \widehat{\mathbf{m}}_{\mathrm{S}})(\mathbf{x}_{i} - \widehat{\mathbf{m}}_{\mathrm{S}})^{\mathrm{T}},
\end{align}
which coincides with expression (9) of \citet{Maronna2005}. 

\subsubsection*{Subspace LTS-estimator}

The subspace {LTS}-estimator $(\widehat{\mathbf{B}}_{\mathrm{LTS}},\widehat{\mathbf{A}}_{\mathrm{LTS}}, \widehat{\mathbf{m}}_{\mathrm{LTS}})$ is obtained by minimizing the LTS-scale of the Euclidean distances $d_{i}(\mathbf{B}_{q},\mathbf{A}_{q}, \mathbf{m})$. That is, $(\widehat{\mathbf{B}}_{\mathrm{LTS}},\widehat{\mathbf{A}}_{\mathrm{LTS}}, \widehat{\mathbf{m}}_{\mathrm{LTS}})$ is a solution  of 
\begin{align}\label{definitionMVLTS}
\min_{\mathbf{B}_{q},\mathbf{A}_{q}, \mathbf{m}}
\widehat{\sigma}_{\mathrm{LTS}}(\mathbf{d}(\mathbf{B}_{q},\mathbf{A}_{q}, \mathbf{m})),
\end{align}
over all $ \mathbf{m} \in \mathbb{R}^{p}$, $\mathbf{A}_{q} \in \mathbb{R}^{n \times q}$ and orthogonal matrices $\mathbf{B}_{q} \in \mathbb{R}^{p \times q}$. For any $\mathbf{d}(\mathbf{B}_{q},\mathbf{A}_{q}, \mathbf{m})
=(d_{1},d_{2},\ldots,d_{n})$, the corresponding LTS scale is defined as
\begin{align}\label{definitionLTSscale}
\widehat{\sigma}^{2}_{\mathrm{LTS}}(\mathbf{d}) = \frac{1}{h}  \sum_{i=1}^{h} d_{(i:n)}^{2},
\end{align}
where $d_{(1:n)} \leq \ldots \leq d_{(n:n)}$ are the ordered Euclidean distances and $h=n-\left\lfloor n\alpha  \right\rfloor$ with $0 \leq \alpha < 0.5$. 
A fraction $\alpha$ of the observations is not taken into account when calculating the LTS estimator and thus 
$\alpha$ determines the robustness of the estimator.

Similarly as in \citet{Maronna2005} it can be shown that the LTS solution satisfies the expressions (\ref{esteqMVS1})-(\ref{esteqMVS3}) with weights now given by
\begin{align}
w_{i}\equiv w(d_i) =\begin{cases}
1 & \text{if  } d_{i}\leq  d_{(h:n)},\\
0 & \text{otherwise.} 
\end{cases} 
\label{LTSweights}
\end{align}
With these weights the LTS estimator $(\widehat{\mathbf{B}}_{\mathrm{LTS}},\widehat{\mathbf{m}}_{\mathrm{LTS}})$ also satisfies~(\ref{preweightedcov}) and~(\ref{weightedcov}) which again coincides with expression (9) in  \citet{Maronna2005}.

\section{The algorithm}\label{Sectionalgorithm}

\citet{Maronna2005} characterizes a $q$-dimensional subspace by an equation $\mathbf{B}_{p-q}^{\mathrm{T}}\mathbf{x}=\mathbf{a}$ with  $\mathbf{a} \in \mathbb{R}^{p-q}$ and $\mathbf{B}_{p-q} \in \mathbb{R}^{p \times (p-q)}$ an orthogonal matrix. 
His algorithm can be summarized by the following steps. First, 50 random orthogonal matrices $\mathbf{B}_{p-q}$ are generated to initialize the algorithm. The corresponding optimal value of $\mathbf{a}$ that minimizes the robust scale can then be calculated easily. A few iterative improvement steps are then applied to each of these starting values. In each iteration the weights $w_i$ of the observations corresponding to the current solution are updated. Based on the updated weights, a new solution is then obtained by calculating the smallest $p-q$ eigenvectors of the weighted covariance matrix in~(\ref{weightedcov}). After two iterations, the 10 best solutions  are selected and these are iterated further until convergence (with a maximum of 10 iterations).

If $p$ is large and the subspace dimension $q$ is small, Maronna's algorithm requires the storage of a high-dimensional covariance matrix and the calculation of a large number (namely, $p-q$) of its eigenvectors which becomes computationally demanding. To improve the computation time and reduce the memory load, we instead propose to directly calculate the $q$ basis directions of the subspace by iterating expressions (\ref{esteqMVS1})-(\ref{esteqMVS3}). Extensive experiments have shown that iterating these expressions only a few times suffices to obtain close approximations to the first eigenvectors of the weighted covariance matrix in~(\ref{weightedcov}). Note that our iterations of these expressions only require operations with $q$-dimensional vectors and matrices and thus will be more suitable for high-dimensional settings.

Algorithm~\ref{alg:MVS_MVLTS} contains a detailed description of the main part of our modified algorithm in pseudo-code.  The algorithm requires initial values of $\mathbf{B}_{q}$ and $\mathbf{m}$ as input. It also depends on tuning parameters $N_1$, $N_2$, $N_3$ and $tol$. 
$tol$ specifies the precision with which the solution is calculated. The tuning parameters $N_1$ and $N_2$ play the same role as in Maronna's original algorithm. That is, for each initial orthogonal matrix $\mathbf{B}_{q}$ first $N_1$ iterations are performed to improve the corresponding estimates of the location $\mathbf{m}$ and the scores matrix $\mathbf{A}_{q}$ while keeping $\mathbf{B}_{q}$ fixed. In the next $N_2$ iterations, the estimates of all three quantities $\mathbf{B}_{q}$, $\mathbf{m}$ and $\mathbf{A}_{q}$ are updated. Since these updates are now calculated by iterating expressions (\ref{esteqMVS1})-(\ref{esteqMVS3}), an additional tuning parameter $N_3$ specifies how often these expressions are iterated.  Using a similar proof as in \citet{Maronna2005}, it can easily be shown that either the M-scale $\widehat{\sigma}_M$ or LTS scale $\widehat{\sigma}_{\mathrm{LTS}}$ decreases in each iteration of our algorithm. 
%\textcolor{blue}{This guarantees that the largest eigenvectors are being estimated by the algorithm}. 
Moreover, the performance of our algorithm is similar to Maronna's algorithm if we use the same settings for the tuning parameters and start from the same initial values for $\mathbf{B}_{q}$ and $\mathbf{m}$. 

\begin{algorithm}[ht!]
\caption{Algorithm for the subspace S and LTS estimators}
\label{alg:MVS_MVLTS}
\footnotesize
Input: $\mathbf{X}$, $\mathbf{B}_{q}$ (with $\mathbf{B}_{q}^{\mathrm{T}}\mathbf{B}_{q}=\mathbf{I}_{q}$), $\mathbf{m}$, $N_1=3$, $N_2=2$, $N_3=3$, $tol=1 \times 10^{-6}$.
 \begin{enumerate}
\item Set $\textit{it} \leftarrow 1$.
\begin{enumerate}[label=\alph*.]
\item Compute $\mathbf{A}_{q}=(\mathbf{a}_{1},\dots \mathbf{a}_{n})^{\mathrm{T}}$ with $\mathbf{a}_{i}
=\mathbf{B}_{q}^{\mathrm{T}}(\mathbf{x}_{i}-\mathbf{m})$, $i=1,\ldots,n$.
\item Compute distances $ d_{i}(\mathbf{B}_{q}, \mathbf{A}_{q},\mathbf{m})  
$, $i=1,\ldots,n$.
\item Compute $\widehat{\sigma}_{0}= \widehat{\sigma}_{\mathrm{M}}(\mathbf{d}(\mathbf{B}_{q},\mathbf{A}_{q},\mathbf{m}))$ or $\widehat{\sigma}_{0}=\widehat{\sigma}_{\mathrm{LTS}}(\mathbf{d}(\mathbf{B}_{q},\mathbf{A}_{q},\mathbf{m}))$.
\end{enumerate}
\item Do until $\textit{it} = N_{1} +  N_{2}$ or $\Delta \leq tol$:
\begin{enumerate}[label=\alph*.]
\item Compute $w_{i}$ from (\ref{weights}) or  (\ref{LTSweights})
and update $\mathbf{m} = \frac{\sum_{i=1}^{n} w_{i} \mathbf{x}_{i}}{\sum_{i=1}^{n} w_{i}}$.
\item If $\textit{it} > N_{1}$:
\begin{enumerate}[label={(\arabic*)}] 
\item Set $\textit{iter} \leftarrow 1$ and $s^{2}_{0} \leftarrow \widehat{\sigma}_0^{2}$.
\item Do until $\textit{iter} = N_{3}$ or $\tilde{\Delta} \leq tol$:
\begin{enumerate}[label=\roman*.] 
\item Compute $\mathbf{A}_{q}=(\mathbf{a}_{1},\dots \mathbf{a}_{n})^{\mathrm{T}}$, 
$\mathbf{B}_{q}=(\mathbf{b}_{1},\dots,\mathbf{b}_{p})^{\mathrm{T}}$ and $\mathbf{m}$ from (\ref{esteqMVS1})-(\ref{esteqMVS3}).
\item Compute distances $ d_{i}( \mathbf{B}_{q}, \mathbf{A}_{q},\mathbf{m})$, $i=1,\ldots,n$.
\item Compute $s=\widehat{\sigma}_{\mathrm{M}}(\mathbf{d}( \mathbf{B}_{q}, \mathbf{A}_{q},\mathbf{m}))$ or $s=\widehat{\sigma}_{\mathrm{LTS}}(\mathbf{d}( \mathbf{B}_{q}, \mathbf{A}_{q},\mathbf{m}))$.
\item Set $\textit{iter} \leftarrow \textit{iter}+1$, $\tilde{\Delta} \leftarrow 1 - s^{2} / s_{0}^{2}$ and $s_{0}^{2} \leftarrow s^{2}$.
\end{enumerate}
\item End do.
\end{enumerate}
\item Compute $\mathbf{A}_{q}=(\mathbf{a}_{1},\dots \mathbf{a}_{n})^{\mathrm{T}}$ using (\ref{esteqMVS1}) and update distances $d_{i}( \mathbf{B}_{q}, \mathbf{A}_{q},\mathbf{m})$, $i=1,\ldots,n$.
\item Compute $\widehat{\sigma}= \widehat{\sigma}_{\mathrm{M}}(\mathbf{d}(\mathbf{B}_{q},\mathbf{A}_{q},\mathbf{m}))$ or $\widehat{\sigma}=\widehat{\sigma}_{\mathrm{LTS}}(\mathbf{d}(\mathbf{B}_{q},\mathbf{A}_{q},\mathbf{m}))$.
\item Set $\Delta \leftarrow 1 - \widehat{\sigma}^{2} / \widehat{\sigma}_{0}^{2}$ and $\widehat{\sigma}_{0}^{2} \leftarrow \widehat{\sigma}^{2}$.
\item Set $\textit{it} \leftarrow \textit{it} + 1$.
\end{enumerate}
\item End do.
\end{enumerate}
Output: $\widehat{\mathbf{B}}_{q}$, $\widehat{\mathbf{A}}_{q}$, $\widehat{\mathbf{m}}$.
\end{algorithm}

For the subspace S-estimator we have used the popular Tukey biweight loss function $\rho(y) = \mathrm{min}(3y^{2} - 3y^{4} + y^{6},1)$ to calculate the M-scales.
Two standard choices for the tuning parameters are %$c=1.54764,b=0.5$ 
$b=0.5$ which yields the maximal breakdown point (BDP) of $50\%$ and %$c=3,b=0.2426$ 
$b=0.2426$ which yields a better compromise between efficiency and robustness (breakdown point $\approx 25\%$). 
For the subspace LTS estimator the  trimming fraction $\alpha$ determines the breakdown point. The most common choices are 
$\alpha=0.5$ ($50\%$ BDP) and $\alpha=0.25$ ($25\%$ BDP).
See~\citet{Maronna2005} for more details on the breakdown point.  

\section{Starting values}\label{globalmin}

It is well-known that the objective functions in (\ref{definitionMVS}) and (\ref{definitionMVLTS}) are nonconvex and thus may have several local minima. Not all of these minima correspond to robust solutions.
A standard approach is to aim for the global minimum in~(\ref{definitionMVS}) or (\ref{definitionMVLTS}) by generating many starting values. The solution that corresponds to the smallest value of the objective function that is reached by iterating these starting values is then the approximation for the global optimum \citep[see e.g.][]{Rousseeuw1999,Salibian-Barrera2006}.
In Maronna's algorithm $50$ random initial orthogonal matrices $\mathbf{B}_{q}$ are generated while the initial estimate for the location $\mathbf{m}$ is the coordinatewise median of the data.
Each of these starting values is iterated $N_{1} = 3$ times with $\mathbf{B}_{q}$ fixed, followed by $N_{2} = 2$ iterations to improve the estimates of $\mathbf{B}_{q}$, $\mathbf{A}$ and $\mathbf{m}$ together. Then, the $10$ best solutions (with the lowest scale) are selected and these are iterated further until convergence ($tol=0.001$) with a maximum of $N^{\prime}_{2} = 10$ iterations. Extensive experiments by \citet{Maronna2005} revealed that the algorithm shows good performance with these choices for the tuning parameters. Following \citet{Maronna2005} we denote the solutions obtained with his algorithm by S-M when the M-scale is minimized and by S-L when the LTS scale is minimized. 

Empirical comparisons have confirmed that if we use the same starting values for our new algorithm as well as the same settings for the tuning parameters, then we obtain a similar performance as Maronna's algorithm. Note that our algorithm requires the additional tuning parameter $N_3$ which determines the number of iterations to calculate the updated estimates by iterating expressions (\ref{esteqMVS1})-(\ref{esteqMVS3}). Extensive experiments showed that it suffices to use $N_3=3$  iterations to obtain good approximations that yield stable results. Note that while Maronna's algorithm uses the fast-to-compute coordinatewise median as starting value for $\mathbf{m}$, we prefer to use as default the orthogonal equivariant spatial median which can also be computed efficiently~\citep{Vardi2000}. Moreover, we generate random orthogonal matrices $\mathbf{B}_{q}$ as starting values by the method of \citet{Stewart1980} which consists of orthogonalizing a matrix of normal random numbers while \citet{Maronna2005} instead orthogonalizes a matrix of uniform random numbers. We denote the solutions of our algorithm with random orthogonal starting values by RsubS for the subspace S-estimator and by RsubLTS for the subspace LTS-estimator.

%When the M-scale is used, we denote the solutions of our algorithm by subS which stands for subspace S-estimator and by subLTS when the LTS scale is used. 

While our adaptation of Maronna's algorithm is indeed faster for high-dimensional data (see results in Section 7), the computation time remains considerable because a sufficient number of random starting values is needed to obtain a robust solution. While the default setting is to use a fixed number of $50$ random orthogonal matrices as starting values, it is clear that the search space increases dramatically with increasing dimension. Hence, it can be expected that for high-dimensional data (many) more random starting values may be needed to find a stable robust solution. Many algorithms for robust estimators encounter the same issue. To address this issue for the calculation of the minimum covariance determinant (MCD) estimator of multivariate location and scatter in high dimensions,  \citet{Hubert2012} introduced a deterministic MCD algorithm~\citep[see also][]{HUBERT201564}. The main idea is to replace the random starting values by a few well-chosen robust starting values. 
These robust starting values should be fast to compute while at the same time they are expected to lie close to a robust minimum of the objective function, which in our case is given by~(\ref{definitionMVS}) or~(\ref{definitionMVLTS}). 
Hence, instead of exploring the whole parameter space to find the optimum, a  few robust starting values should point us to that part of the parameter space where robust solutions can be found. 
Since convergence from the robust starting values to their closest local minimum is generally fast as well, this results in an algorithm with a much lower computation time which allows us to handle problems in higher dimensions. 

For the deterministic version of our algorithm, we could use as starting values for $\mathbf{B}_{q}$ the $q$ largest eigenvectors corresponding to the robust starting values for the $p$-dimensional scatter matrix proposed by~\citet{Hubert2012}. However, we want to avoid having to calculate $p$-dimensional scatter matrices which may be unstable for large $p$ and consumes a lot of memory.
Therefore, inspired by~\citet{Hubert2012} we propose five robust starting values for $\mathbf{B}_{q}$ which are obtained by the following procedure:

\begin{description}
\item[Step 1.]
Robustly standardize each variable $X_{j}$ by subtracting its median and dividing by its $Q_{n}$ scale estimate~\citep{Rousseeuw1993}. 
Let 
$\mathbf{Z} = (\mathbf{z}_{1},\dots,\mathbf{z}_{n})^{\mathrm{T}}$ denote the standardized data matrix and $Z_{1},\dots, Z_{p}$ the standardized variables. Consider the following transformations of the data:
\begin{enumerate}[label=\arabic*)]
\item Hyperbolic tangent (sigmoid) transformation: let $H_{j} = \mathrm{tanh}(Z_{j}), j=1,\ldots,p$.\\ 
Set $\mathbf{U}_{1}^{\prime}=(H_{1},\dots H_{p})$.
\item Rank transformation: let $R_{j}$ be the vector of ranks of column $X_{j}, j=1,\ldots,p$.\\
Set $\mathbf{U}_{2}=(R_{1},\dots,R_{p})$.
\item Normal scores: compute normal scores from the ranks $R_{j}$, i.e. $T_{j} = \Phi^{-1} \left[(R_{j} - 1/3)/(n+1/3)  \right]$, where $\Phi$(.) is the normal cumulative distribution function. \\ 
Set $\mathbf{U}_{3}=(T_{1},\dots,T_{p})$.
\item Spatial signs: let $\mathbf{s}_{i}= \mathbf{z}_{i}/ \left\| \mathbf{z}_{i} \right\|$, $i=1,\ldots,n$.\\
 Set $\mathbf{U}_{4}^{\prime}=(\mathbf{s}_{1}, \dots,\mathbf{s}_{n})^{\mathrm{T}}$. 
\end{enumerate}
Robustly standardize the columns of $\mathbf{U}_{1}^{\prime}$ and $\mathbf{U}_{4}^{\prime}$ obtained above by subtracting their median and dividing by their $Q_{n}$ scale estimate. Denote these standardized transformed data matrices by $\mathbf{U}_{1}$ and $\mathbf{U}_{4}$  and take $\mathbf{U}_{5}=\mathbf{Z}$.
\item[Step 2.] Calculate the classical $q$-dimensional PCA subspace using the alternating least squares algorithm (see Algorithm~\ref{PCAsub} below) for $\mathbf{Y}=\mathbf{U}_{k}$.  Let 
$\widetilde{\mathbf{B}}_{q}^{(k)}$ be the resulting estimate of the $q$ PC directions for $k=1,\dots,5$. 
\item[Step 3.] 
For each of the score matrices $\widetilde{\mathbf{A}}_{q}^{(k)} = \mathbf{Z} \widetilde{\mathbf{B}}_{q}^{(k)}$, $k=1,\dots,5$, select the $l=\left\lceil n/2  \right\rceil$ rows with smallest Euclidean norm. Let $\mathcal{I}_k \subset \{1,\dots,n\}$ denote the indices of the selected rows. 
\item[Step 4.] Set  $\widetilde{\mathbf{X}}^{(k)}=\mathbf{X}_{\mathcal{I}_k}$ for $k=1,\dots,5$.
For each of these five reduced data matrices, calculate the classical $q$-dimensional PCA subspace by applying Algorithm~\ref{PCAsub} on $\mathbf{Y}=\widetilde{\mathbf{X}}^{(k)} - \widetilde{\mathbf{m}}^{(k)}$ where $\widetilde{\mathbf{m}}^{(k)}= \frac{1}{l} \sum_{i=1}^{l} \widetilde{\mathbf{x}}_{i}^{(k)}$. The resulting estimates of $\mathbf{B}_{q}$ together with $\widetilde{\mathbf{m}}^{(k)}$ are the starting values for Algorithm \ref{alg:MVS_MVLTS}.
\end{description}

\begin{algorithm}[h!]
\caption{PCA subspace algorithm}
\label{PCAsub}
\scriptsize
Input: $\mathbf{Y}$,  $\mathbf{B}_{q}$ (with $\mathbf{B}_{q}^{\mathrm{T}}\mathbf{B}_{q}=\mathbf{I}_{q}$), $N_3=3$, $tol=1 \times 10^{-6}$.
\begin{enumerate}
\item Compute $\mathbf{A}_{q}=(\mathbf{a}_{1},\dots \mathbf{a}_{l})^{\mathrm{T}}$ with
$\mathbf{a}_{i}=\mathbf{B}_{q}^{\mathrm{T}} \mathbf{y}_{i}$.
\item Compute distances $ d_{i} (\mathbf{B}_{q}, \mathbf{A}_{q})$, $i=1,\ldots,l$.
\item Set $iter \leftarrow 1$ and compute $ \widehat{s}^{2}_{0} = \frac{1}{l}\sum_{i=1}^{l} d_{i}^{2}$.
\item Do until $iter  = N_{3}$ or $\tilde{\Delta} \leq tol$:
\begin{enumerate}[label=\roman*.] 
\item Compute $\mathbf{A}_{q}=(\mathbf{a}_{1},\dots \mathbf{a}_{l})^{\mathrm{T}}$ and $\mathbf{B}_{q}=(\mathbf{b}_{1},\dots \mathbf{b}_{p})^{\mathrm{T}}$ from (\ref{esteqMVS1})-(\ref{esteqMVS2}) with weights $w_{i}=1$, $i=1,\ldots,l$.
\item Compute distances $ d_{i}(\mathbf{B}_{q}, \mathbf{A}_{q})$, $i=1,\ldots,l$.
\item Compute $ \widehat{s}^{2} = \frac{1}{l}\sum_{i=1}^{l} d_{i}^{2}$.
\item Set $iter \leftarrow iter +1$, $\tilde{\Delta}\leftarrow 1 - \widehat{s}^{2} / \widehat{s}_{0}^{2}$ and $\widehat{s}_{0}^{2} \leftarrow \widehat{s}^{2}$.
\end{enumerate}
\item End do.
\end{enumerate}
Output: $\widehat{\mathbf{B}}_{q}$, $\widehat{\mathbf{A}}_{q}$.
\end{algorithm}

The alternating least squares algorithm for classical PCA, summarized in Algorithm~\ref{PCAsub}, calculates the $q$-dimensional PCA subspace without performing singular value decomposition of a high-dimensional matrix, and has the advantage that it only requires operations with $q$-dimensional vectors and matrices in the iterative updating of the estimates in step 4.i. 
Note that for simplicity Algorithm~\ref{PCAsub} uses a centered data matrix as input data. 
Similarly as for Algorithm~\ref{alg:MVS_MVLTS}, we found that $N_{3} = 3$ iterations in step 4 of Algorithm~\ref{PCAsub} were enough to obtain close approximations to the first eigenvectors of the classical covariance matrix. The $q-$dimensional canonical basis or a random orthogonal matrix can be used to initialize $\mathbf{B}_{q}$. In fact, we found that Algorithm~\ref{PCAsub} converges to the global solution independent of the initial value that is chosen for $\mathbf{B}_{q}$. 

The procedure to obtain deterministic starting values can be summarized in two parts. The first part  (first three steps) aims to identify subsets of the data with only regular observations. Transformations of the standardized data are used in Step 1 to reduce the effect of outliers. In Step 2 naive robust estimates of the subspace are calculated from the transformed data which are used to project the data in Step 3 before obtaining clean subsets. This projection makes the algorithm better suitable for correlated data. In Section \ref{equivariance} we investigate by means of a simulation study the effect of orthogonal transformations of the data on the estimates obtained with these deterministic algorithms. The second part (last step) calculates the classical PCA subspace corresponding to each of the obtained subsets of the data matrix to obtain promising starting values for Algorithm \ref{alg:MVS_MVLTS}. Similarly as for orthogonal starting values, each of these five starting values is iterated $N_{2} = 2$ times to improve the estimates of $\mathbf{B}_{q}$, $\mathbf{A}$ and $\mathbf{m}$. Then, the best solution (with the lowest scale) is selected and iterated further until convergence ($tol=1 \times 10^{-6}$) with a maximum of $N^{\prime}_{2} = 10$ iterations. We denote our algorithm with deterministic starting values by DsubS when the M-scale is minimized and by DsubLTS when the LTS scale is minimized. 

\section{Performance comparison}\label{sec:simulation}

To compare our new algorithms to Maronna's original algorithms, we repeat the simulations of \citet{Maronna2005}. We use our algorithms with random orthogonal matrices as starting values as well as with deterministic starting values. For the simulations we choose the 
tuning parameters for the M-scale and LTS scale that yield the maximal breakdown point.

We also consider some other computationally efficient methods for robust subspace estimation. These are robust projection pursuit (PP) \citep{CrouxChristopheRuizGazen}, Spherical PCA (SPC) \citep{Locantore1999} and ROBPCA \citep{Hubert2005}. For robust PP we use the algorithm of  \citet{Croux2005} and maximize the LTS scale with $\alpha=0.5$. Note that PP based on the M-scale was already considered in~\citet{Maronna2005}. For ROBPCA we include the reweighting step as in \citet{Engelen2005}. The number of random directions through two data points to compute the measure of outlyingness was fixed to 250 in all experiments as in \citet{Hubert2005}. The parameter $\alpha$ in ROBPCA was set equal to $\alpha=0.5$ for maximal robustness. 
%Note that PPLTS and ROBPCA can be considered as competitors for the LTS subspace estimator.

Following \citet{Maronna2005} we generated $M=200$ samples of size $n=100$ and dimension $p=10$. The regular observations are generated according to $N(\mathbf{0},\mathbf{\Sigma})$ with $\mathbf{\Sigma} = \mathrm{diag}(\lambda_{1},\lambda_{2},\ldots,\lambda_{p})$ where the following two designs are considered for the diagonal elements.
\begin{enumerate}[label=\alph*)]
\item An abrupt increase of the eigenvalues: $\lambda_{j}=1+0.1j$ for $1 \leq j \leq (p-q)$ and $\lambda_{j} = 20(1+0.5(j-p+q))$ for $(p-q+1) \leq j \leq p$.
\item A smooth increase of the eigenvalues: $\lambda_{j}=2^{j-1}$ for $1 \leq j \leq p$.
\end{enumerate}
A fraction $\epsilon=0\%,\,10\%$ or $20\%$ of outlying observations is generated from $N(k\mathbf{x}_{0}, 0.25\mathbf{\Sigma})$, where $\mathbf{x}_{0}$ is a vector of length $p$  with $x_{0j} = 1$ for $j \leq (p-q) $ and 0 otherwise. The value of $k$ runs between $0$ and $20$ with steps of 0.5. 
All methods are applied for $q=2$, so the best two-dimensional linear subspace approximation is estimated. 

To compare the methods we measure their prediction performance as in \citet{Maronna2005}. Hence, we measure the proportion of variance in independent regular data that remains unexplained by their approximation according to the estimated subspace. More formally, let $\mathbf{x}$ be a $N(\mathbf{0},\mathbf{\Sigma})$ vector independent of the random sample used to obtain $\widehat{\mathbf{B}}_{q}$. Then, the variability of $\mathbf{x}$ around the subspace generated by $\widehat{\mathbf{B}}_{q}$ equals
\begin{align*}
E\lVert\mathbf{x} - \widehat{\mathbf{B}}_{q}\widehat{\mathbf{B}}_{q}^{\mathrm{T}}\mathbf{x} \rVert^{2} = \mathrm{tr}\big[ \mathbf{\Sigma} \big] - \mathrm{tr}\big[\widehat{\mathbf{B}}_{q}^{\mathrm{T}} \mathbf{\Sigma} \widehat{\mathbf{B}}_{q} \big],
\end{align*}
and the prediction proportion of unexplained variance is given by
\begin{align}\label{uqpred1}
u_{q}^{\mathrm{pred}} = \frac{E\lVert\mathbf{x} - \widehat{\mathbf{B}}_{q}\widehat{\mathbf{B}}_{q}^{\mathrm{T}}\mathbf{x} \rVert^{2}}{\mathrm{tr}\big[ \mathbf{\Sigma} \big]}= 1 - \frac{\mathrm{tr}\big[\widehat{\mathbf{B}}_{q}^{\mathrm{T}} \mathbf{\Sigma} \widehat{\mathbf{B}}_{q} \big]}{\mathrm{tr}\big[ \mathbf{\Sigma} \big]}.
\end{align}
Since Maronna's algorithms characterize the subspace by an estimate $\widehat{\mathbf{B}}_{p-q} \in \mathbb{R}^{p \times (p-q)}$ of its orthogonal complement, the corresponding prediction proportion of unexplained variability in this case becomes:
\begin{align}\label{uqpred2}
u_{q}^{\mathrm{pred}} = \frac{\mathrm{tr}\big[\widehat{\mathbf{B}}_{p-q}^{\mathrm{T}} \, \mathbf{\Sigma} \, \, \widehat{\mathbf{B}}_{p-q} \big]}{\mathrm{tr}\big[ \mathbf{\Sigma} \big]}.
\end{align}
Comparing this prediction error with the optimal value, given by
\begin{align}\label{uqopt}
u_{q}^{\mathrm{opt}} = \frac{\sum_{j=1}^{p-q} \lambda_{j}}{\sum_{j=1}^{p} \lambda_{j}},
\end{align}
yields the relative prediction error
\begin{align}\label{relativeprederror}
e_{\mathrm{pred}} = \frac{u_{q}^{\mathrm{pred}}}{u_{q}^{\mathrm{opt}}} - 1.
\end{align}

%When the subspace dimension $q$ is unknown, it is often selected based on an estimate $\widehat{u}_{q}$ of the proportion of unexplained variability. To measure the accuracy of this estimate  \citet{Maronna2005} proposed to use the relative estimation error, given by
%\begin{align*}
%e_{\mathrm{est}} = \max \left(\frac{\widehat{u}_{q}}{u_{q}^{\mathrm{pred}}}, \frac{u_{q}^{\mathrm{pred}}}{\widehat{u}_{q}}\right) - 1.
%\end{align*}
%Since PPLTS and spherical PCA yield eigenvalue estimates $\widehat{\lambda}_{j}$, $j=1,\ldots,p$,   the estimator 
%\begin{align*}
%\widehat{u}_{q} = 
%\frac{\sum_{j=q+1}^{p} \widehat{\lambda}_{j}}{\sum_{j=1}^{p} \widehat{\lambda}_{j}},
%\end{align*}
%can be used for these methods. 
%For the robust subspace estimators,  \citet{Maronna2005} proposed the estimator
%\begin{align}\label{uqhat}
%\widehat{u}_{q} = \frac{\widehat{\sigma}_{q}^{2}}{\widehat{\sigma}_{0}^{2}}.
%\end{align}
%where $\widehat{\sigma}_{q}$ is the minimal scale in~(\ref{definitionMVS}) or~(\ref{definitionMVLTS}), respectively.  
%$\widehat{\sigma}_{0}$ is the corresponding minimal scale that is obtained for $q=p$ and thus yields a robust estimate of the total variability in the data.   

\begin{table*}[htb!]
\renewcommand*{\arraystretch}{1.2}
\centering
\setlength{\abovecaptionskip}{0pt} 
\scriptsize
\begin{center}
\caption{Mean of relative prediction errors: $\overline{e}_{\mathrm{pred}}$. }
\begin{tabular}{*{3}{p{0.2cm}} *{9}{>{\centering\arraybackslash} p{0.82cm}}} %{cc cccccccccccccccccc}
  \hline
Des. & $\epsilon$ & $k$ & SPC & S-M & RsubS & DsubS & PPLTS & S-L & RsubLTS & DsubLTS & ROBPCA \\ 
  \hline
a) & 0\% & & 0.04 & 0.02 & 0.02 & 0.02 & 0.32 & 0.07 & 0.07 & 0.06 & 0.03 \\ 
  &10\% & 1 & 0.04 & 0.03 & 0.03 & 0.03 & 0.41 & 0.10 & 0.11 & 0.08 & 0.03 \\ 
  & & 3 & 0.05 & 0.03 & 0.03 & 0.03 & 0.72 & 0.07 & 0.07 & 0.06 & 0.03 \\ 
  &  & 6 & 0.05 & 0.03 & 0.03 & 0.03 & 0.93 & 0.07 & 0.07 & 0.06 & 0.03 \\ 
  &20\% & 1 & 0.06 & 0.03 & 0.03 & 0.03 & 0.48 & 0.28 & 0.28 & 0.15 & 0.04 \\ 
  & & 1.5 & 0.09 & 0.03 & 0.03 & 0.03 & 0.78 & 1.18 & 0.96 & 0.09 & 0.05 \\ 
 & & 2 & 0.15 & 0.03 & 0.03 & 0.03 & 1.34 & 1.31 & 0.84 & 0.06 & 0.04 \\ 
  & & 3 & 0.44 & 0.17 & 0.03 & 0.03 & 1.86 & 0.11 & 0.11 & 0.06 & 0.03 \\ 
  & & 3.5 & 0.57 & 0.11 & 0.03 & 0.03 & 1.92 & 0.09 & 0.08 & 0.06 & 0.03 \\ 
  & & 4.5 & 0.77 & 0.03 & 0.03 & 0.03 & 1.80 & 0.07 & 0.07 & 0.06 & 0.03 \\ 
b) &   0\% & & 0.05 & 0.04 & 0.04 & 0.04 & 0.27 & 0.14 & 0.14 & 0.12 & 0.05 \\ 
  &10\% & 1 & 0.09 & 0.07 & 0.07 & 0.06 & 0.34 & 0.23 & 0.22 & 0.21 & 0.10 \\ 
  & & 1.5 & 0.13 & 0.09 & 0.08 & 0.07 & 0.40 & 0.24 & 0.20 & 0.16 & 0.14 \\ 
  & & 2 & 0.16 & 0.10 & 0.09 & 0.08 & 0.46 & 0.15 & 0.16 & 0.14 & 0.11 \\ 
  & & 4 & 0.19 & 0.05 & 0.05 & 0.05 & 0.49 & 0.11 & 0.12 & 0.12 & 0.06 \\ 
  & & 5 & 0.20 & 0.05 & 0.05 & 0.05 & 0.45 & 0.11 & 0.11 & 0.12 & 0.06 \\ 
  &20\% & 1.5 & 0.46 & 0.45 & 0.44 & 0.35 & 0.67 & 0.71 & 0.70 & 0.40 & 0.42 \\ 
  & & 2 & 0.55 & 0.67 & 0.66 & 0.45 & 0.78 & 0.75 & 0.74 & 0.28 & 0.33 \\ 
  & & 3 & 0.60 & 0.71 & 0.71 & 0.31 & 0.73 & 0.49 & 0.44 & 0.12 & 0.16 \\ 
 & & 3.5 & 0.61 & 0.71 & 0.70 & 0.24 & 0.68 & 0.24 & 0.19 & 0.11 & 0.11 \\ 
  & & 5 & 0.62 & 0.69 & 0.69 & 0.07 & 0.57 & 0.11 & 0.11 & 0.11 & 0.07 \\ 
   \hline
\end{tabular}
\label{table1}
\end{center}
\end{table*}
%We present the results of the simulations as in \citet{Maronna2005}.
Table \ref{table1} shows the mean of the relative prediction errors over $M=200$ samples. The values of $k$ included in Table \ref{table1} are those values at which some estimator attains its maximal error. Standard errors are not shown because they were small (below 0.08) in all cases.
\begin{figure*}[h!]
\vspace{-0.1cm}
\setlength{\abovecaptionskip}{0pt} 
\center
\begin{minipage}{.5\textwidth}
\centering
\includegraphics[width=\textwidth]{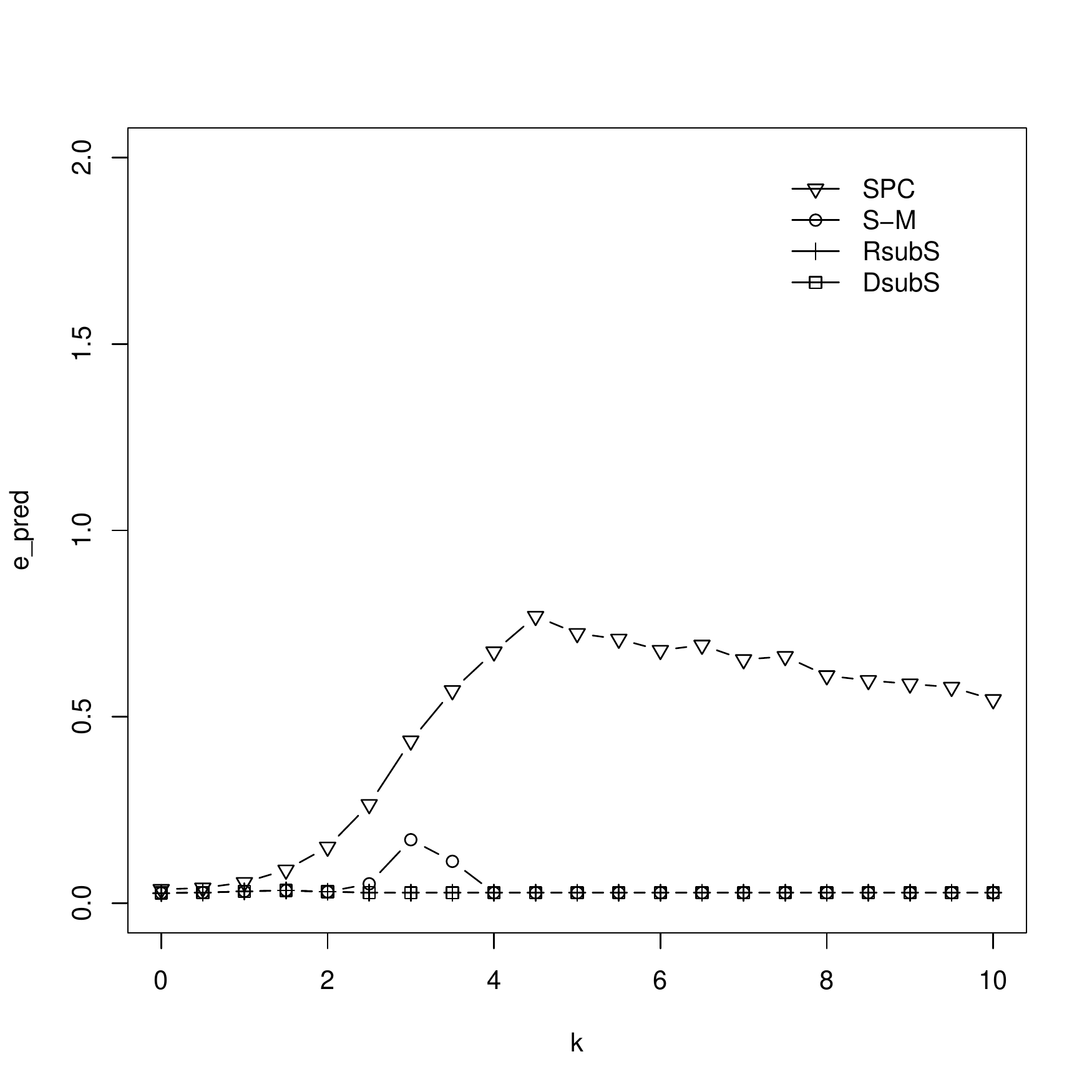}
\end{minipage}%
\begin{minipage}{0.5\textwidth}
\centering
\includegraphics[width=\textwidth]{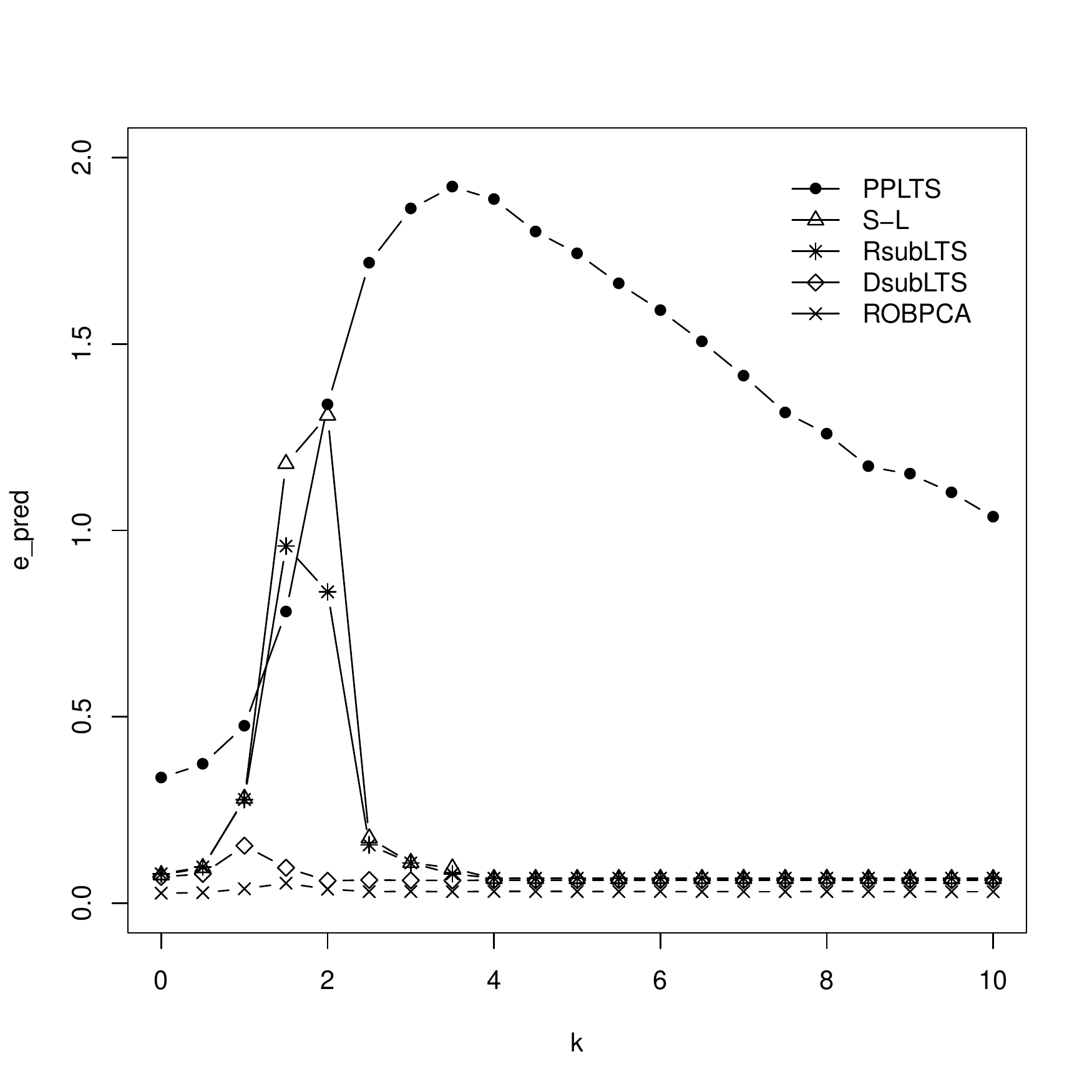}
\end{minipage}
\caption{Mean relative unexplained variance ($\overline{e}_{\mathrm{pred}}$) corresponding to the estimated 2 dimensional subspace  as a function of $k$ for eigenvalue configuration a) and $\epsilon=20\%$.}
\label{fig1chap1}
\end{figure*}
\begin{figure*}[h!]
\setlength{\abovecaptionskip}{0pt} 
\center
\begin{minipage}{.5\textwidth}
\centering
\includegraphics[width=\textwidth]{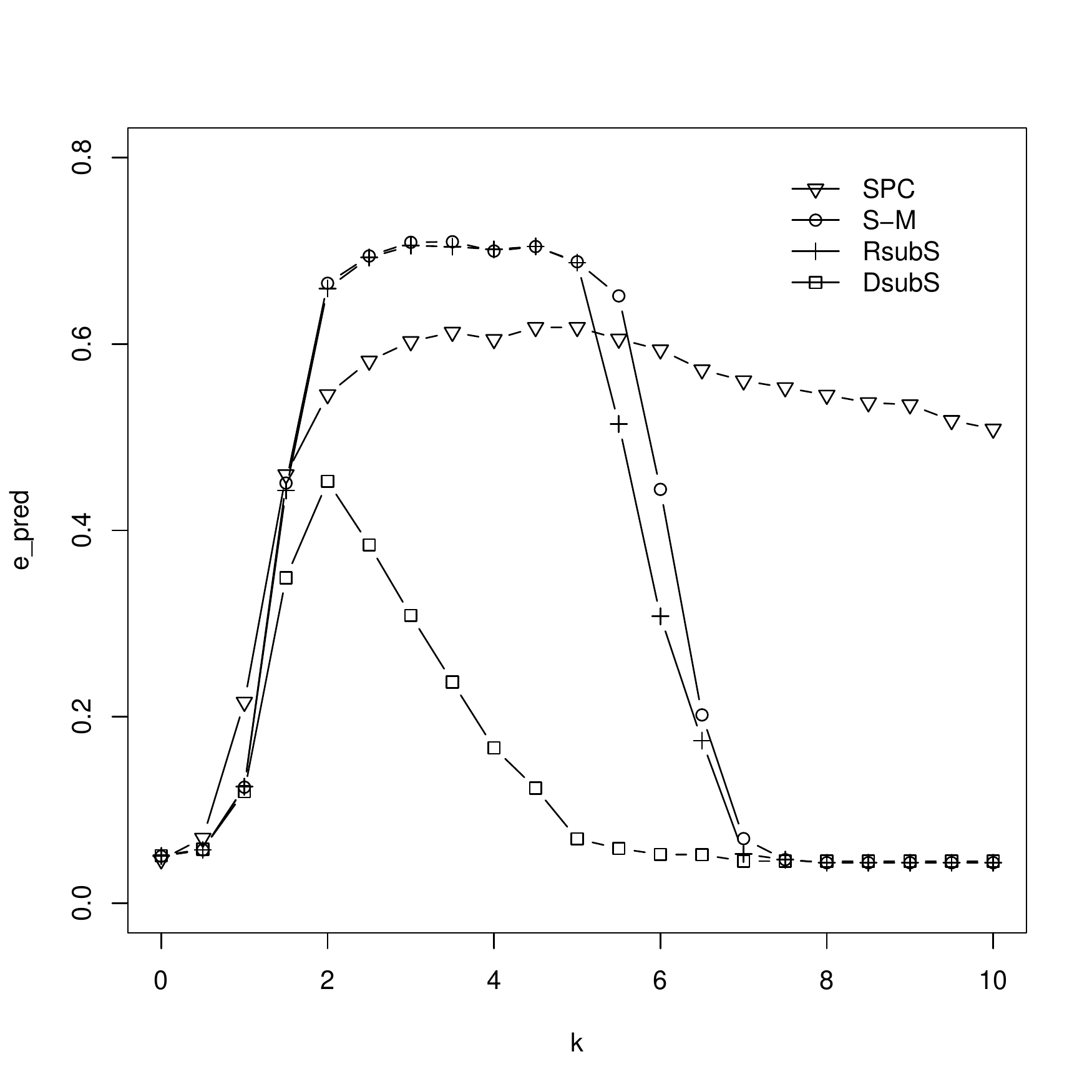}
\end{minipage}%
\begin{minipage}{0.5\textwidth}
\centering
\includegraphics[width=\textwidth]{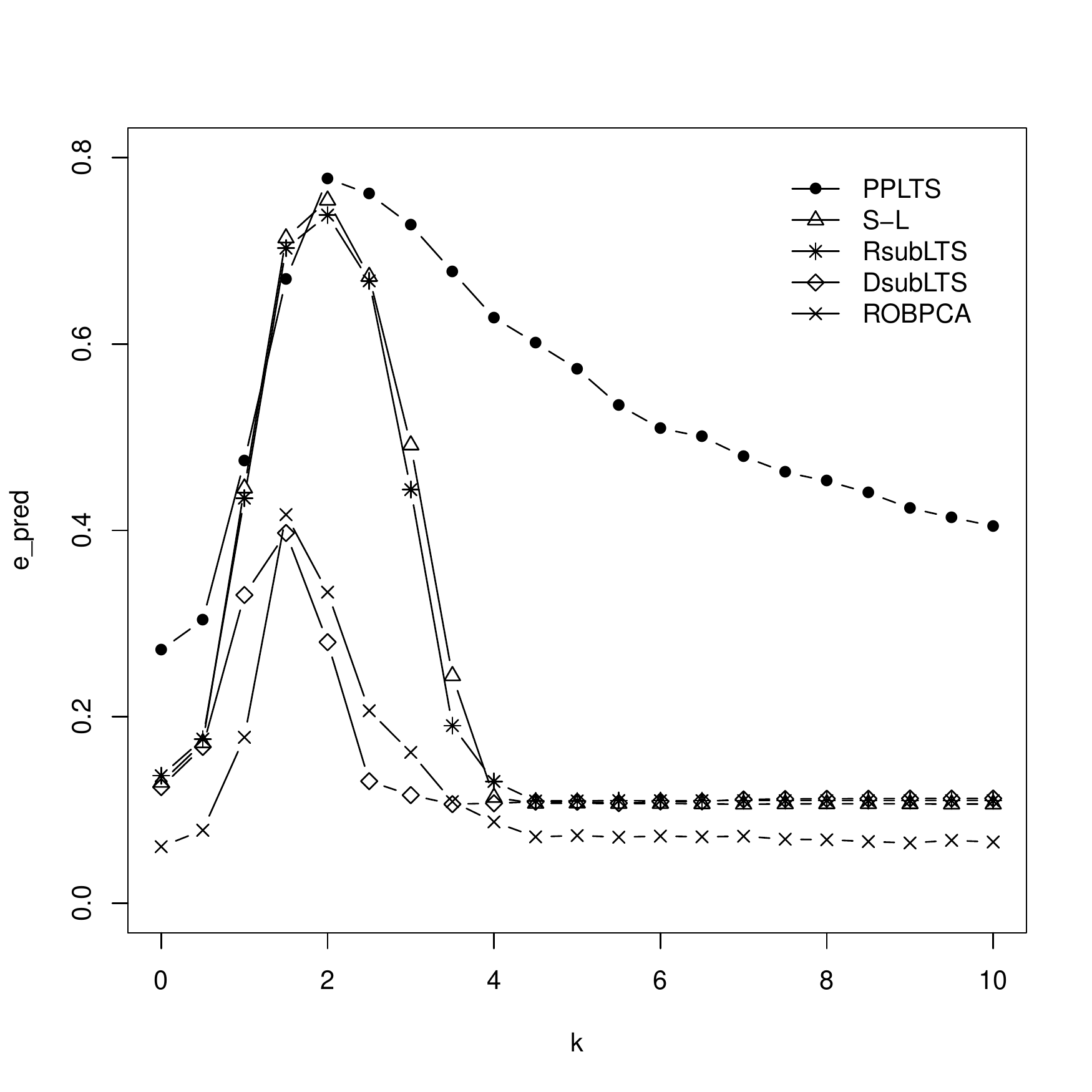}
\end{minipage}
\caption{Mean relative unexplained variance ($\overline{e}_{\mathrm{pred}}$) corresponding to the estimated 2 dimensional subspace  as a function of $k$ for eigenvalue configuration b) and $\epsilon=20\%$.}
\label{fig2chap1}
\end{figure*}
For $\epsilon=20\%$ of contamination, Figures \ref{fig1chap1} and \ref{fig2chap1} show the mean relative prediction errors as a function of $k$. From $k=10$ onwards these prediction errors stabilize so we only show results up to $k=10$. 
%\textcolor{red}{One can see as a first finding that the algorithms based on M-scales (i.e. S-M, RsubS, DsubS) are more efficient than the algorithms based on LTS-scales (i.e. PPLTS, S-L, RsubLTS, DsubLTS) when no contamination is present. ROBPCA, which is based on the MCD scatter estimator, also shows slightly lower efficiency than the algorithms based on M-scales for the case with only clean data. This result was expected as S-estimators yield in general more efficient estimates than those obtained by LTS estimators.} 
These results show that the estimates obtained by our new RsubS and RsubLTS algorithms are very similar to the results of Maronna's S-M and S-L algorithms, as expected. Moreover, the DsubS and DsubLTS algorithms often succeed in obtaining more robust estimates than their counterparts based on random starts. The advantage of using robust starting values is most pronounced when the outliers are at a small to moderate distance of the majority ($k \leq 6$) as can be seen from Figures \ref{fig1chap1} and \ref{fig2chap1}. Looking at the robust competitors, we can see that PPLTS turns out to be inefficient for $\epsilon=0\%$, similarly as for the PP methods considered by \citet{Maronna2005}. Moreover, also for contaminated data PPLTS gives worse results in these settings. Spherical PCA is generally better than PPLTS but its performance decreases considerably for $\epsilon=20\%$. ROBPCA performs well in these settings, which is in accordance with findings in \citet{Hubert2005,Engelen2005, robusttrimmedsub}. 
%Our deterministic algorithms show competitive results compared to ROBPCA. One can see that there is no consistent winner between DsubS, DsubLTS and ROBPCA across the different scenarios. However, DsubS often shows a better compromise between robustness and efficiency in these simulation settings.
Overall, ROBPCA and DsubS show the best performance in these simulations.

\section{Orthogonal equivariance}\label{equivariance}
In the previous section we showed that the RsubS/RsubLTS algorithms yield similar estimates as the original S-M/S-L algorithms while the DsubS/DsubLTS often performed even better. 
%Our deterministic algorithms also showed similar performance compared to ROBPCA, which is consider to be a strong robust competitor. 
Since orthogonal equivariance is important in the context of PCA, we now investigate the effect of orthogonal transformations on our deterministic algorithms by means of a simulation study. Similarly to \citet{MaronnaZamar2002}, let $\mathbf{X} \in \mathbb{R}^{n\times p}$ denote the data matrix and let $\mathbf{P} \in \mathbb{R}^{p\times p}$ be a random orthogonal matrix, i.e. $\mathbf{P}^{\mathrm{T}}\mathbf{P}=\mathbf{P}\mathbf{P}^{\mathrm{T}}=\mathbf{I}_{p}$. Let $ \widehat{\mathbf{B}}_{q}(\mathbf{X})$ denote an estimate of the first $q$ principal component directions for the data matrix $\mathbf{X}$. If the estimator is orthogonal equivariant, then it must hold that:
\begin{align*}
\widehat{\mathbf{B}}_{q}(\mathbf{X}\mathbf{P}^{\mathrm{T}}) = \mathbf{P} \widehat{\mathbf{B}}_{q}(\mathbf{X}).
\end{align*}
To investigate orthogonal equivariance, we therefore compare $\widehat{\mathbf{B}}_{q}(\mathbf{X})$ to
\begin{align*}
\widehat{\mathbf{B}}_{q}^{\prime}(\mathbf{X}) = \mathbf{P}^{\mathrm{T}} \widehat{\mathbf{B}}_{q}(\mathbf{X}\mathbf{P}^{\mathrm{T}}).
\end{align*}
To measure the effect of the transformation we calculate the standardized last principal angle between the subspaces generated by the columns of $\widehat{\mathbf{B}}_{q}$ and $\widehat{\mathbf{B}}_{q}^{\prime}(\mathbf{X})$, using the algorithm of \citet{BjorckGolub1973}. We divide this angle by $\pi/2$ to obtain values between 0 and 1. Values closer to 0 indicate that the subspaces are more comparable. 

\begin{figure*}[h!]
\vspace{-0.1cm}
\setlength{\abovecaptionskip}{0pt} 
\center
\begin{minipage}{.5\textwidth}
\centering
\includegraphics[width=\textwidth]{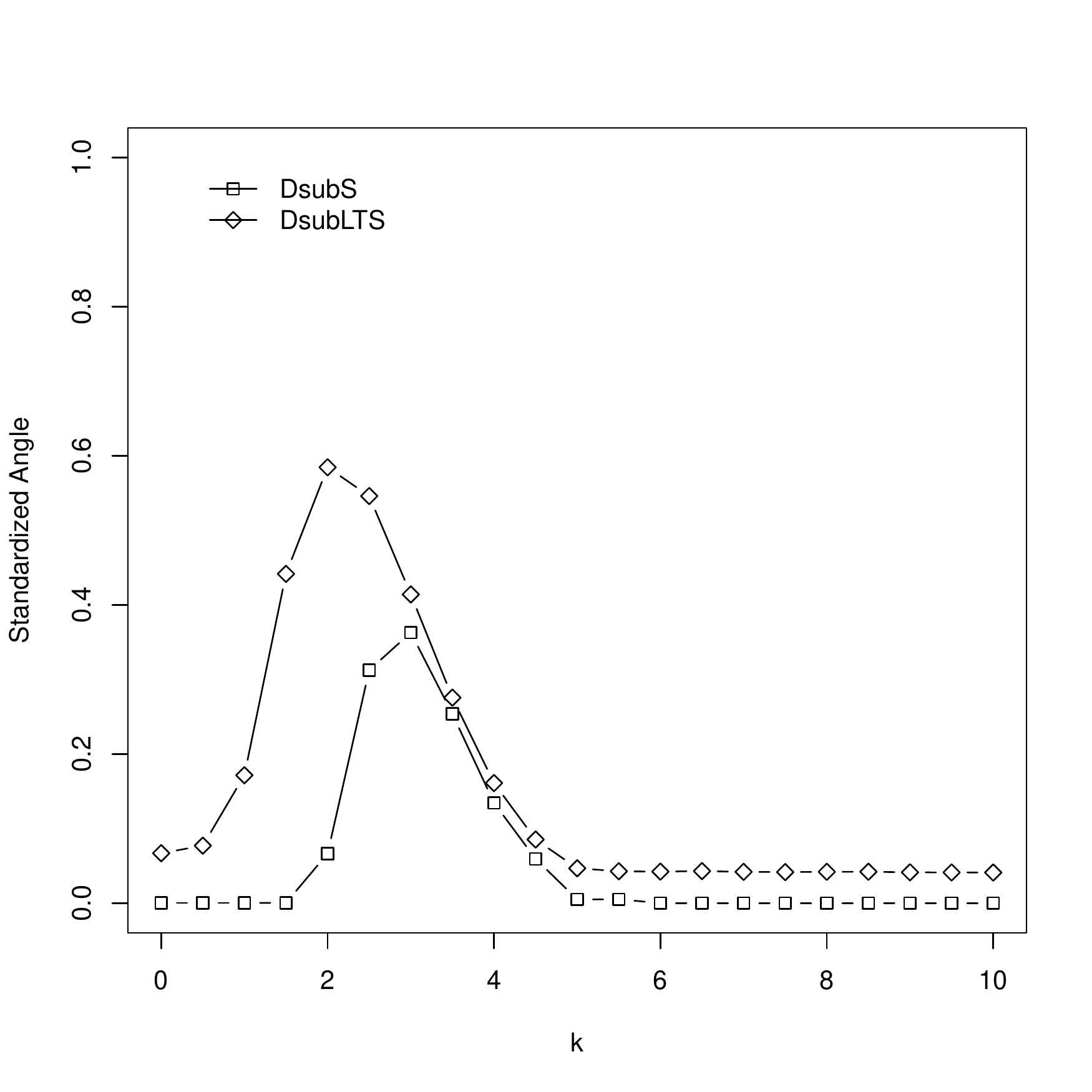}
\end{minipage}%
\begin{minipage}{0.5\textwidth}
\centering
\includegraphics[width=\textwidth]{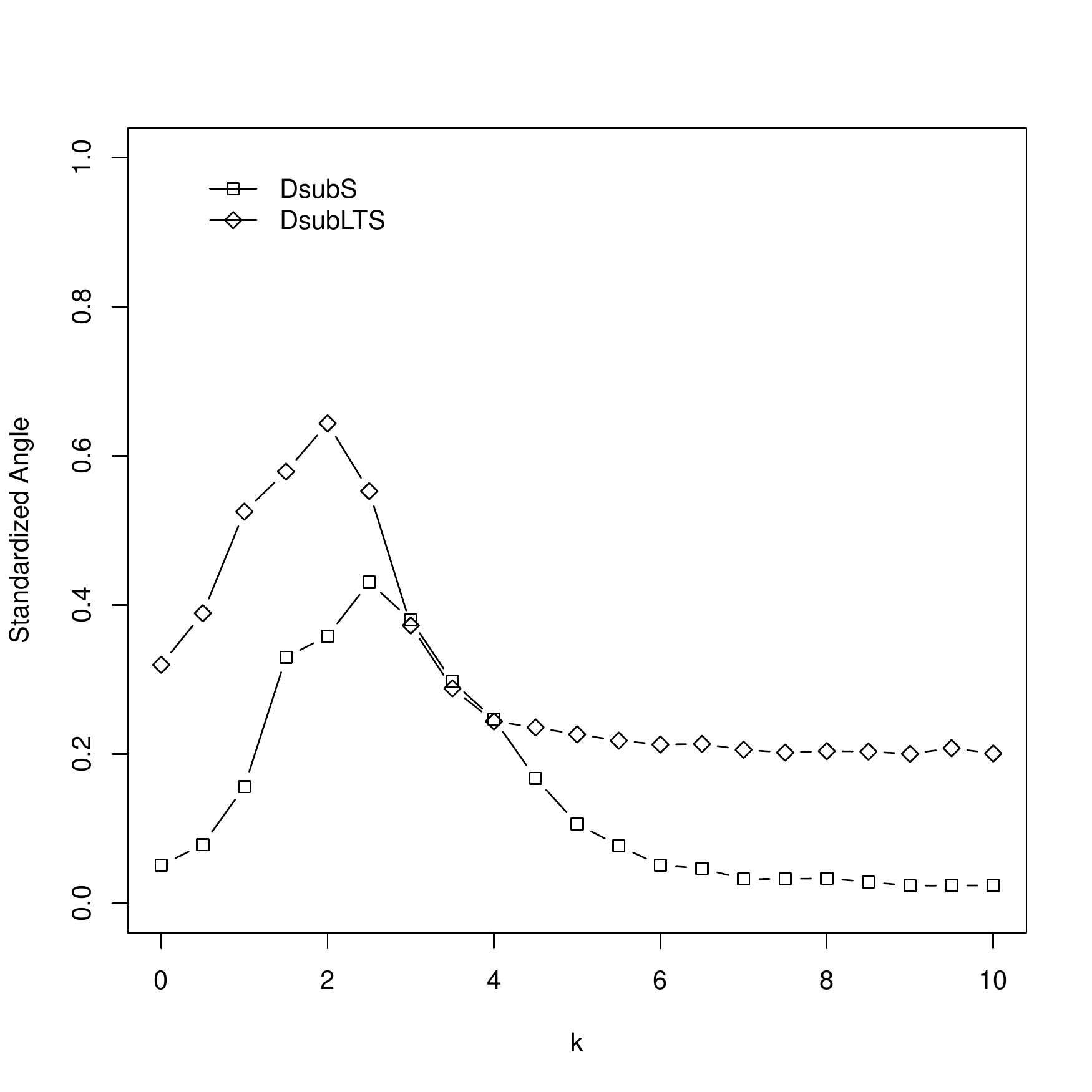}
\end{minipage}
\caption{Mean standardized angle corresponding to the estimated 2 dimensional subspace as a function of $k$ for eigenvalue configuration a) in the left panel and configuration b) in the right panel. In both cases $\epsilon=20\%$ of contamination was introduced.}
\label{orthoequiv}
\end{figure*}

We generated data matrices $\mathbf{X}$ according to designs a) and b) in Section \ref{sec:simulation} with $\epsilon=20\%$ and estimate the best $q=2$ linear dimensional subspace. We generated $M=200$ replicates and for each data replicate we generated a random orthogonal matrix $\mathbf{P}$ using the method of \citet{Stewart1980}. Figure \ref{orthoequiv} displays the mean standardized angles as a function of $k$ for both designs.
We can see that for both DsubS and DsubLTS the effect of the transformation is largest when the outliers are at a small to moderate distance of the regular data. In these settings the DsubS algorithm is also clearly less affected by orthogonal transformations than DsubLTS and the effect quickly disappears when the outlier distance increases.

\section{Computation time}\label{simcomptime}

We now compare the computation times of our algorithms to those of the original algorithms of Maronna  and the other methods considered in Section \ref{sec:simulation}. For this purpose we generalize the data generating model in design a). More specifically, we generate clean data from the $p$-dimensional normal distribution with center zero and diagonal covariance matrix. The eigenvalues on the diagonal are set equal to
\begin{equation}
\lambda_{j} =
\begin{cases}
1+c_{p,q}\,j  & \text{ for } 1 \leq j \leq (p-q), \\ 
20(1+0.5(j-p+q)) & \text{ for } (p-q+1) \leq j \leq p.
\end{cases}
\label{HD_eigenvalues}
\end{equation}

The constant $c_{p,q}$ is chosen such that the $q$ main directions explain about $80\%$ of the total variability.
We consider sample sizes $n=1000$ or $5000$ in dimension $p=10, 500, 1000$ with $q=2, 5$. The data contain $\epsilon=20\%$ of contamination, generated as explained in Section \ref{sec:simulation} with $k=15$. 

All algorithms are implemented in \verb|R|~\citep{Rcore}. 
For the S-M, S-L, RsubS, RsubLTS, DsubS, DsubLTS and PPLTS algorithms we used our own implementation. For S-M and S-L we use the \verb|R| function \verb|eigen()| to calculate the smallest $p-q$ eigenvectors of the weighted covariance matrix in~(\ref{weightedcov}). This function uses LAPACK~\citep{lapack} routines which are written in FORTRAN. Alternatives such as the \verb|R| function \verb|svd()| for instance could be used, but the speed difference is generally small. For spherical PCA we use the function \verb|PcaLocantore()| and for ROBPCA we use the function \verb|PcaHubert()|, both in \verb|R| package \verb|rrcov|~\citep{rrcov}. \verb|PcaLocantore()| also uses \verb|svd()| or \verb|eigen()| to compute classical PCA on the projected data onto the unit sphere while \verb|PcaHubert()| uses a FORTRAN implementation to compute the MCD estimator of location and scatter on the projected data and \verb|eigen()| to compute its decomposition. To allow a fair comparison of computation times between these algorithms and our new algorithms, for our algorithms we implemented the iterative updating of the estimates based on expressions (\ref{esteqMVS1})-(\ref{esteqMVS3}) in C++,  using \verb|R| package \verb|RcppArmadillo|~\citep{RcppArmadillo}. Note that the implementations of all methods in the comparison use plain \verb|R|, except for the calculation of covariance matrices and/or its decomposition. S-M, S-L, spherical PCA and ROBPCA use FORTRAN implementations for this purpose while our algorithms perform the iterative updates in C++. The algorithms were compared on a single Intel i7 CPU (3.4GHz) machine running Windows 7.  

Table \ref{comptime} shows the computation time of the algorithms in seconds, averaged over $M=50$ replications. 
\begin{table*}[h!]
\centering
\caption{Computation times in seconds}
\begin{small}
\begin{tabular}{llrrrcrrr}
  \hline
  \hline
& & \multicolumn{3}{c}{$n=1000$} & & \multicolumn{3}{c}{$n=5000$} \\ \cline{3-5}\cline{7-9}
&$p$ & 10 & 500 & 1000 && 10  & 500 & 1000\\   \cline{2-9}  
& SPC & 0.04 & 1.34 & 4.19 && 0.16 & 5.67 & 21.59 \\
 & S-M & 2.47 & 121.55 & 567.92&& 39.78 & 1178.71 & 3195.05 \\ 
 & RsubS & 0.61 & 25.32 & 154.42 && 3.06 & 57.18 & 250.19 \\ 
 $q=2$  & DsubS & 0.19 & 3.33 & 6.70 && 1.00 & 16.46 & 33.87 \\ 
 & PPLTS & 5.75 & 25.91 & 52.97 && 142.07 & 703.32 & 1372.31 \\ 
 & S-L & 0.60 & 95.46 & 502.53 && 2.66 & 288.50 & 1383.55 \\ 
  & RsubLTS & 0.90 & 24.70 & 152.38 && 4.90 & 52.32 & 243.24 \\ 
& DsubLTS & 0.27 & 3.01 & 6.02 && 1.31 & 15.04 & 31.29 \\ 
& ROBPCA & 1.02 & 3.08 & 6.67 && 4.47 & 13.16 & 40.37 \\
\hline   
& SPC & 0.04 & 1.34 & 4.21 &&  0.18 & 5.74 & 22.01 \\ 
 & S-M & 2.33 & 119.93 & 569.61 && 35.00 & 1170.10 & 3206.02 \\ 
 & RsubS & 0.94 & 28.70 & 161.39 && 4.63 & 74.42 & 281.46 \\ 
 $q=5$ & DsubS & 0.30 & 4.53 & 9.08 &&  1.53 & 21.88 & 44.95 \\ 
 & PPLTS & 14.61 & 64.40 & 122.36 && 356.52 & 1774.19 & 3385.82 \\ 
 & S-L & 0.58 & 95.11 & 507.15 &&  2.68 & 300.09 & 1369.12 \\  
  & RsubLTS & 1.70 & 28.80 & 161.07 &&  9.25 & 75.68 & 282.14 \\
 & DsubLTS & 0.45 & 4.12 & 8.29 &&  2.25 & 20.35 & 41.36 \\ 
 & ROBPCA & 1.04 & 3.07 & 6.65 && 4.65 & 13.55 & 40.78 \\
\hline   
\hline
\end{tabular}
\end{small}
\label{comptime}
\end{table*}
Not surprisingly, we see that spherical PCA is the fastest to compute because it does not require any iterative 
process~\citep[see also][]{Maronna2005,Wilcox2008}.  
For $n=1000$ and $q=2$ projection-pursuit is relatively fast to compute regardless of the dimension of the data, as can be seen from Table \ref{comptime}. However, the computation time grows quickly with increasing subspace dimension (case $q=5$) and/or sample size (case $n=5000$).
 The computation time of Maronna's algorithms is reasonable for low dimensional data, but increases quickly when the dimension grows. This result was expected because these algorithms need to compute a large number ($p-q$) of eigenvectors of a high-dimensional covariance matrix which is very time-consuming. 
Clearly, our implementations of the new algorithms with random starting values are faster than Maronna's original algorithms, but computation time still increases quickly with growing dimension. 
The new algorithms with deterministic starting values are even faster and their computation time increases much slower when the dimension grows. 

When using 250 random directions, the computation time of ROBPCA is comparable to our deterministic algorithms for the different sample sizes and subspace dimensions considered. Increasing the number of directions considered to compute the measure of outlyingness makes  ROBPCA much slower (results not shown here), especially for the case $n=5000$ and $q=5$. Thus, computation time of ROBPCA critically depends on the number of directions considered. These results indicate that the algorithms with deterministic starting values do not only show good performance but are also computationally attractive. They make it possible to compute robust subspace estimates for a large scale of problems.

\section{Performance for high dimensional data}\label{highdimsim}

In Sections \ref{sec:simulation} and \ref{simcomptime} we have shown that DsubS and DsubLTS show good performance for low-dimensional data and remain computationally attractive for high-dimensional data. We now investigate whether their good performance also carries over to the high-dimensional setting. 
To this end we consider the model of the previous section with eigenvalues according to~(\ref{HD_eigenvalues}) for  $q=2$ or $q=5$, and $\epsilon=20\%$ of contamination generated as in Section~\ref{sec:simulation}. We let $k$ range between 0 and $20$ as before and generated data of size $n=1000$ in dimension $p=1000$ or $p=10000$. 

 For $p=1000$ we did not include the S-L and S-M algorithms in the comparison because their performance was similar to the RsubS and RsubLTS algorithms, but computationally more demanding. Moreover, we did not consider the RsubS and RsubLTS algorithms in the case $p=10000$ because these algorithms require too much computation time in these settings. %The reason is that the algorithms of Maronna require the computation of a high-dimensional covariance followed by its decomposition while our RsubS and RsubLTS algorithms needs a reasonable number of starting values to attempt a good approximation to the global minimum. 
SPC implemented with the \verb|R| function \verb|PcaLocantore()| also required excessive computation time for the case $p=10000$. Therefore, we wrote an alternative SPC \verb|R| function that computes classical PCA on the projected data using Algorithm \ref{PCAsub}. 

\begin{figure*}[h!]
\setlength{\abovecaptionskip}{0pt} 
\center
\begin{minipage}{.52\textwidth}
\centering
\includegraphics[width=\textwidth]{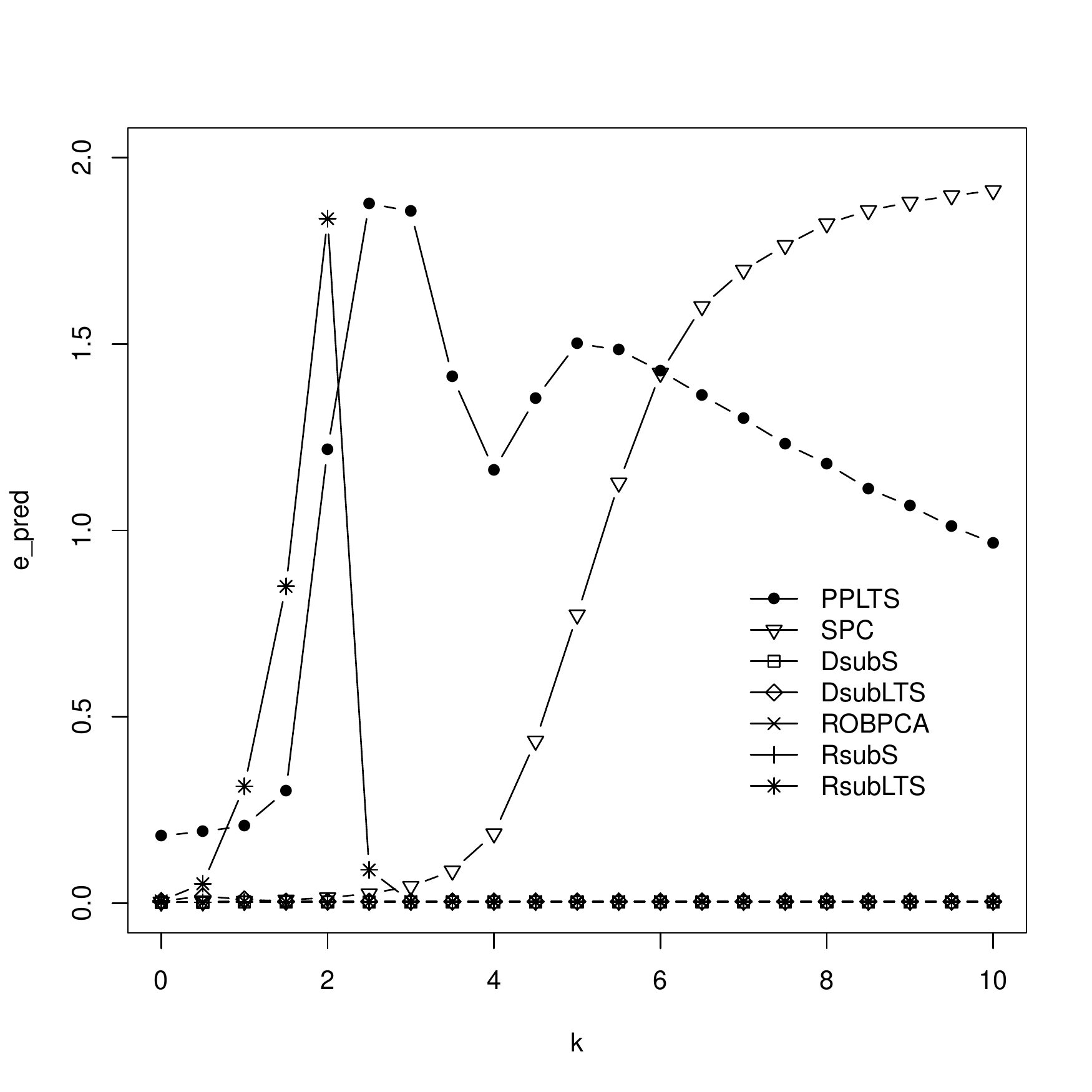}
\end{minipage}%
\begin{minipage}{0.52\textwidth}
\centering
\includegraphics[width=\textwidth]{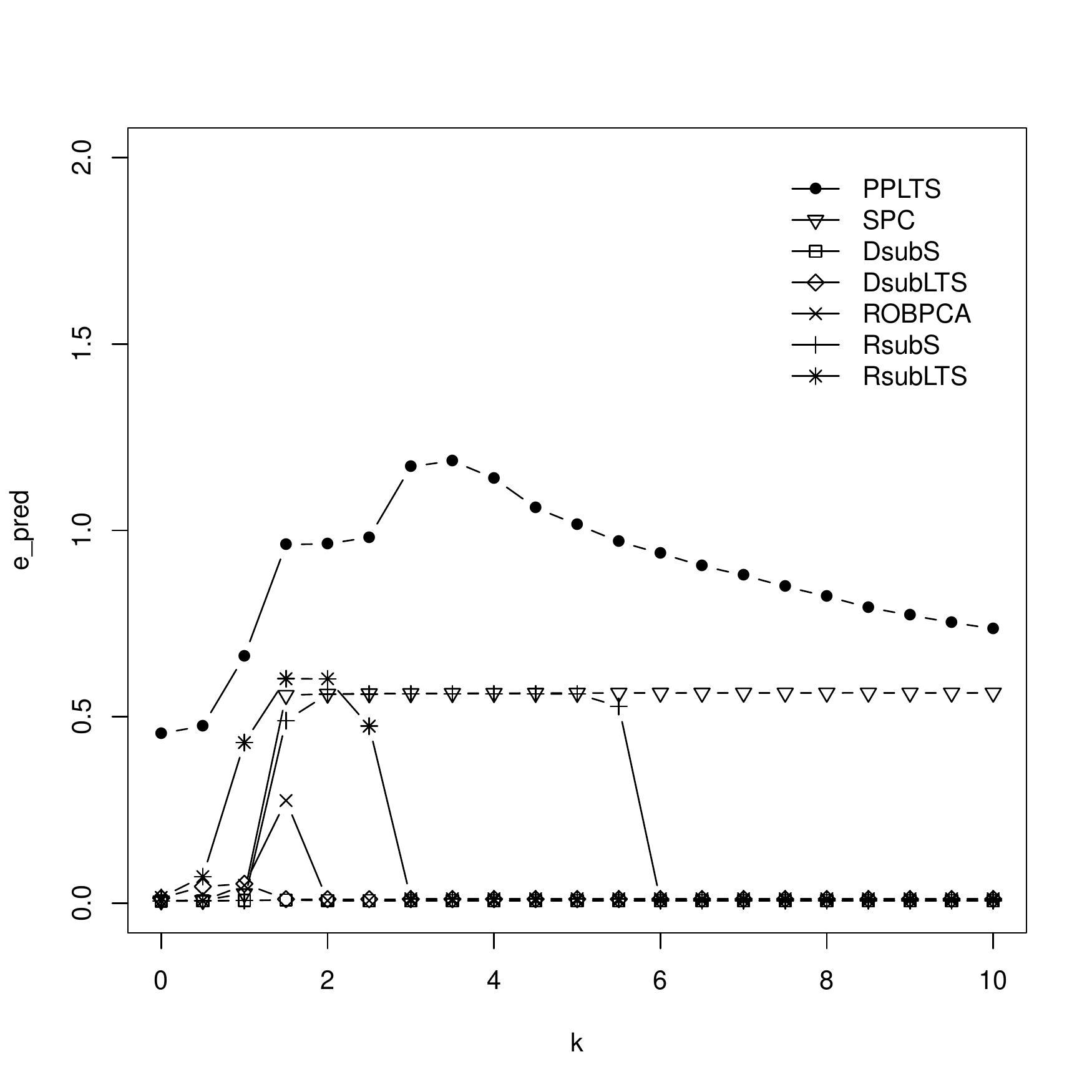}
\end{minipage}
\caption{Mean relative prediction errors $\overline{e}_{pred}$ for the case $n=1000$ and $p=1000$ with $q=2$ subspace estimation (left panel) and $q=5$ subspace estimation (right panel). In both cases the level of contamination is $\epsilon=20\%$.}
\label{fig4chap1}
\end{figure*}

\begin{figure*}[h!]
\setlength{\abovecaptionskip}{0pt} 
\center
\begin{minipage}{.52\textwidth}
\centering
\includegraphics[width=\textwidth]{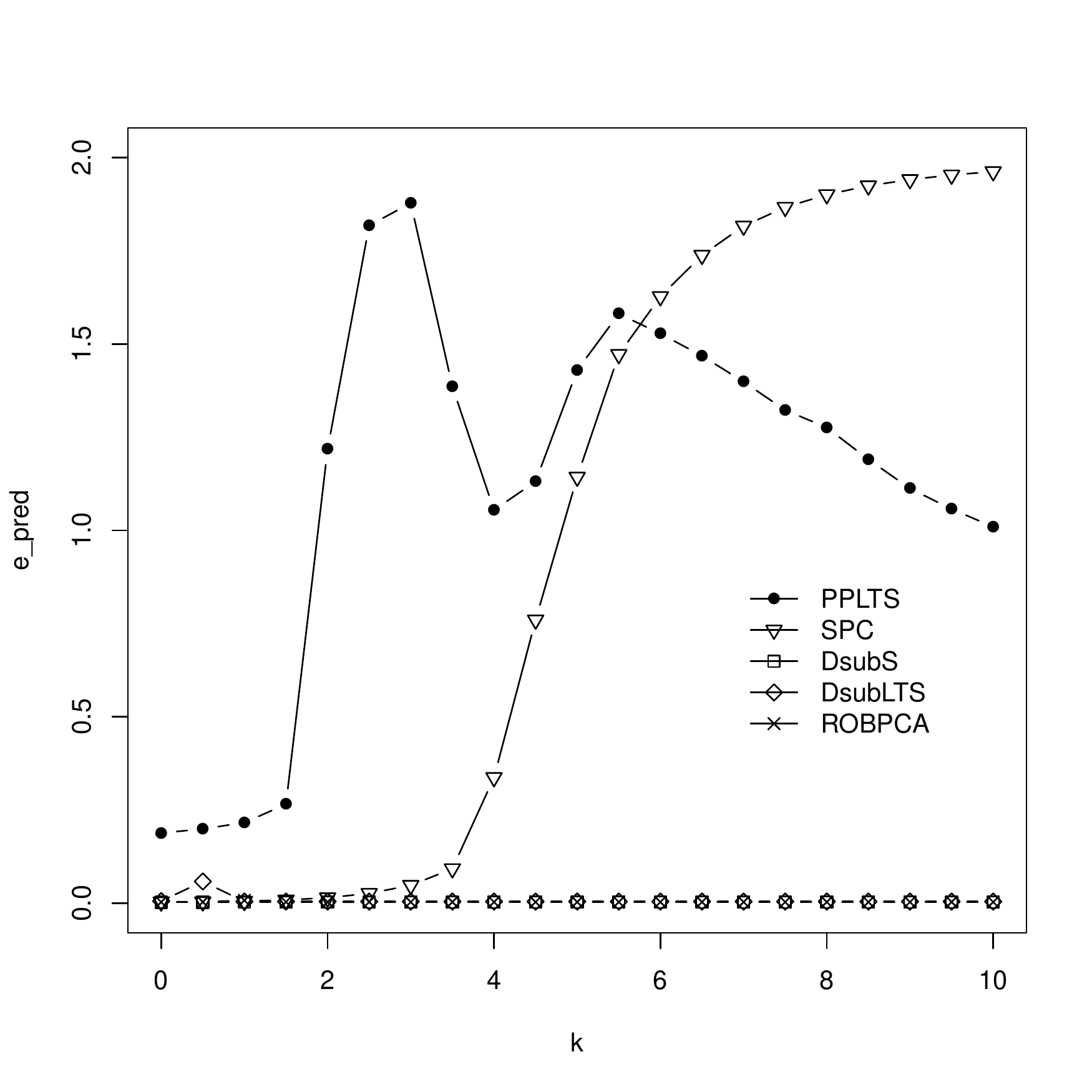}
\end{minipage}%
\begin{minipage}{0.52\textwidth}
\centering
\includegraphics[width=\textwidth]{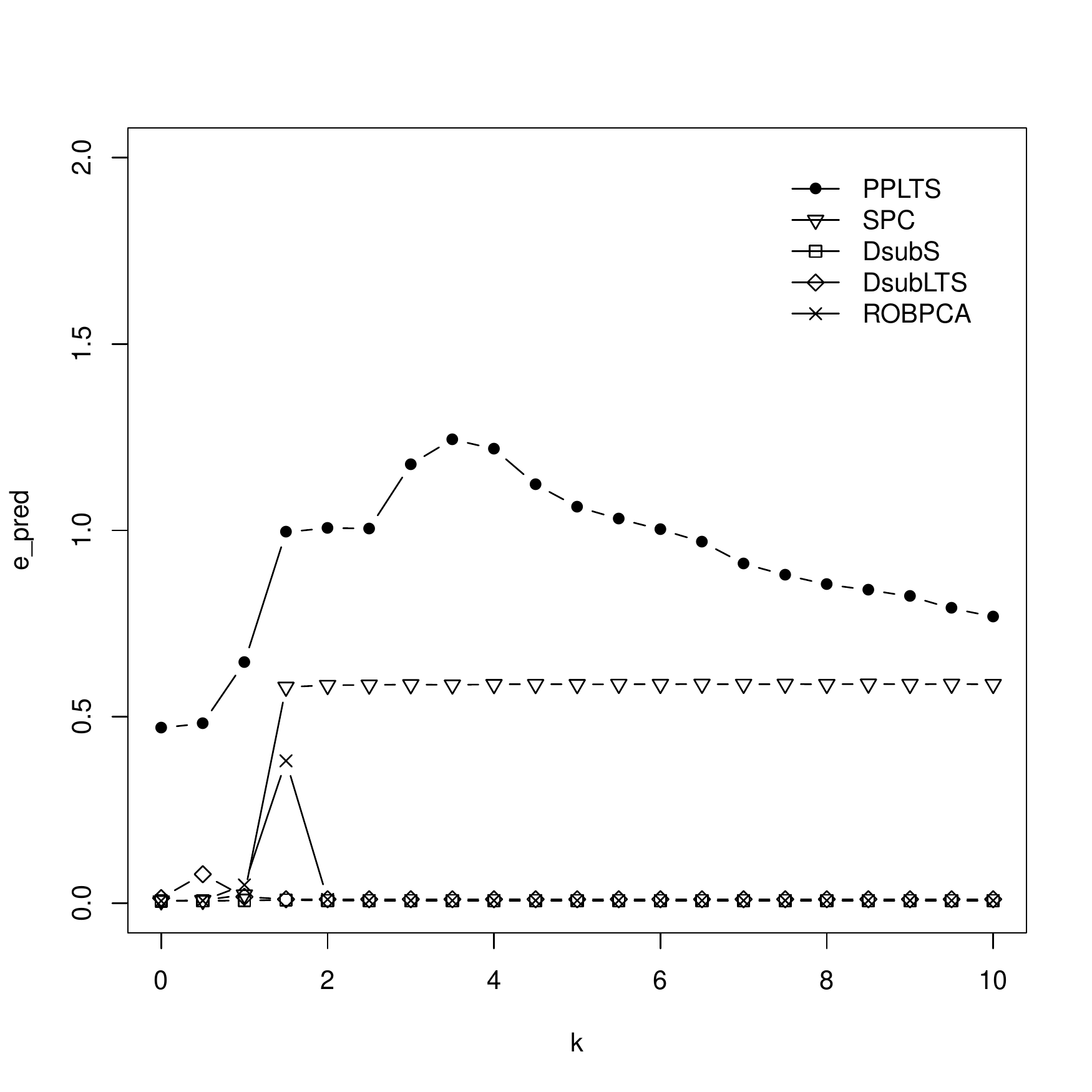}
\end{minipage}
\caption{Mean relative prediction errors $\overline{e}_{pred}$ for the case $n=1000$ and $p=10000$ with $q=2$ subspace estimation (left panel) and $q=5$ subspace estimation (right panel). In both cases the level of contamination is $\epsilon=20\%$.}
\label{fig4chap2}
\end{figure*}

Figures~\ref{fig4chap1} and \ref{fig4chap2} show the mean relative prediction errors, averaged over $M=200$ samples, when $p=1000$ and $p=10000$, respectively. Relative prediction errors again stabilized from $k=10$ onwards, so we only present results up to $k=10$. From Figure~\ref{fig4chap1} we can see that in higher dimensions DsubS and DsubLTS behave similar as for low-dimensional data and clearly perform better than SPC and PPLTS. Figure~\ref{fig4chap2} shows that DsubS and DsubLTS keep their excellent performance when the dimension is increased further to $p=10000$. The performance of ROBPCA is mostly similar to our deterministic algorithms in this case, but it has slightly more difficulty to estimate the best $q=5$ dimensional subspace when the outliers are at a small distance ($k=1.5$) from the regular data. Similar conclusions were obtained for sample size $n=100$ so these results are not shown.
\section{Real data example}\label{Examplech1}
\begin{figure}[h!]
\begin{center}$
\begin{array}{ccccc}
  \includegraphics[width=2cm]{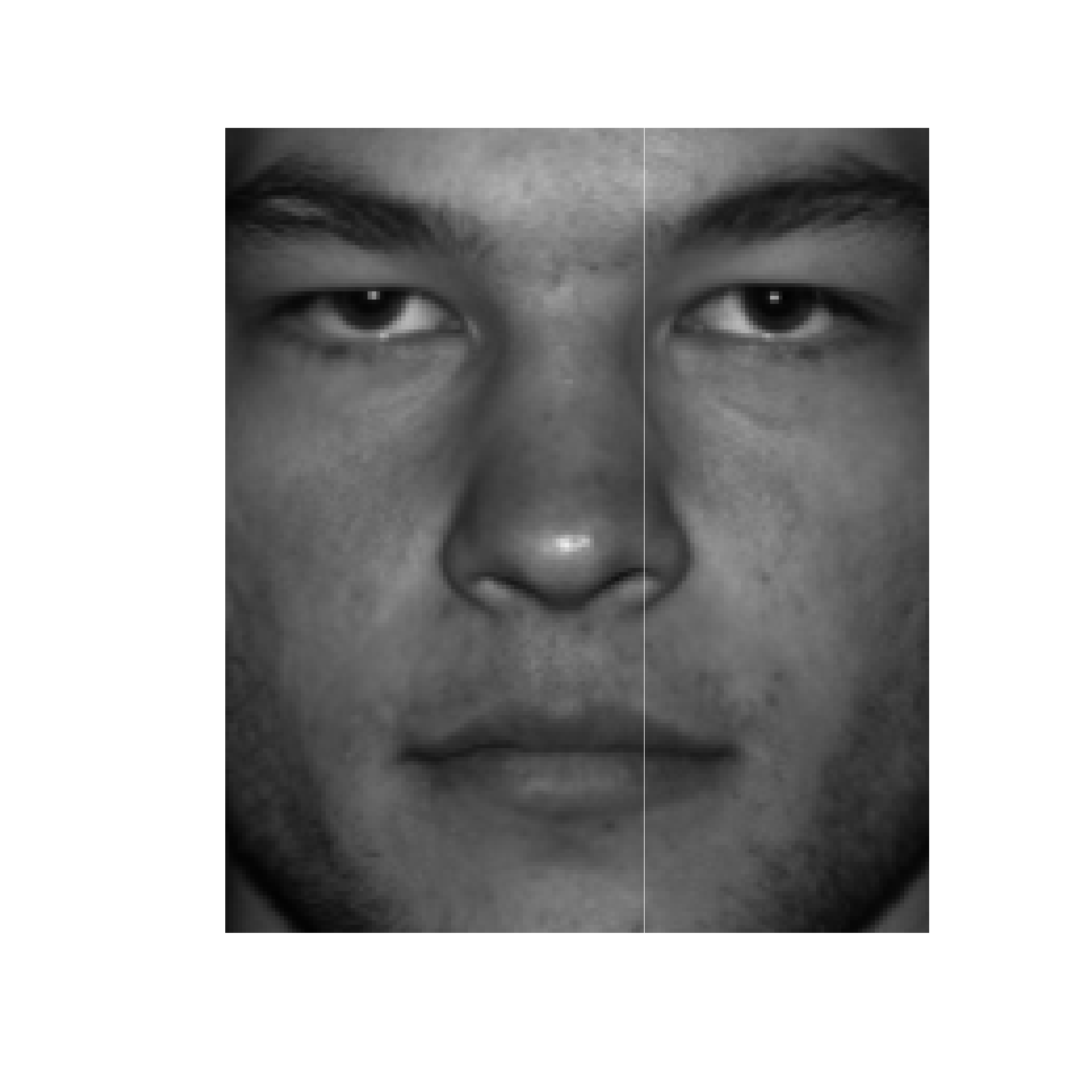} \hspace*{-0.5cm} &   
  \includegraphics[width=2cm]{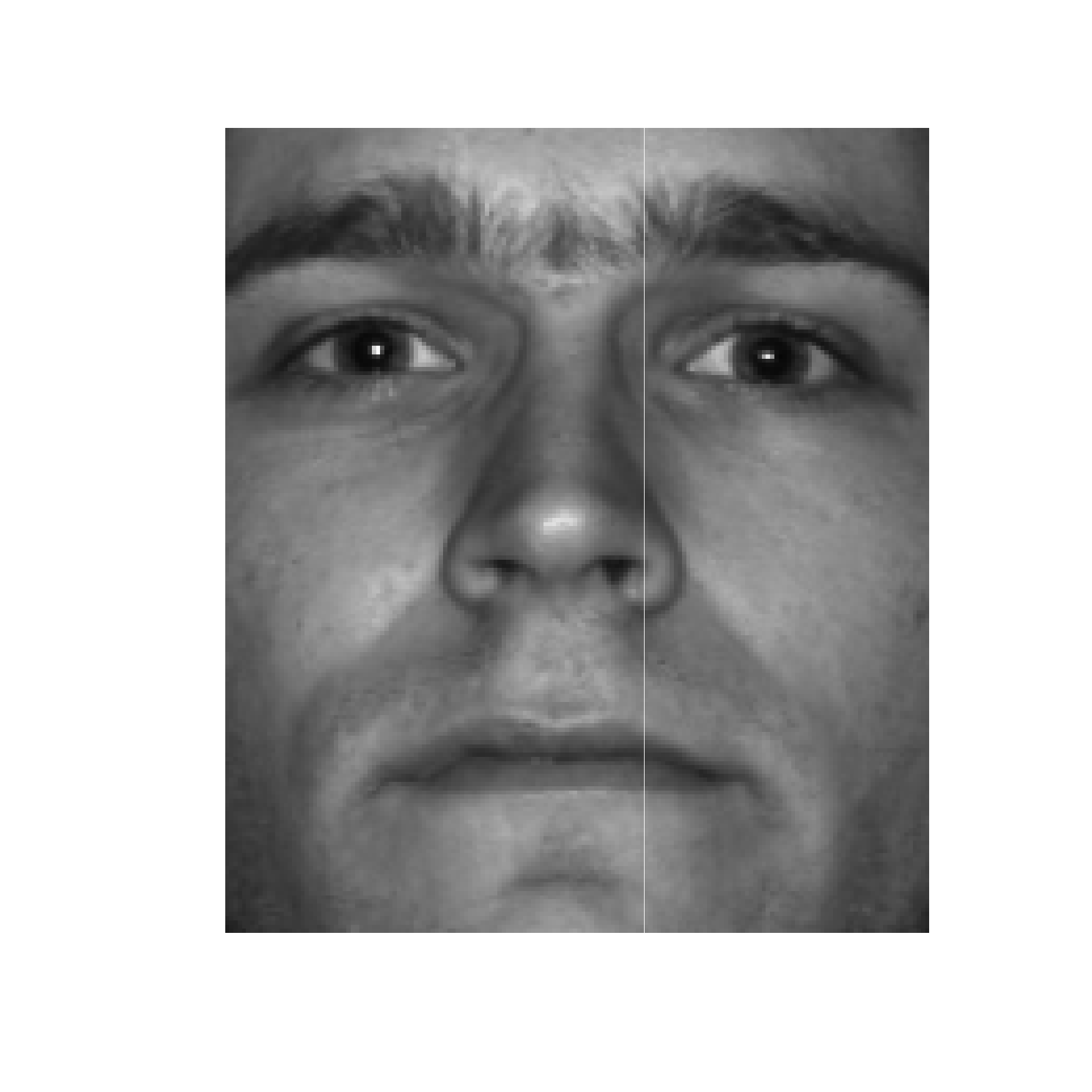} \hspace*{-0.5cm} &
 \includegraphics[width=2cm]{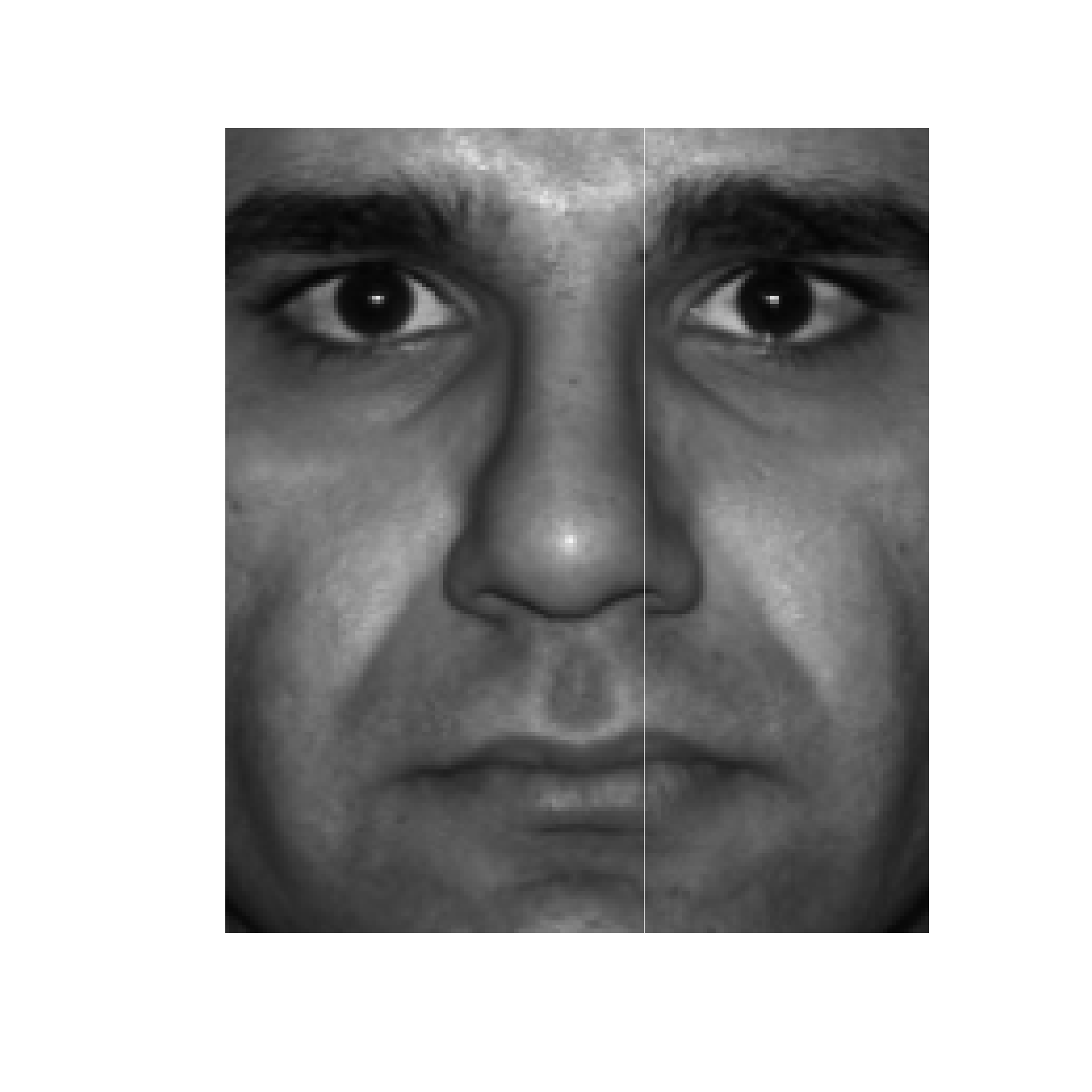} \hspace*{-0.5cm} &   
 \includegraphics[width=2cm]{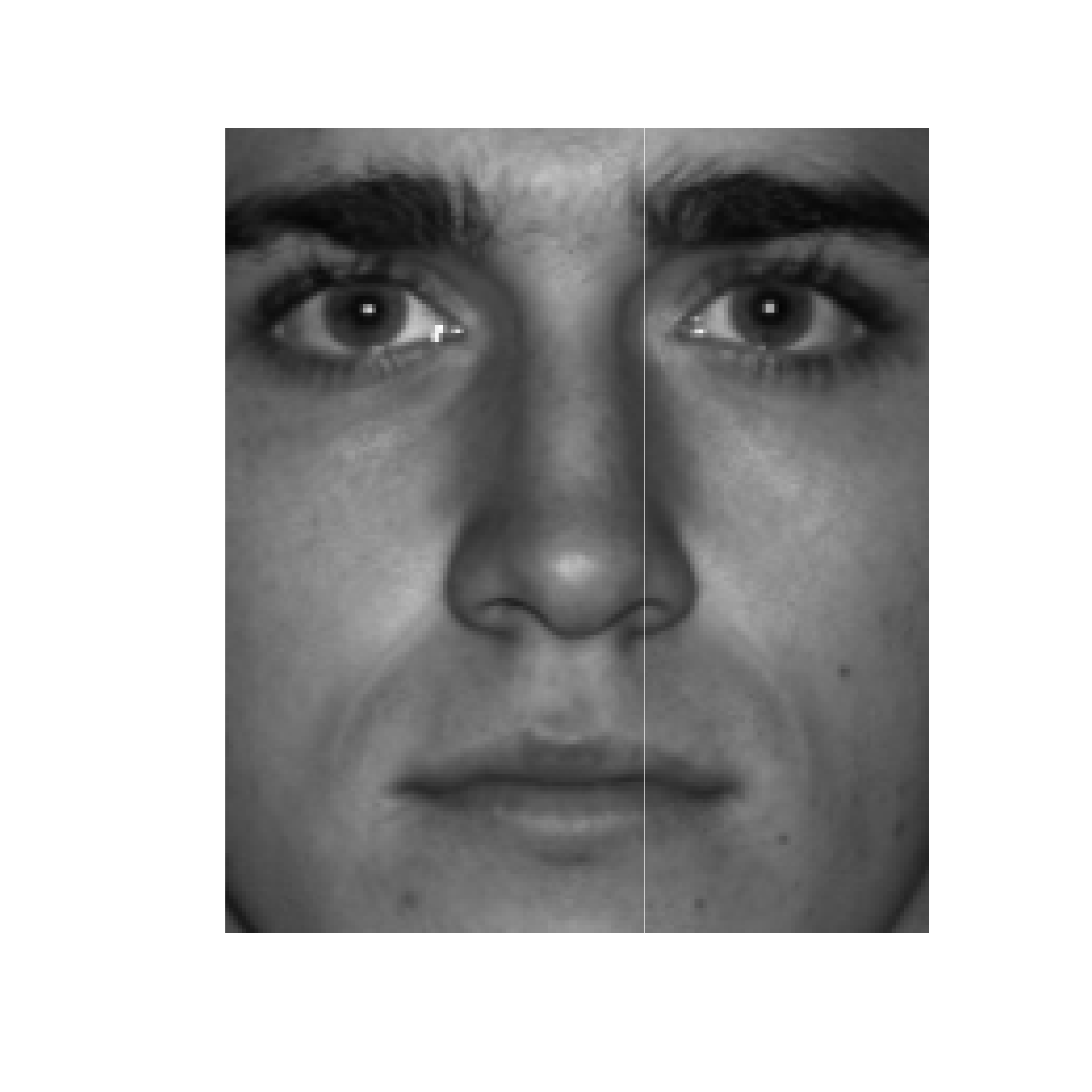} \hspace*{-0.5cm} &
\includegraphics[width=2cm]{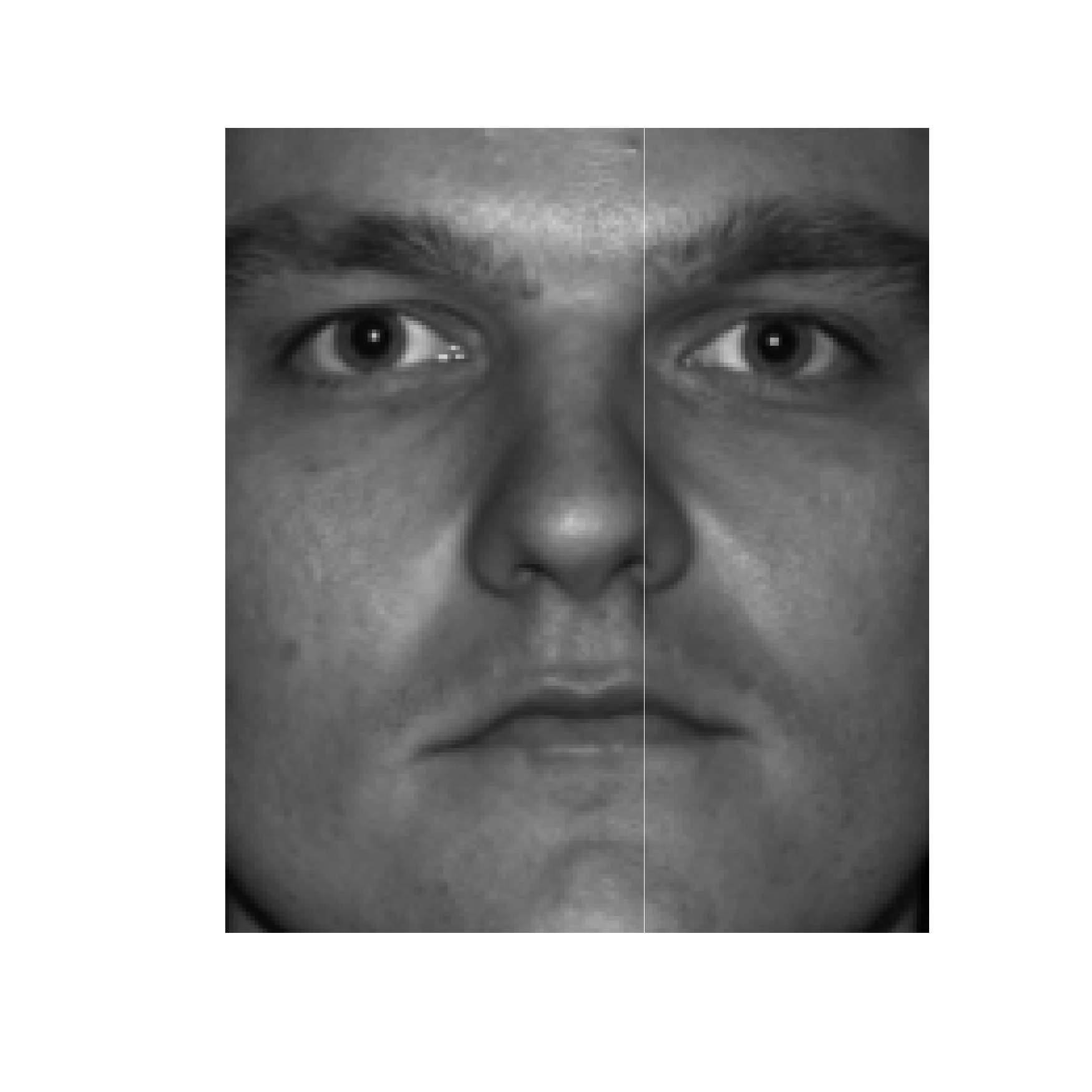} \hspace*{-0.5cm} \\
 \includegraphics[width=2cm]{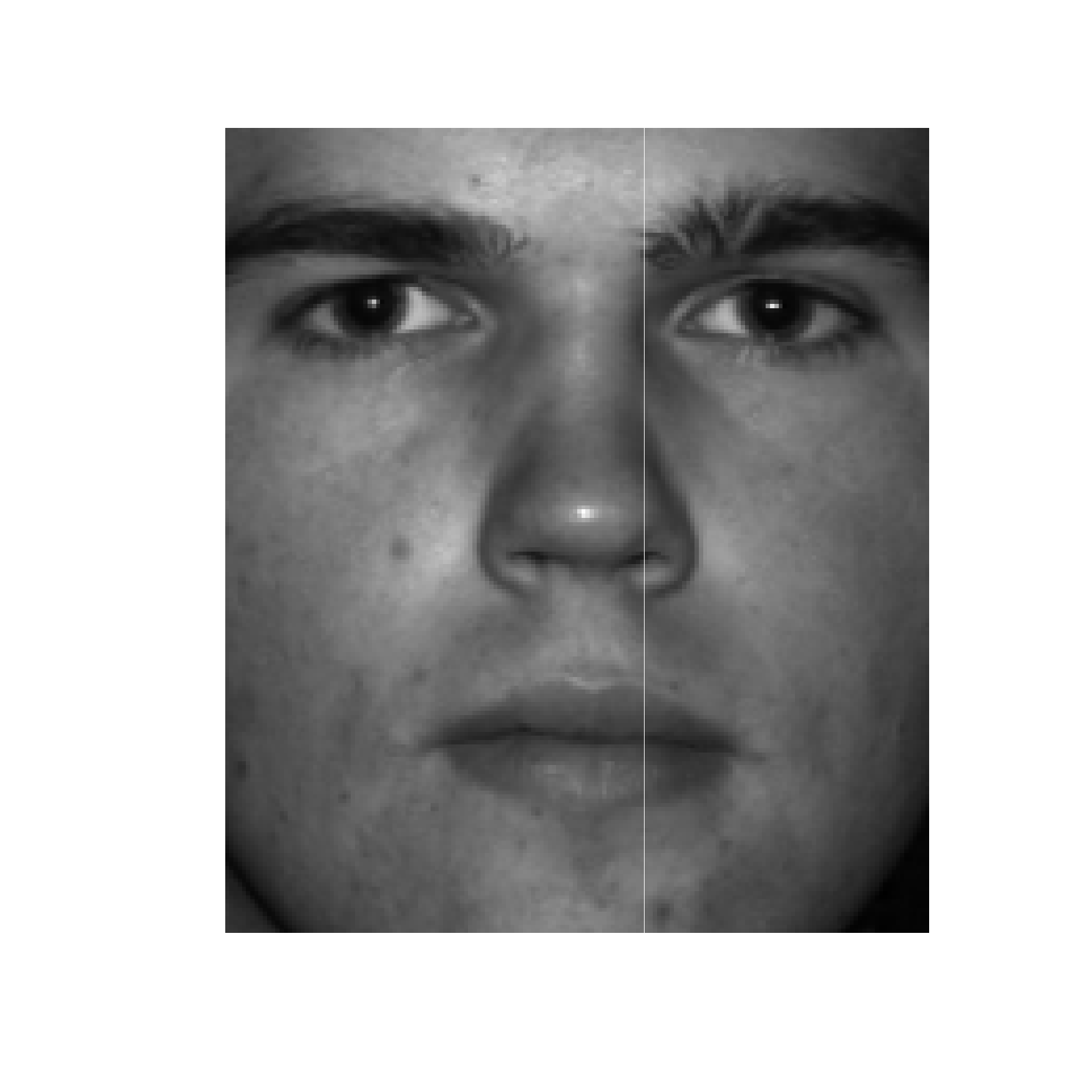} \hspace*{-0.5cm}&  
 \includegraphics[width=2cm]{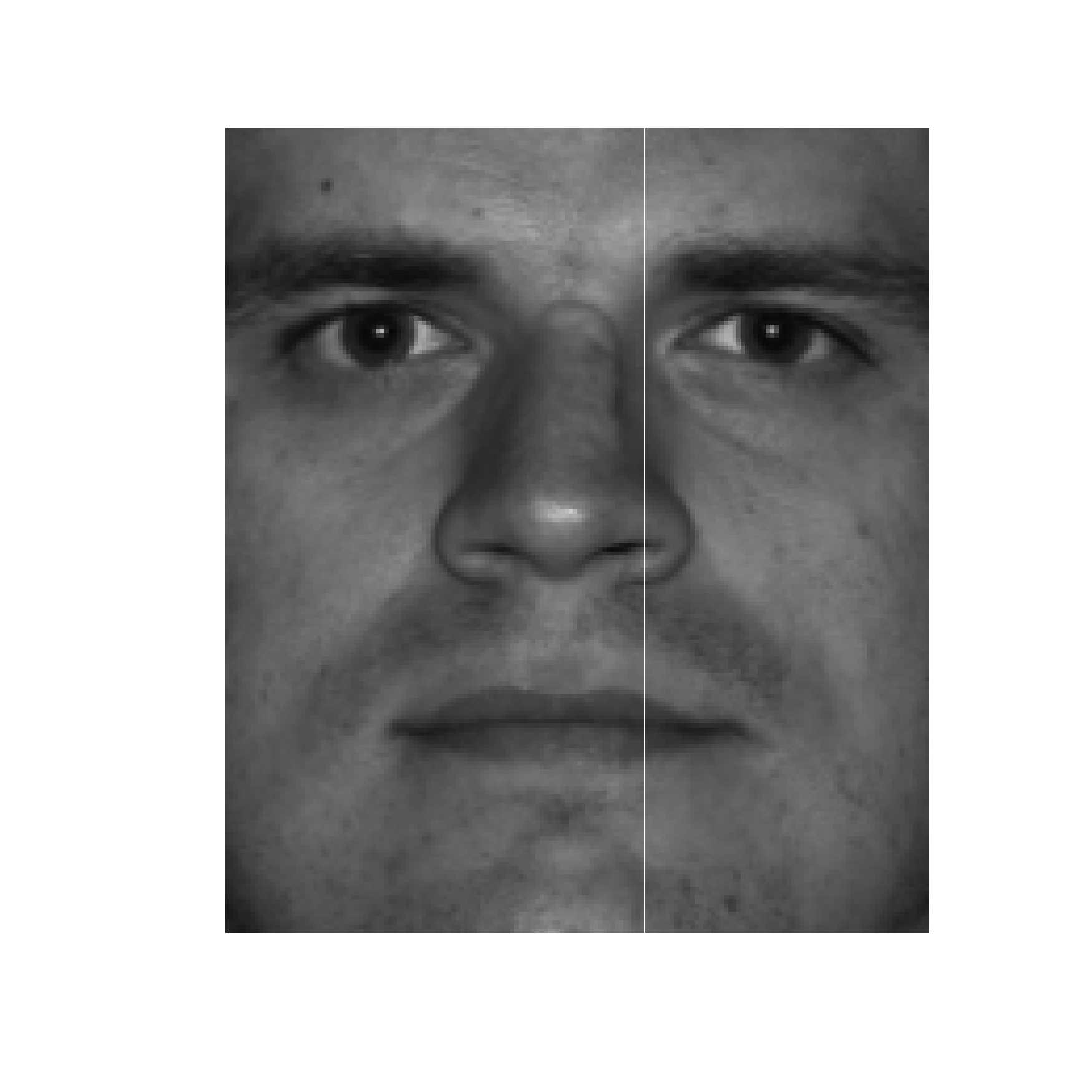} \hspace*{-0.5cm} & 
 \includegraphics[width=2cm]{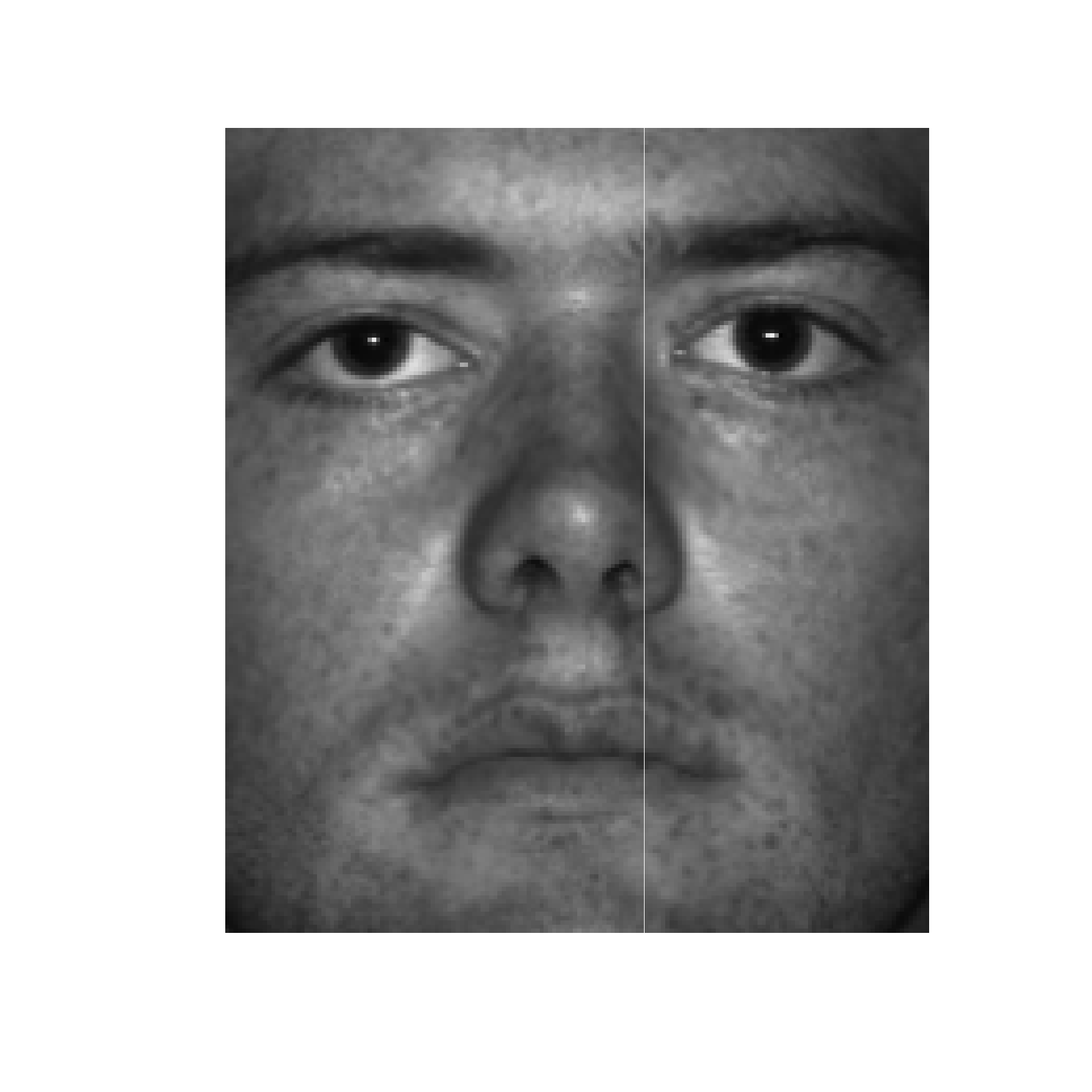}\hspace*{-0.5cm} &  
 \includegraphics[width=2cm]{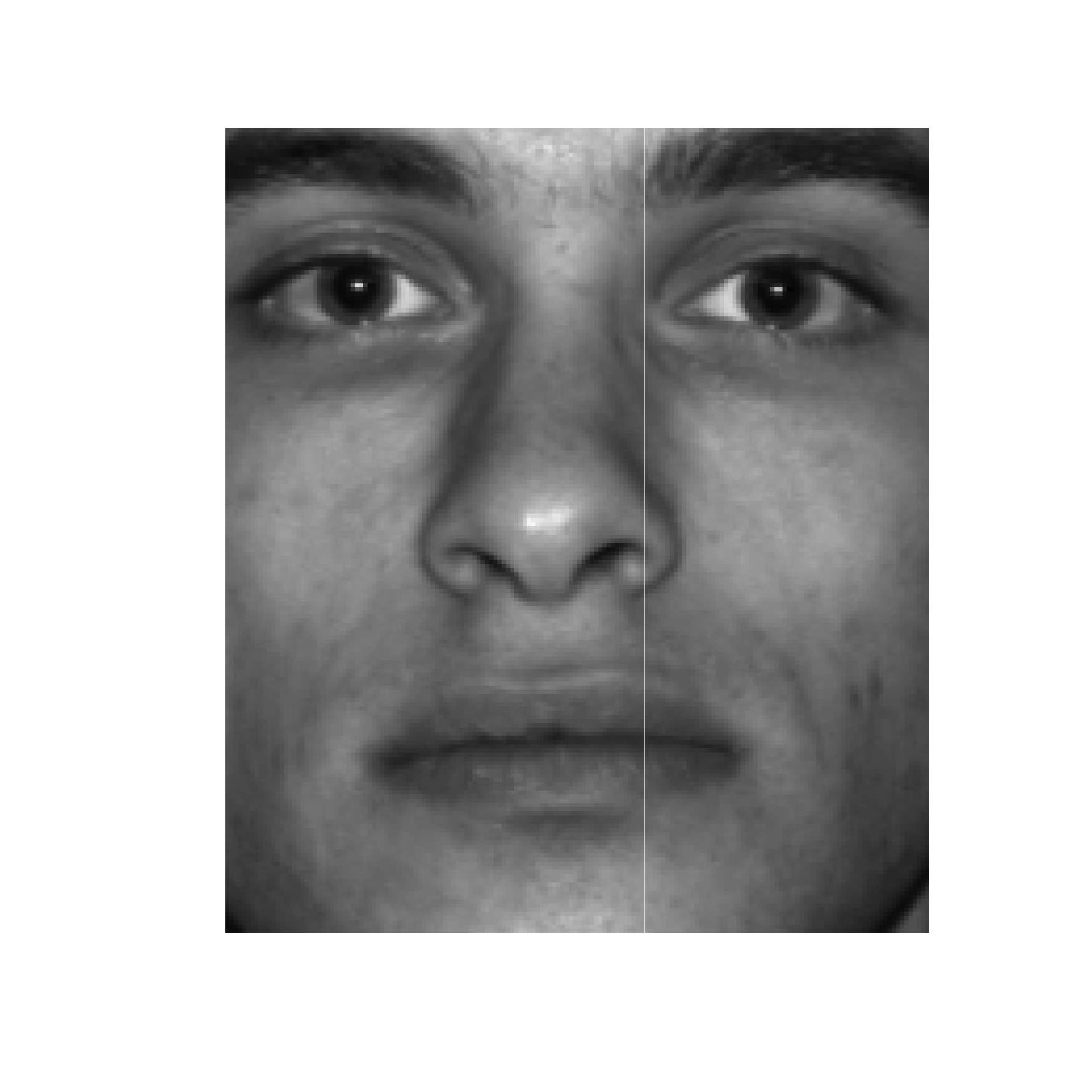} \hspace*{-0.5cm}&
 \includegraphics[width=2cm]{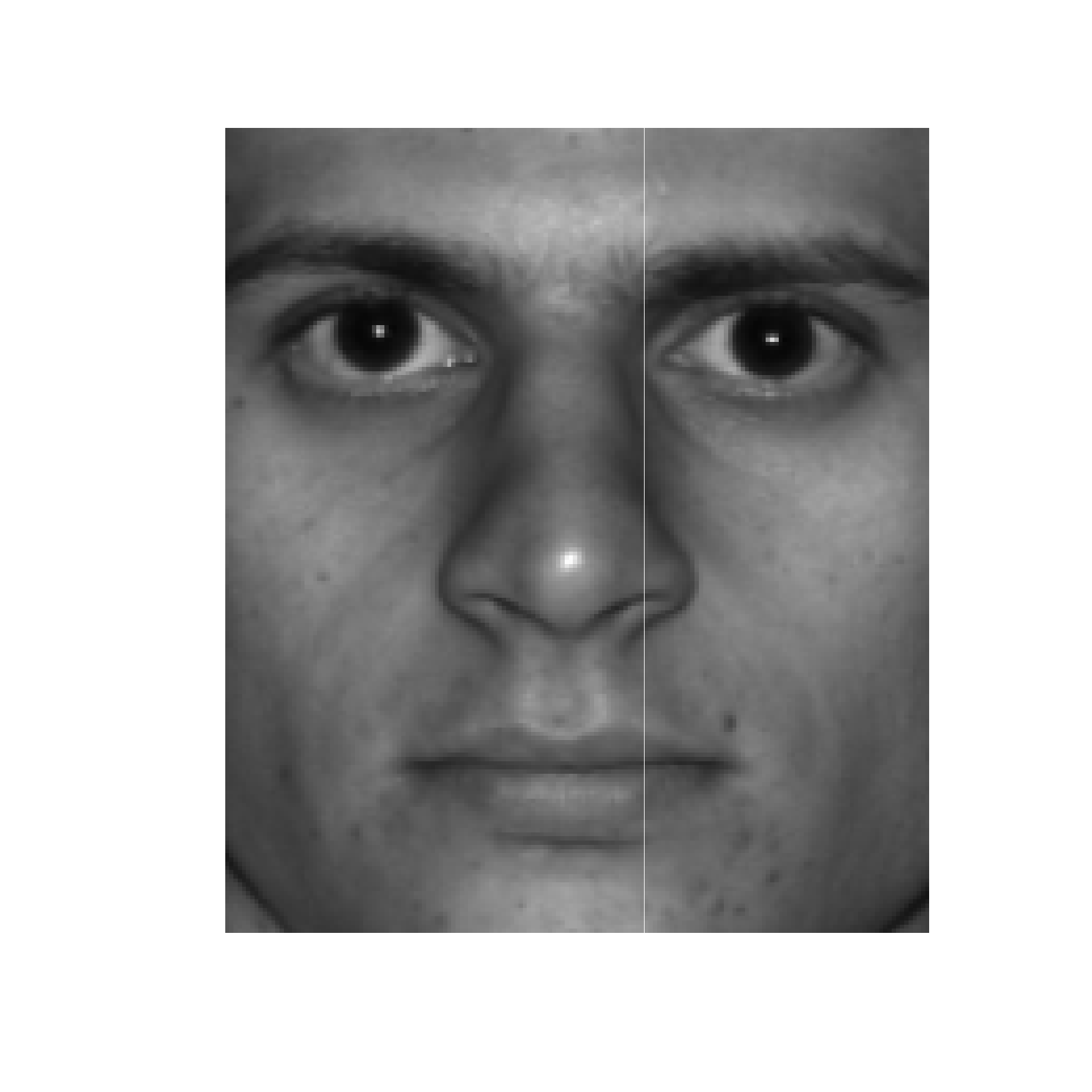} \hspace*{-0.5cm}\\
\end{array}$
\end{center}
\caption{
        Random subset of the face images in the image data example.
     }
   \label{subsetFaces}
\end{figure}

\begin{figure}[h!]
\scriptsize
\begin{center}$
\begin{array}{ccccc} \hspace*{-0.3cm}
  \includegraphics[width=2cm]{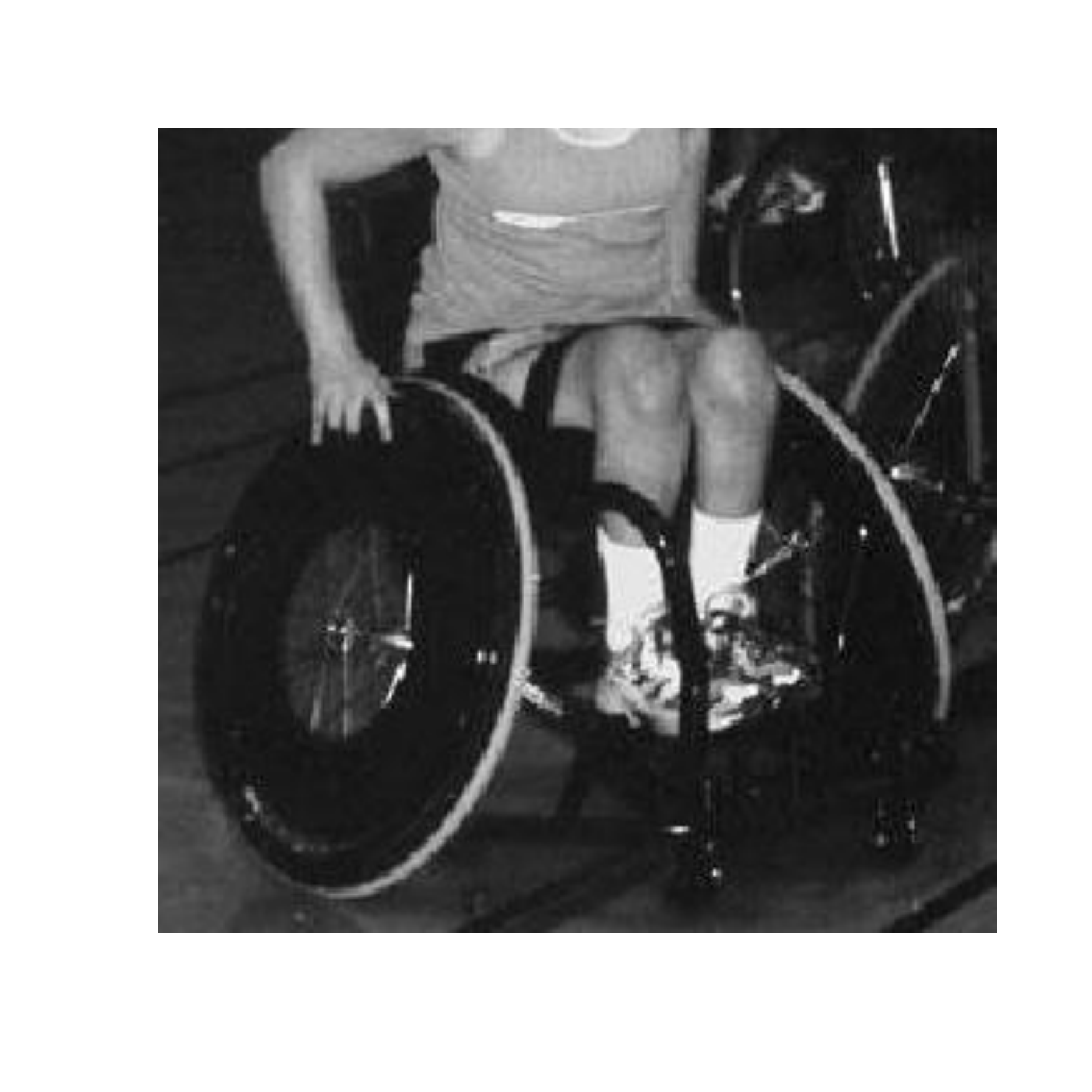} \hspace*{-0.5cm} &   
  \includegraphics[width=2cm]{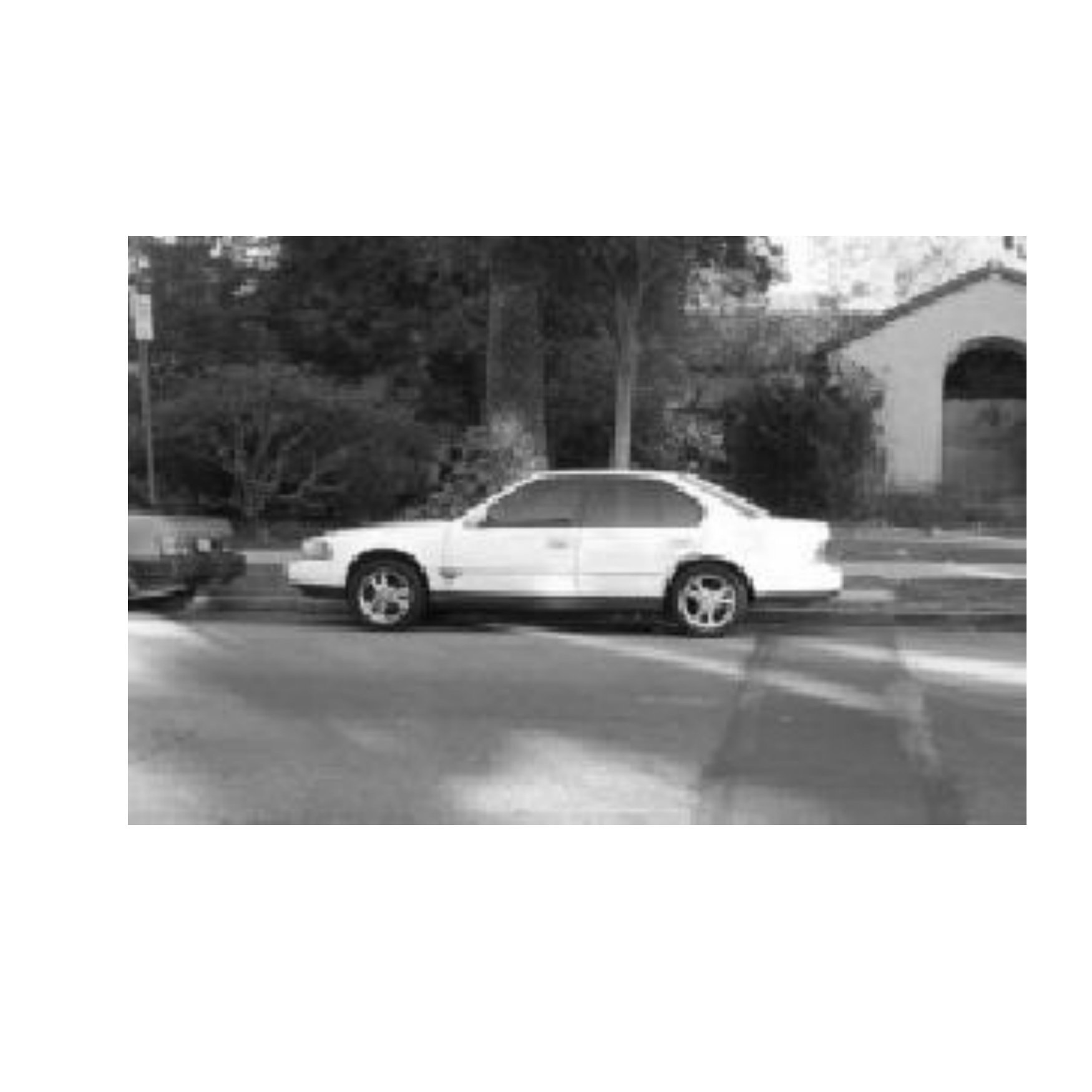} \hspace*{-0.3cm} &
 \includegraphics[width=2cm]{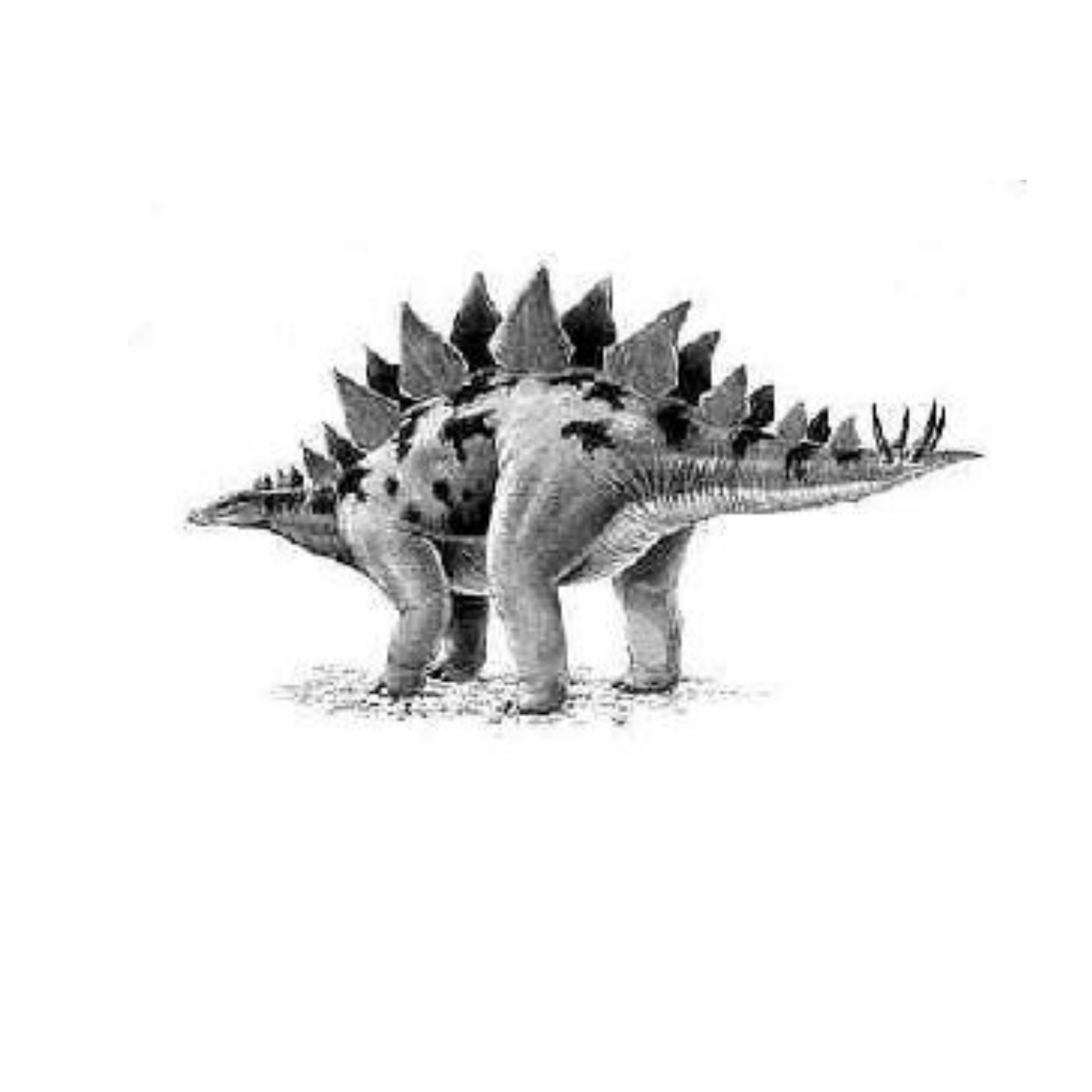} \hspace*{-0.3cm} &   
 \includegraphics[width=2cm]{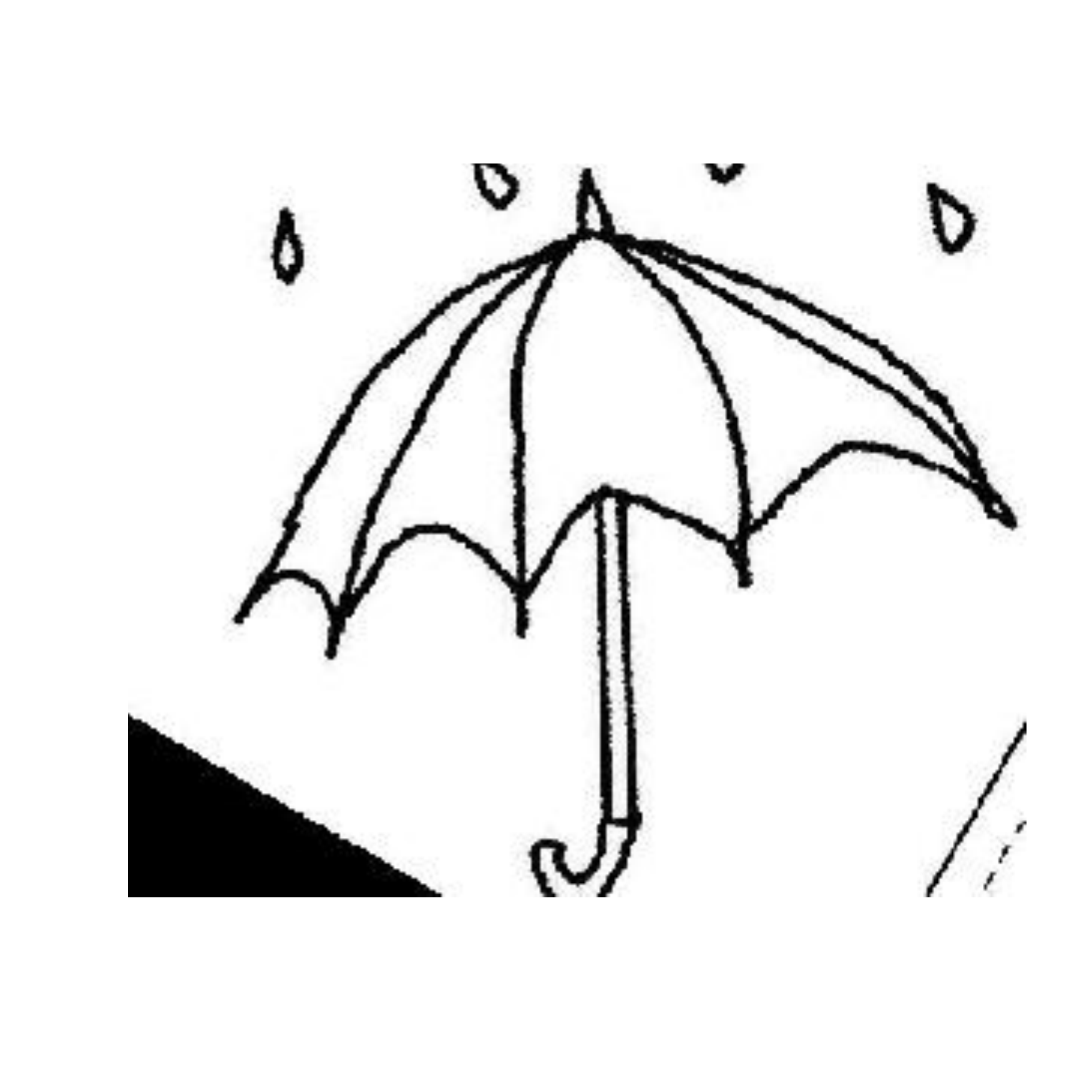} \hspace*{-0.3cm} &
\includegraphics[width=2cm]{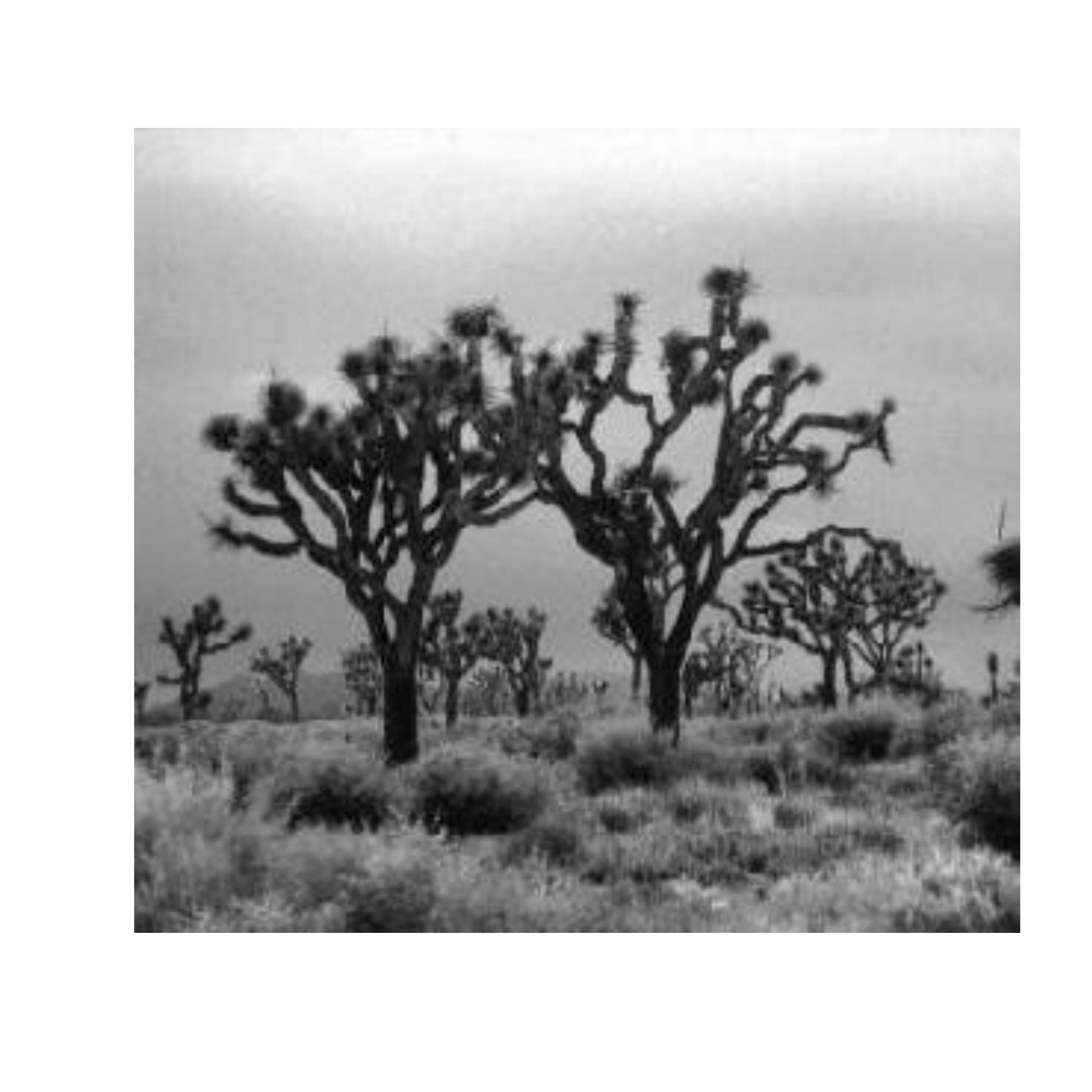} \hspace*{-0.3cm} \\
 67 \hspace*{-0.5cm} & 68 \hspace*{-0.5cm} & 69 \hspace*{-0.5cm} & 70 \hspace*{-0.5cm} & 71 \hspace*{-0.5cm}\\
 \includegraphics[width=2cm]{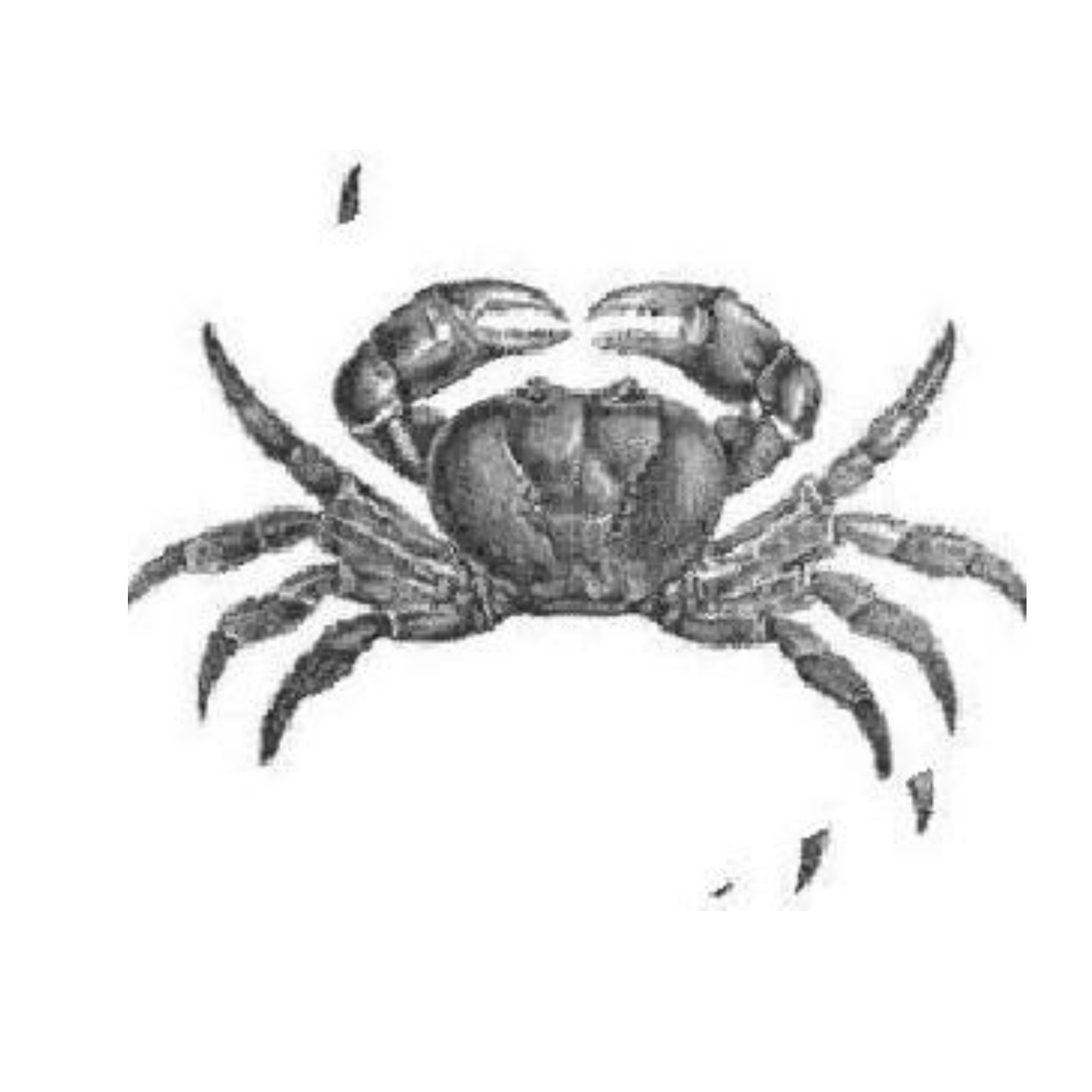} \hspace*{-0.3cm}&  
 \includegraphics[width=2cm]{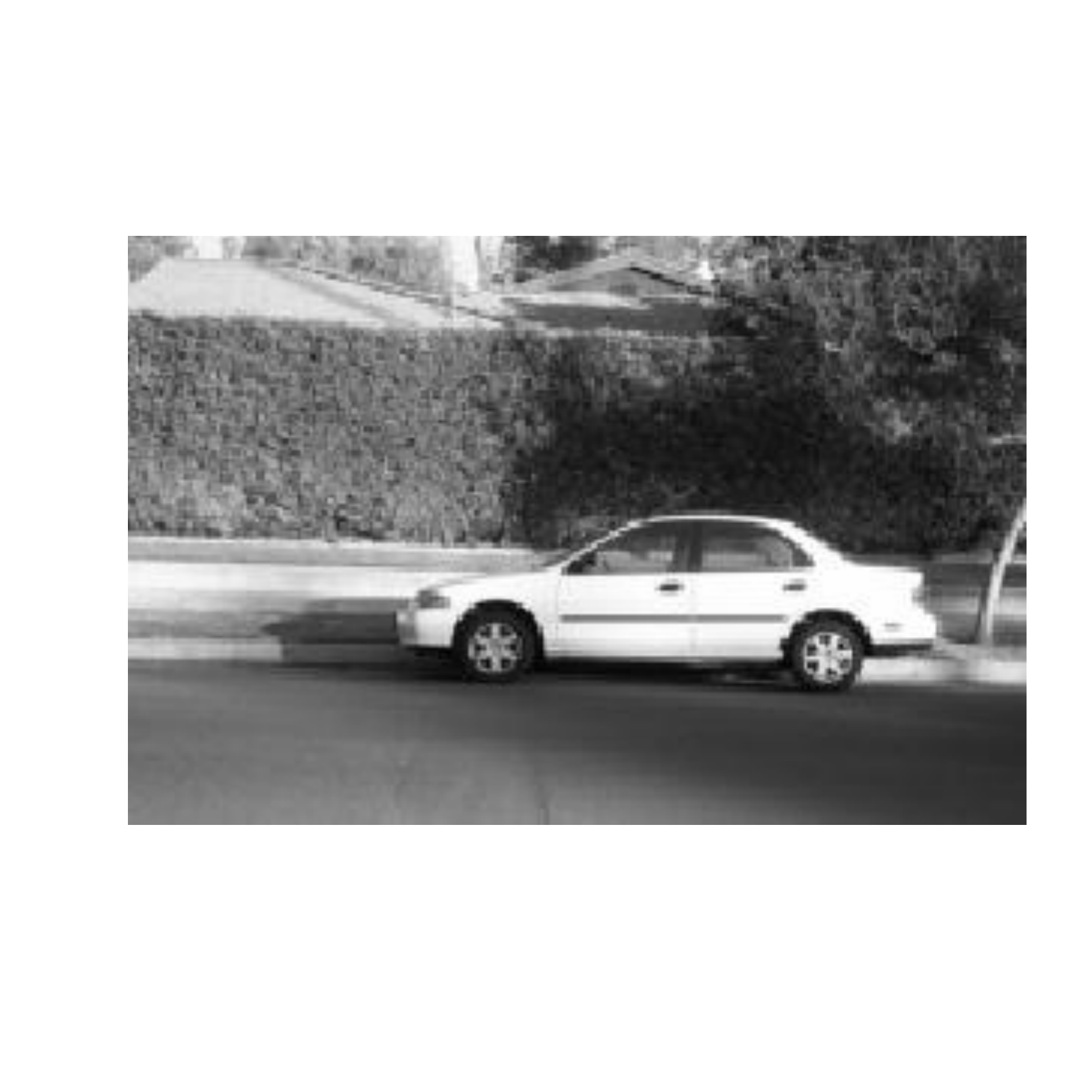} \hspace*{-0.3cm} & 
 \includegraphics[width=2cm]{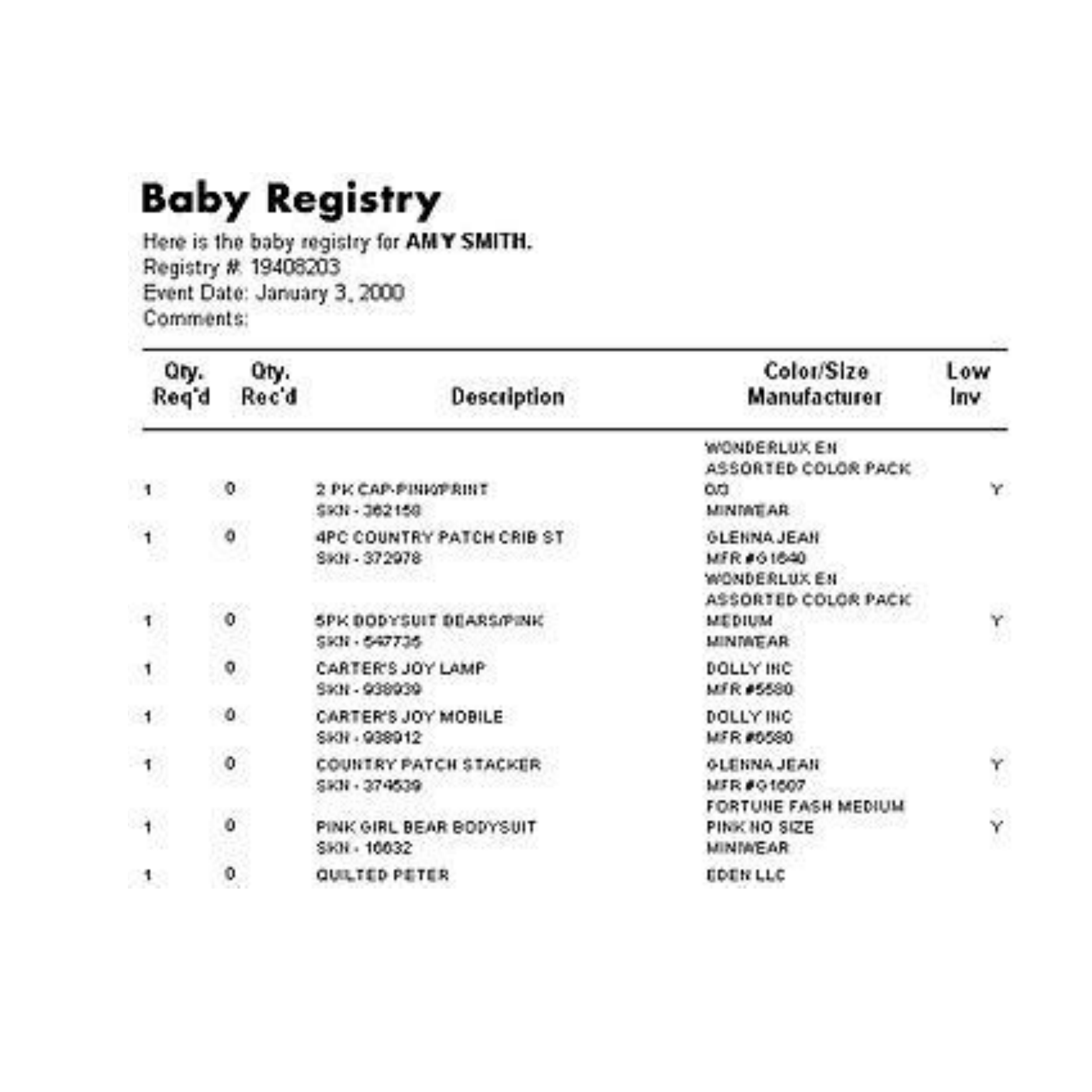}\hspace*{-0.3cm} &  
 \includegraphics[width=2cm]{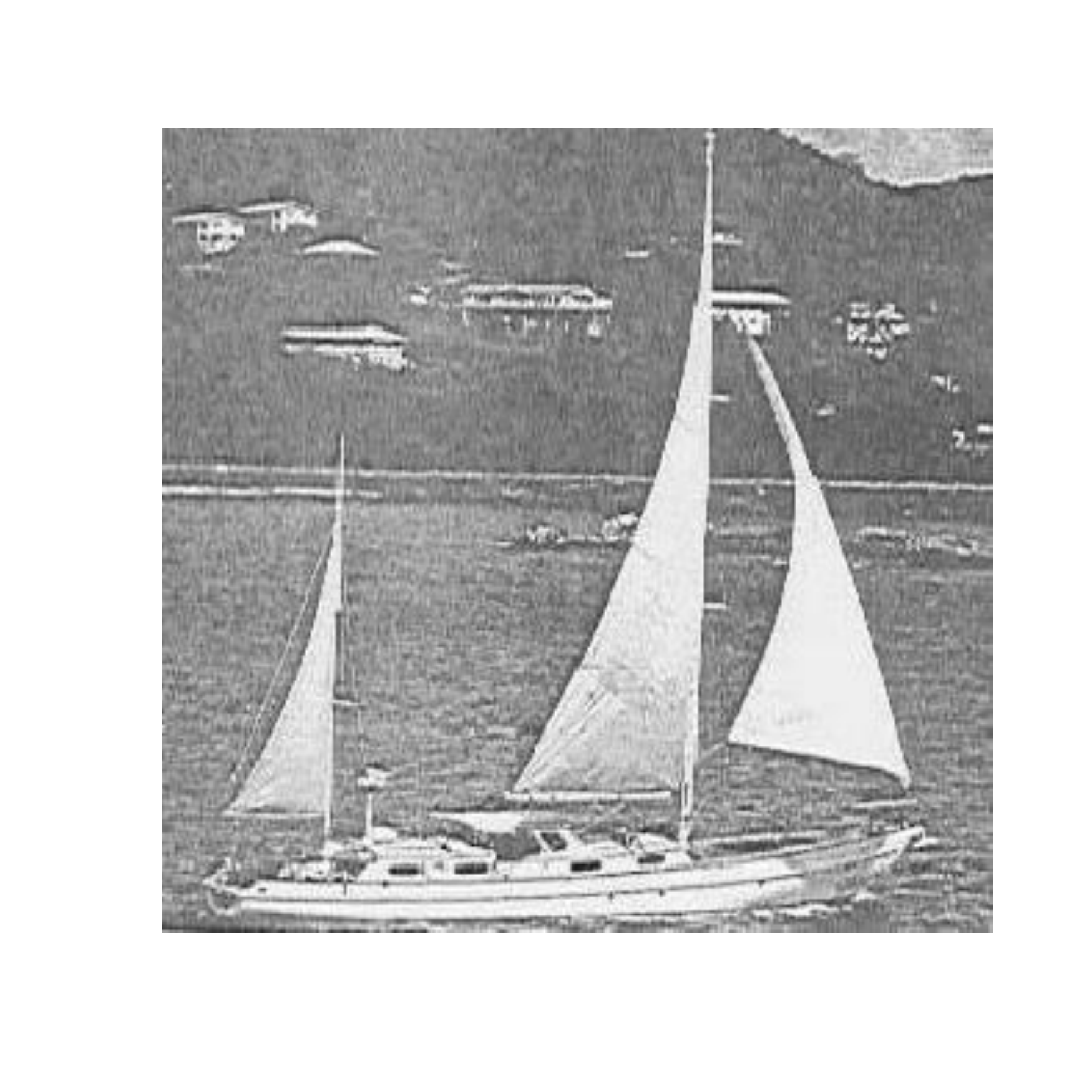} \hspace*{-0.3cm}\\
72 \hspace*{-0.5cm} & 73 \hspace*{-0.5cm} & 74 \hspace*{-0.5cm} & 75 \hspace*{-0.3cm} \\
\end{array}$
\end{center}
\caption{
        Object images in the image data example.
     }
   \label{Objects}
\end{figure}

We illustrate our deterministic algorithms on real image data that have been vectorized to form a high-dimensional dataset with $p>>n$. We use the Extended Yale Face Database B \citep{Georghiades2001,Lee2005} which contains aligned grayscale face images of 38 subjects under the same frontal pose and 64 different illumination conditions. This database contains cropped face images which have resolution $192 \times 168$ (=32256) pixels each. For the analysis we randomly sampled 11 subjects and selected for each subject the 6 images with the highest illumination contrasts, so that the face characteristics are clearly identifiable. In particular, we used the images with the following codes for light conditions: "P00A+000E+00.pgm", "P00A+010E-20.pgm", "P00A-010E-20.pgm", "P00A+000E-20.pgm",
"P00A-005E-10.pgm" and "P00A+005E-10.pgm". Similarly to 
\citet{Rahmani2016}, next to these 66 face images we also sampled 9 non-face images from the Caltech 101 database \citep{FeiFei2007} which constitute potential outliers. The Caltech 101 database contains 9144 color and grayscale images from 102 object categories that includes vehicles, plants, animals and cartoon characters. We only considered grayscale images with a height of at least 192 pixels and a width of at least 168 pixels. We randomly sampled 9 object images from this subset and cropped the sampled images to 192 pixel height and 168 pixel width when necessary. Combining the vectorized face and object images resulted in $75$ observations in $p=192\times 168= 32256$ dimensions. The last 9 observations (rows 67-75 in the data matrix) correspond to the object images. Hence, $12\%$ of the observations in the dataset are contamination. Figure \ref{subsetFaces} displays a random subset of the face images while Figure \ref{Objects} displays the 9 object images in the data matrix. The goal of our analysis is to recognize the faces and thus to flag the object images as outliers. We compare the results obtained by DsubLTS and DsubS to those of SPC, PPLTS and ROBPCA. We estimate the $q=2$ dimensional subspace which explains about 80\% of the robustly estimated total variability. 

To identify outliers, we use the diagnostic plot introduced by \citet{Hubert2005}. It plots the (robust) orthogonal distances between the observations and the estimated subspace versus (robust) score distances of the projected observations in the subspace with respect to their center. To identify outlying observations cutoff values for both the robust orthogonal distances and the robust score distances were proposed by \citet{Hubert2005}. To compute the robust score distances robust estimates for the variances according to the basis directions within the estimated subspace are required. 
Since our algorithms do not yield such estimates, we estimate these variances robustly using univariate LTS or M-scales of the scores corresponding to the estimated basis directions. Also SPC does not yield such variance estimates, so we computed the MCD estimator on the scores to compute robust
distances in the PCA subspace, similarly to \citet{Hubert2005}. Note that as in Section \ref{highdimsim}, we used our SPC implementation in \verb|R| based on Algorithm \ref{PCAsub} for these high-dimensional data. %Also here the S-L, S-M, RsubS and RsubLTS algorithms could not be computed in a reasonable amount of time. 

\begin{figure*}[htbp]
\setlength{\abovecaptionskip}{0pt} 
\center
\begin{minipage}{.53\textwidth}
\centering
\includegraphics[width=.9\textwidth]{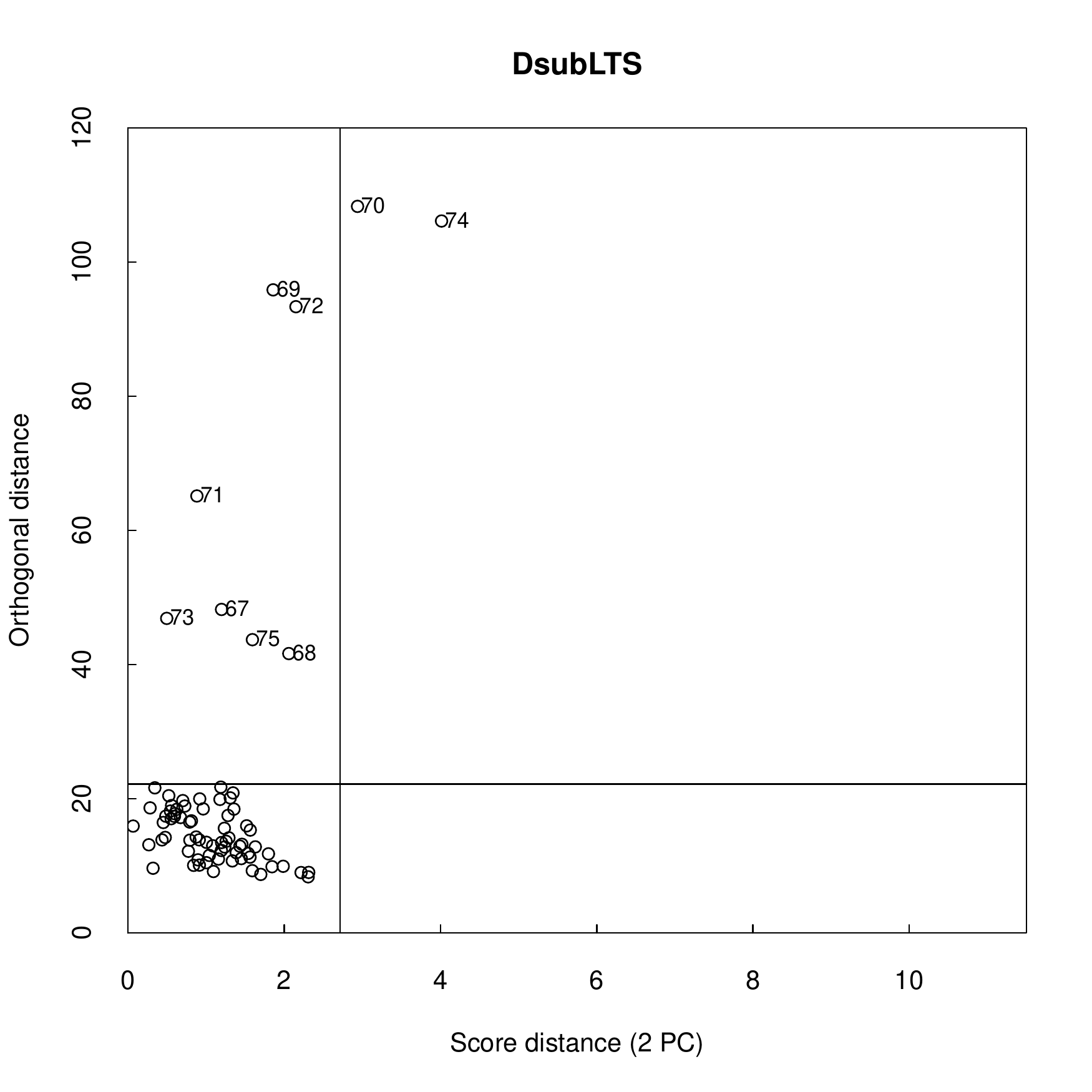}
\end{minipage}%
\begin{minipage}{0.53\textwidth}
\centering
\includegraphics[width=.9\textwidth]{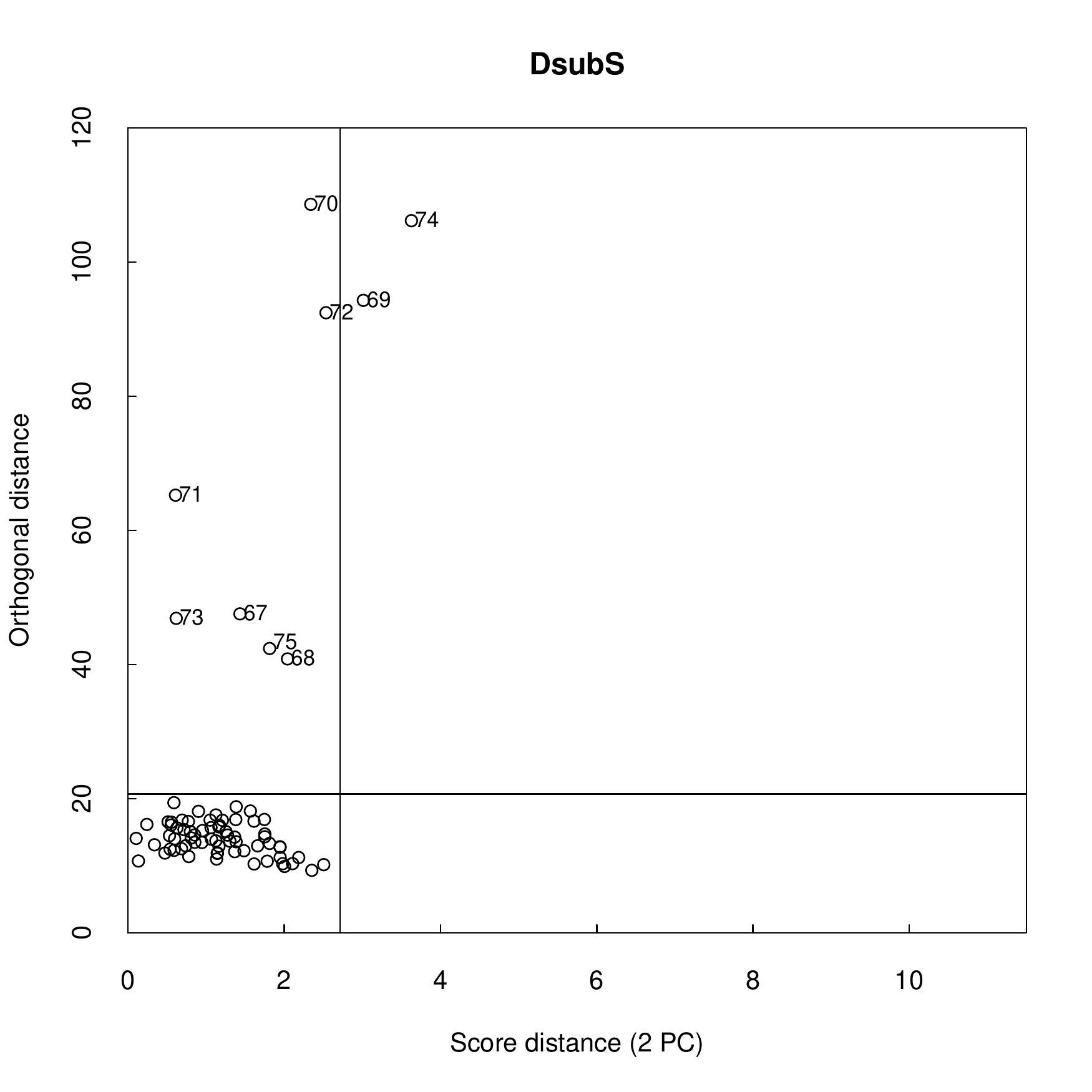}
\end{minipage}
\begin{minipage}{0.53\textwidth}
\centering
\includegraphics[width=.9\textwidth]{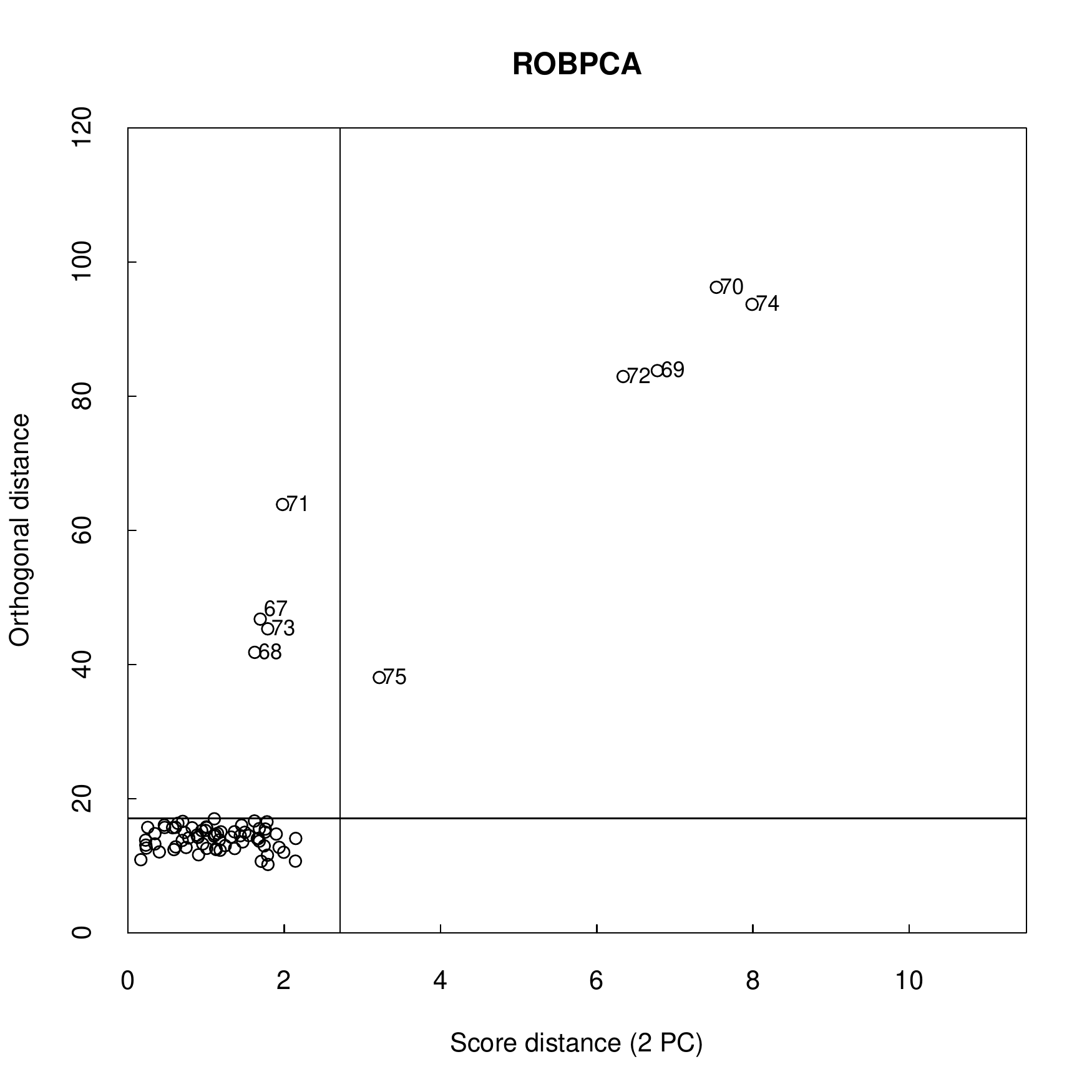}
\end{minipage}%
\begin{minipage}{0.53\textwidth}
\centering
\includegraphics[width=.9\textwidth]{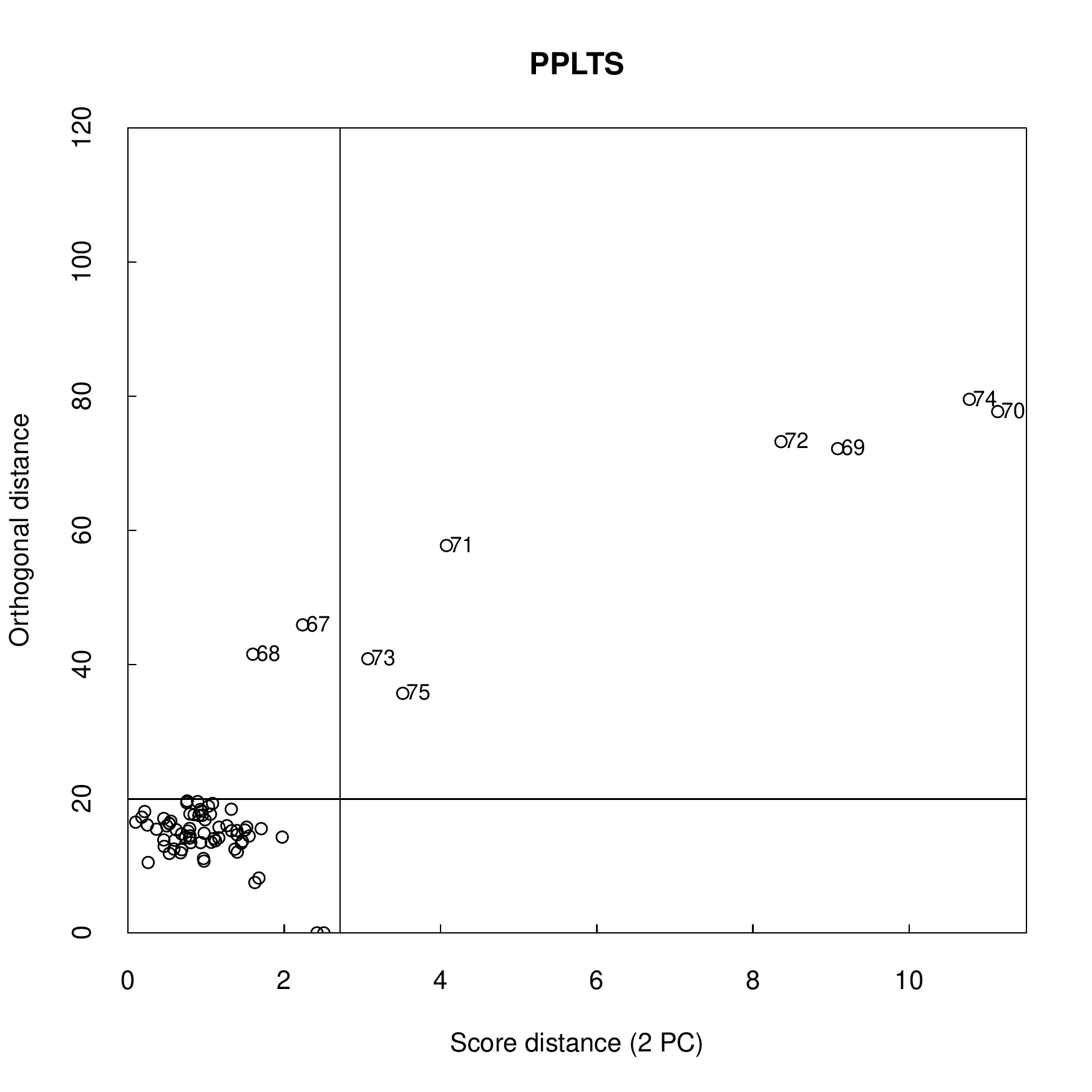}
\end{minipage}
\begin{minipage}{0.53\textwidth}
\centering
\includegraphics[width=.9\textwidth]{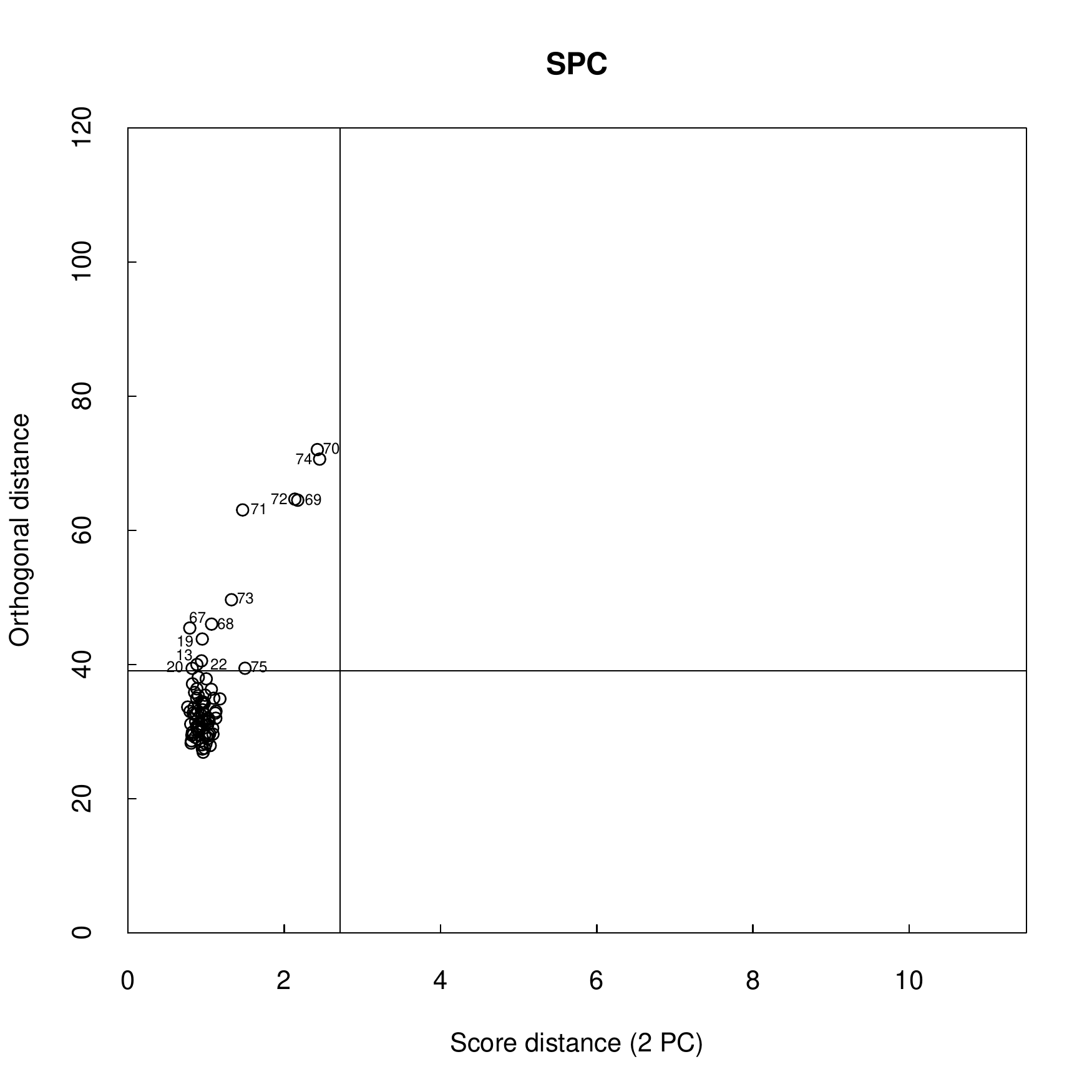}
\end{minipage}
\caption{Diagnostic plots for the Face Recognition example}
\label{diagnosticplotsFaceRecognition}
\end{figure*}

Figure \ref{diagnosticplotsFaceRecognition} shows the diagnostic plots of the solutions obtained by DsubLTS, DsubS, SPC, PPLTS and ROBPCA. All robust methods are able to flag the 9 object images as outliers with respect to the estimated 2-dimensional subspace. Object images are flagged as either orthogonal outliers or bad leverage points by the different methods. Note that SPC also detects four faces as outliers, but these false positives are merely borderline cases.
%Objects 67 (wheelchair), 68 (car side), 71 (Joshua tree), 73 (car side) and 75 (ketch) are flagged as orthogonal outliers, while objects 69 (Stegosaurus), 70 (umbrella), 72 (crab) and 74 (google background) are flagged as bad leverage points. 
This example illustrates that also in very high-dimensional settings our deterministic algorithms are able to robustly estimate a low-dimensional subspace that best represents the regular observations and allows to identify outliers. 

\section{Conclusions} \label{sec:conclusions}

We proposed new algorithms for the robust subspace estimation methods proposed by~\citet{Maronna2005}. These algorithms directly estimate principal directions of the subspace, which makes them more suitable for high-dimensional problems. For the starting values of the algorithm we considered random orthogonal matrices, as well as five deterministic starting values.  
These well-chosen deterministic starting values can be computed fast in high-dimensional settings because they avoid the need to calculate a high-dimensional scatter matrix and only use simple manipulations of the data. 
%\textcolor{blue}{and because they are expected to lie close to a robust local minimum.} 
Our experiments show that the deterministic algorithms yield results that are as good as or better than the results for the algorithms with random starting values, while having a much lower computation time.  
Moreover, the excellent performance of our deterministic algorithms carries over to high-dimensional settings. Maronna's algorithms and our algorithms with random orthogonal matrices attempt to find the global minimum of their objective function by using a sufficient number of starting values. On the other hand, 
our algorithms with deterministic starting values aim to find this minimum by using a few well-chosen robust but rough initial estimates of the subspace. While this may decrease the probability of obtaining the global minimum, it does often lead to a robust local minimum. 
This explains why these deterministic algorithms show a better performance. 

For the deterministic algorithms orthogonal equivariance is not guaranteed anymore. However, our experiments have shown that in most cases there is not a serious loss of equivariance. 
Note that the computation time of our deterministic algorithms can be reduced further on multi-core machines by calculating each of the five starting solutions in parallel on different cores. Finally, the subspace S-estimator usually shows a better compromise between robustness and efficiency than the subspace LTS-estimator.
Implementations of our algorithms in R~\citep{Rcore} are available from the website \url{http://wis.kuleuven.be/stat/robust/software}.

\section*{Acknowledgments}
The research by Van Aelst was supported by the Internal Funds KU Leuven under Grant C16-15-068  and 
COST Action IC1408 CRoNoS. Their support is gratefully acknowledged.

%\section*{References}

\bibliography{RobSubspaceComp_CSDA}

\begin{thebibliography}{43}
\expandafter\ifx\csname natexlab\endcsname\relax\def\natexlab#1{#1}\fi
\providecommand{\url}[1]{\texttt{#1}}
\providecommand{\href}[2]{#2}
\providecommand{\path}[1]{#1}
\providecommand{\DOIprefix}{doi:}
\providecommand{\ArXivprefix}{arXiv:}
\providecommand{\URLprefix}{URL: }
\providecommand{\Pubmedprefix}{pmid:}
\providecommand{\doi}[1]{\href{http://dx.doi.org/#1}{\path{#1}}}
\providecommand{\Pubmed}[1]{\href{pmid:#1}{\path{#1}}}
\providecommand{\bibinfo}[2]{#2}
\ifx\xfnm\relax \def\xfnm[#1]{\unskip,\space#1}\fi
%Type = Book
\bibitem[{Anderson et~al.(1999)Anderson, Bai, Bischof, Blackford, Demmel,
  Dongarra, Du~Croz, Hammarling, Greenbaum, McKenney \& Sorensen}]{lapack}
\bibinfo{author}{Anderson, E.}, \bibinfo{author}{Bai, Z.},
  \bibinfo{author}{Bischof, C.}, \bibinfo{author}{Blackford, L.~S.},
  \bibinfo{author}{Demmel, J.}, \bibinfo{author}{Dongarra, J.~J.},
  \bibinfo{author}{Du~Croz, J.}, \bibinfo{author}{Hammarling, S.},
  \bibinfo{author}{Greenbaum, A.}, \bibinfo{author}{McKenney, A.}, \&
  \bibinfo{author}{Sorensen, D.} (\bibinfo{year}{1999}).
\newblock {\it \bibinfo{title}{LAPACK Users' Guide (Third Ed.)}\/}.
\newblock \bibinfo{address}{Philadelphia, PA, USA}: \bibinfo{publisher}{Society
  for Industrial and Applied Mathematics}.
\newblock \URLprefix \url{http://www.netlib.org/lapack/lug/lapack_lug.html}.
%Type = Article
\bibitem[{Bj{\"o}rck \& Golub(1973)}]{BjorckGolub1973}
\bibinfo{author}{Bj{\"o}rck, {\AA}.}, \& \bibinfo{author}{Golub, G.~H.}
  (\bibinfo{year}{1973}).
\newblock \bibinfo{title}{Numerical methods for computing angles between linear
  subspaces}.
\newblock {\it \bibinfo{journal}{Mathematics of Computation}\/},  {\it
  \bibinfo{volume}{27}\/}, \bibinfo{pages}{579--594}. \URLprefix
  \url{http://www.jstor.org/stable/2005662}.
%Type = Article
\bibitem[{Boente \&
  Salibian-Barrera(2015)}]{BoenteGracielaSalibian-Barrera2015}
\bibinfo{author}{Boente, G.}, \& \bibinfo{author}{Salibian-Barrera, M.}
  (\bibinfo{year}{2015}).
\newblock \bibinfo{title}{S-estimators for functional principal component
  analysis}.
\newblock {\it \bibinfo{journal}{Journal of the American Statistical
  Association}\/},  {\it \bibinfo{volume}{110}\/}, \bibinfo{pages}{1100--1111}.
  \URLprefix \url{http://dx.doi.org/10.1080/01621459.2014.946991}.
  \DOIprefix\doi{10.1080/01621459.2014.946991}.
  \href{http://arxiv.org/abs/http://dx.doi.org/10.1080/01621459.2014.946991}{\tt
  arXiv:http://dx.doi.org/10.1080/01621459.2014.946991}.
%Type = Article
\bibitem[{Brahma et~al.(2018)Brahma, She, Li, Li \& Wu}]{Brahma2017}
\bibinfo{author}{Brahma, P.~P.}, \bibinfo{author}{She, Y.},
  \bibinfo{author}{Li, S.}, \bibinfo{author}{Li, J.}, \& \bibinfo{author}{Wu,
  D.} (\bibinfo{year}{2018}).
\newblock \bibinfo{title}{Reinforced robust principal component pursuit}.
\newblock {\it \bibinfo{journal}{IEEE Transactions on Neural Networks and
  Learning Systems}\/},  {\it \bibinfo{volume}{PP}\/}, \bibinfo{pages}{1--14}.
  \DOIprefix\doi{10.1109/TNNLS.2017.2671849}.
%Type = Article
\bibitem[{Campbell(1980)}]{Campbell1980}
\bibinfo{author}{Campbell, N.~A.} (\bibinfo{year}{1980}).
\newblock \bibinfo{title}{Robust procedures in multivariate analysis i: Robust
  covariance estimation}.
\newblock {\it \bibinfo{journal}{Applied Statistics}\/},  {\it
  \bibinfo{volume}{29}\/}, \bibinfo{pages}{231--237}.
%Type = Article
\bibitem[{Cand\'{e}s et~al.(2011)Cand\'{e}s, Li, Ma \& Wright}]{Candes2011}
\bibinfo{author}{Cand\'{e}s, E.~J.}, \bibinfo{author}{Li, X.},
  \bibinfo{author}{Ma, Y.}, \& \bibinfo{author}{Wright, J.}
  (\bibinfo{year}{2011}).
\newblock \bibinfo{title}{Robust principal component analysis?}
\newblock {\it \bibinfo{journal}{J. ACM}\/},  {\it \bibinfo{volume}{58}\/},
  \bibinfo{pages}{11:1--11:37}. \URLprefix
  \url{http://doi.acm.org/10.1145/1970392.1970395}.
  \DOIprefix\doi{10.1145/1970392.1970395}.
%Type = Inproceedings
\bibitem[{Chiang et~al.(2016)Chiang, Hsieh \& Dhillon}]{chiang2016}
\bibinfo{author}{Chiang, K.-Y.}, \bibinfo{author}{Hsieh, C.-J.}, \&
  \bibinfo{author}{Dhillon, I.~S.} (\bibinfo{year}{2016}).
\newblock \bibinfo{title}{Robust principal component analysis with side
  information}.
\newblock In {\it \bibinfo{booktitle}{International Conference on Machine
  Learning (ICML)}\/}.
%Type = Article
\bibitem[{Croux et~al.(2003)Croux, Filzmoser, Pison \& Rousseeuw}]{Croux2003}
\bibinfo{author}{Croux, C.}, \bibinfo{author}{Filzmoser, P.},
  \bibinfo{author}{Pison, G.}, \& \bibinfo{author}{Rousseeuw, P.~J.}
  (\bibinfo{year}{2003}).
\newblock \bibinfo{title}{Fitting multiplicative models by robust alternating
  regressions}.
\newblock {\it \bibinfo{journal}{Statistics and Computing}\/},  {\it
  \bibinfo{volume}{13}\/}, \bibinfo{pages}{23--36}. \URLprefix
  \url{http://dx.doi.org/10.1023/A:1021979409012}.
  \DOIprefix\doi{10.1023/A:1021979409012}.
%Type = Article
\bibitem[{Croux et~al.(2017)Croux, Garc\'{i}a-Escudero, Gordaliza, Ruwet \&
  Mart\'{i}n}]{robusttrimmedsub}
\bibinfo{author}{Croux, C.}, \bibinfo{author}{Garc\'{i}a-Escudero, L.~A.},
  \bibinfo{author}{Gordaliza, A.}, \bibinfo{author}{Ruwet, C.}, \&
  \bibinfo{author}{Mart\'{i}n, R.~S.} (\bibinfo{year}{2017}).
\newblock \bibinfo{title}{{Robust principal component analysis based on
  trimming around affine subspaces}}.
\newblock {\it \bibinfo{journal}{Statistica Sinica}\/},  {\it
  \bibinfo{volume}{27}\/}, \bibinfo{pages}{1437--1459}.
%Type = Article
\bibitem[{Croux \& Haesbroeck(2000)}]{Croux2000}
\bibinfo{author}{Croux, C.}, \& \bibinfo{author}{Haesbroeck, G.}
  (\bibinfo{year}{2000}).
\newblock \bibinfo{title}{{Principal component analysis based on robust
  estimators of the covariance or correlation matrix: influence functions and
  efficiencies}}.
\newblock {\it \bibinfo{journal}{Biometrika}\/},  {\it \bibinfo{volume}{87}\/},
  \bibinfo{pages}{603--618}. \URLprefix
  \url{http://biomet.oxfordjournals.org/content/87/3/603.short}.
  \DOIprefix\doi{10.1093/biomet/87.3.603}.
%Type = Inproceedings
\bibitem[{Croux \& Ruiz-Gazen(1996)}]{CrouxChristopheRuizGazen}
\bibinfo{author}{Croux, C.}, \& \bibinfo{author}{Ruiz-Gazen, A.}
  (\bibinfo{year}{1996}).
\newblock \bibinfo{title}{A fast algorithm for robust principal components
  based on projection pursuit}.
\newblock In \bibinfo{editor}{A.~Prat} (Ed.), {\it
  \bibinfo{booktitle}{COMPSTAT: Proceedings in Computational Statistics 12th
  Symposium held in Barcelona, Spain, 1996}\/} (pp. \bibinfo{pages}{211--216}).
\newblock \bibinfo{address}{Heidelberg}: \bibinfo{publisher}{Physica-Verlag
  HD}.
\newblock \URLprefix \url{http://dx.doi.org/10.1007/978-3-642-46992-3\_22}.
  \DOIprefix\doi{10.1007/978-3-642-46992-3\_22}.
%Type = Article
\bibitem[{Croux \& Ruiz-Gazen(2005)}]{Croux2005}
\bibinfo{author}{Croux, C.}, \& \bibinfo{author}{Ruiz-Gazen, A.}
  (\bibinfo{year}{2005}).
\newblock \bibinfo{title}{{High breakdown estimators for principal components:
  The projection-pursuit approach revisited}}.
\newblock {\it \bibinfo{journal}{Journal of Multivariate Analysis}\/},  {\it
  \bibinfo{volume}{95}\/}, \bibinfo{pages}{206--226}.
  \DOIprefix\doi{10.1016/j.jmva.2004.08.002}.
%Type = Article
\bibitem[{Devlin et~al.(1981)Devlin, Gnanadesikan \& Kettenring}]{Devlin1981}
\bibinfo{author}{Devlin, S.~J.}, \bibinfo{author}{Gnanadesikan, R.}, \&
  \bibinfo{author}{Kettenring, J.~R.} (\bibinfo{year}{1981}).
\newblock \bibinfo{title}{{Robust Estimation of Dispersion Matrices and
  Principal Components}}.
\newblock {\it \bibinfo{journal}{Journal of the American Statistical
  Association}\/},  {\it \bibinfo{volume}{76}\/}, \bibinfo{pages}{354--362}.
  \URLprefix \url{http://www.jstor.org/stable/2287836}.
  \DOIprefix\doi{10.2307/2287836}.
%Type = Article
\bibitem[{Eddelbuettel \& Sanderson(2014)}]{RcppArmadillo}
\bibinfo{author}{Eddelbuettel, D.}, \& \bibinfo{author}{Sanderson, C.}
  (\bibinfo{year}{2014}).
\newblock \bibinfo{title}{Rcpparmadillo: Accelerating r with high-performance
  c++ linear algebra}.
\newblock {\it \bibinfo{journal}{Computational Statistics and Data
  Analysis}\/},  {\it \bibinfo{volume}{71}\/}, \bibinfo{pages}{1054--1063}.
  \URLprefix \url{http://dx.doi.org/10.1016/j.csda.2013.02.005}.
%Type = Article
\bibitem[{Engelen et~al.(2005)Engelen, Hubert \& Vanden~Branden}]{Engelen2005}
\bibinfo{author}{Engelen, S.}, \bibinfo{author}{Hubert, M.}, \&
  \bibinfo{author}{Vanden~Branden, K.} (\bibinfo{year}{2005}).
\newblock \bibinfo{title}{{A Comparison of Three Procedures for Robust PCA in
  High Dimensions}}.
\newblock {\it \bibinfo{journal}{Austrian Journal of Statistics}\/},  {\it
  \bibinfo{volume}{34}\/}, \bibinfo{pages}{117--126}.
%Type = Article
\bibitem[{Fei-Fei et~al.(2007)Fei-Fei, Fergus \& Perona}]{FeiFei2007}
\bibinfo{author}{Fei-Fei, L.}, \bibinfo{author}{Fergus, R.}, \&
  \bibinfo{author}{Perona, P.} (\bibinfo{year}{2007}).
\newblock \bibinfo{title}{Learning generative visual models from few training
  examples: An incremental bayesian approach tested on 101 object categories}.
\newblock {\it \bibinfo{journal}{Comput. Vis. Image Underst.}\/},  {\it
  \bibinfo{volume}{106}\/}, \bibinfo{pages}{59--70}. \URLprefix
  \url{http://dx.doi.org/10.1016/j.cviu.2005.09.012}.
  \DOIprefix\doi{10.1016/j.cviu.2005.09.012}.
%Type = Article
\bibitem[{Gabriel \& Zamir(1979)}]{RubenGabriel1979}
\bibinfo{author}{Gabriel, K.~R.}, \& \bibinfo{author}{Zamir, S.}
  (\bibinfo{year}{1979}).
\newblock \bibinfo{title}{Lower rank approximation of matrices by least squares
  with any choice of weights}.
\newblock {\it \bibinfo{journal}{Technometrics}\/},  {\it
  \bibinfo{volume}{21}\/}, \bibinfo{pages}{489--498}. \URLprefix
  \url{http://www.jstor.org/stable/1268288}.
%Type = Article
\bibitem[{Georghiades et~al.(2001)Georghiades, Belhumeur \&
  Kriegman}]{Georghiades2001}
\bibinfo{author}{Georghiades, A.~S.}, \bibinfo{author}{Belhumeur, P.~N.}, \&
  \bibinfo{author}{Kriegman, D.~J.} (\bibinfo{year}{2001}).
\newblock \bibinfo{title}{From few to many: illumination cone models for face
  recognition under variable lighting and pose}.
\newblock {\it \bibinfo{journal}{IEEE Transactions on Pattern Analysis and
  Machine Intelligence}\/},  {\it \bibinfo{volume}{23}\/},
  \bibinfo{pages}{643--660}. \DOIprefix\doi{10.1109/34.927464}.
%Type = Article
\bibitem[{Hubert et~al.(2015)Hubert, Rousseeuw, Vanpaemel \&
  Verdonck}]{HUBERT201564}
\bibinfo{author}{Hubert, M.}, \bibinfo{author}{Rousseeuw, P.},
  \bibinfo{author}{Vanpaemel, D.}, \& \bibinfo{author}{Verdonck, T.}
  (\bibinfo{year}{2015}).
\newblock \bibinfo{title}{The dets and detmm estimators for multivariate
  location and scatter}.
\newblock {\it \bibinfo{journal}{Computational Statistics \& Data Analysis}\/},
   {\it \bibinfo{volume}{81}\/}, \bibinfo{pages}{64 -- 75}. \URLprefix
  \url{http://www.sciencedirect.com/science/article/pii/S0167947314002175}.
  \DOIprefix\doi{https://doi.org/10.1016/j.csda.2014.07.013}.
%Type = Article
\bibitem[{Hubert et~al.(2005)Hubert, Rousseeuw \& {Vanden
  Branden}}]{Hubert2005}
\bibinfo{author}{Hubert, M.}, \bibinfo{author}{Rousseeuw, P.~J.}, \&
  \bibinfo{author}{{Vanden Branden}, K.} (\bibinfo{year}{2005}).
\newblock \bibinfo{title}{{ROBPCA: A New Approach to Robust Principal Component
  Analysis}}.
\newblock {\it \bibinfo{journal}{Technometrics}\/},  {\it
  \bibinfo{volume}{47}\/}, \bibinfo{pages}{64--79}.
  \DOIprefix\doi{10.1198/004017004000000563}.
%Type = Article
\bibitem[{Hubert et~al.(2012)Hubert, Rousseeuw \& Verdonck}]{Hubert2012}
\bibinfo{author}{Hubert, M.}, \bibinfo{author}{Rousseeuw, P.~J.}, \&
  \bibinfo{author}{Verdonck, T.} (\bibinfo{year}{2012}).
\newblock \bibinfo{title}{{A deterministic algorithm for robust location and
  scatter}}.
\newblock {\it \bibinfo{journal}{Journal of Computational and Graphical
  Statistics}\/},  {\it \bibinfo{volume}{21}\/}, \bibinfo{pages}{618--637}.
  \URLprefix
  \url{http://dx.doi.org/10.1080/10618600.2012.672100$\backslash$nhttp://www.tandfonline.com/doi/abs/10.1080/10618600.2012.672100\#.UdeZC0A-nCM$\backslash$nhttp://www.tandfonline.com/doi/abs/10.1080/10618600.2012.672100}.
  \DOIprefix\doi{10.1080/10618600.2012.672100}.
%Type = Article
\bibitem[{Lee et~al.(2005)Lee, Ho \& Kriegman}]{Lee2005}
\bibinfo{author}{Lee, K.-C.}, \bibinfo{author}{Ho, J.}, \&
  \bibinfo{author}{Kriegman, D.~J.} (\bibinfo{year}{2005}).
\newblock \bibinfo{title}{Acquiring linear subspaces for face recognition under
  variable lighting}.
\newblock {\it \bibinfo{journal}{IEEE Transactions on Pattern Analysis and
  Machine Intelligence}\/},  {\it \bibinfo{volume}{27}\/},
  \bibinfo{pages}{684--698}. \DOIprefix\doi{10.1109/TPAMI.2005.92}.
%Type = Article
\bibitem[{Li \& Chen(1985)}]{Li1985}
\bibinfo{author}{Li, G.}, \& \bibinfo{author}{Chen, Z.} (\bibinfo{year}{1985}).
\newblock \bibinfo{title}{{Projection-pursuit approach to robust dispersion
  matrices and principal components: primary theory and Monte Carlo}}.
\newblock {\it \bibinfo{journal}{Journal of the American Statistical
  Association}\/},  {\it \bibinfo{volume}{80}\/}, \bibinfo{pages}{759--766}.
  \URLprefix
  \url{http://www.tandfonline.com/doi/abs/10.1080/01621459.1985.10478181$\backslash$npapers3://publication/uuid/E7A6D092-1420-4117-A47C-C6F759DA3121}.
  \DOIprefix\doi{10.1080/01621459.1985.10478181}.
%Type = Article
\bibitem[{Liu et~al.(2003)Liu, Hawkins, Ghosh \& Young}]{Liu2003}
\bibinfo{author}{Liu, L.}, \bibinfo{author}{Hawkins, D.~M.},
  \bibinfo{author}{Ghosh, S.}, \& \bibinfo{author}{Young, S.~S.}
  (\bibinfo{year}{2003}).
\newblock \bibinfo{title}{{Robust singular value decomposition analysis of
  microarray data.}}
\newblock {\it \bibinfo{journal}{Proceedings of the National Academy of
  Sciences of the United States of America}\/},  {\it \bibinfo{volume}{100}\/},
  \bibinfo{pages}{13167--72}. \URLprefix
  \url{http://www.pubmedcentral.nih.gov/articlerender.fcgi?artid=263735\&tool=pmcentrez\&rendertype=abstract}.
  \DOIprefix\doi{10.1073/pnas.1733249100}.
%Type = Article
\bibitem[{Locantore et~al.(1999)Locantore, Marron, Simpson, Tripoli, Zhang \&
  Cohen}]{Locantore1999}
\bibinfo{author}{Locantore, N.}, \bibinfo{author}{Marron, J.~S.},
  \bibinfo{author}{Simpson, D.~G.}, \bibinfo{author}{Tripoli, N.},
  \bibinfo{author}{Zhang, J.~T.}, \& \bibinfo{author}{Cohen, K.~L.}
  (\bibinfo{year}{1999}).
\newblock \bibinfo{title}{Robust principal component analysis for functional
  data}.
\newblock {\it \bibinfo{journal}{Test}\/},  {\it \bibinfo{volume}{8}\/},
  \bibinfo{pages}{1--73}. \URLprefix
  \url{http://dx.doi.org/10.1007/BF02595862}.
  \DOIprefix\doi{10.1007/BF02595862}.
%Type = Article
\bibitem[{Maronna(2005)}]{Maronna2005}
\bibinfo{author}{Maronna, R.~A.} (\bibinfo{year}{2005}).
\newblock \bibinfo{title}{{Principal Components and Orthogonal Regression Based
  on Robust Scales}}.
\newblock {\it \bibinfo{journal}{Technometrics}\/},  {\it
  \bibinfo{volume}{47}\/}, \bibinfo{pages}{264--273}. \URLprefix
  \url{http://pubs.amstat.org/doi/abs/10.1198/004017005000000166}.
  \DOIprefix\doi{10.1198/004017005000000166}.
%Type = Book
\bibitem[{Maronna et~al.(2006)Maronna, Martin \& Yohai}]{Maronna2006}
\bibinfo{author}{Maronna, R.~A.}, \bibinfo{author}{Martin, R.~D.}, \&
  \bibinfo{author}{Yohai, V.~J.} (\bibinfo{year}{2006}).
\newblock {\it \bibinfo{title}{{Robust Statistics: Theory and Methods}}\/}.
\newblock \bibinfo{publisher}{Wiley}.
%Type = Article
\bibitem[{Maronna et~al.(2015)Maronna, M\'{e}ndez \& Yohai}]{Maronna2015}
\bibinfo{author}{Maronna, R.~A.}, \bibinfo{author}{M\'{e}ndez, F.}, \&
  \bibinfo{author}{Yohai, V.~J.} (\bibinfo{year}{2015}).
\newblock \bibinfo{title}{Robust nonlinear principal components}.
\newblock {\it \bibinfo{journal}{Statistics and Computing}\/},  {\it
  \bibinfo{volume}{25}\/}, \bibinfo{pages}{439--448}. \URLprefix
  \url{http://dx.doi.org/10.1007/s11222-013-9442-0}.
  \DOIprefix\doi{10.1007/s11222-013-9442-0}.
%Type = Article
\bibitem[{Maronna \& Zamar(2002)}]{MaronnaZamar2002}
\bibinfo{author}{Maronna, R.~A.}, \& \bibinfo{author}{Zamar, R.~H.}
  (\bibinfo{year}{2002}).
\newblock \bibinfo{title}{Robust estimates of location and dispersion for
  high-dimensional datasets}.
\newblock {\it \bibinfo{journal}{Technometrics}\/},  {\it
  \bibinfo{volume}{44}\/}, \bibinfo{pages}{307--317}. \URLprefix
  \url{http://www.jstor.org/stable/1271538}.
%Type = Article
\bibitem[{Naga \& Antille(1990)}]{Naga1990}
\bibinfo{author}{Naga, R.~A.}, \& \bibinfo{author}{Antille, G.}
  (\bibinfo{year}{1990}).
\newblock \bibinfo{title}{Stability of robust and non-robust principal
  components analysis}.
\newblock {\it \bibinfo{journal}{Comput. Stat. Data Anal.}\/},  {\it
  \bibinfo{volume}{10}\/}, \bibinfo{pages}{169--174}. \URLprefix
  \url{http://dx.doi.org/10.1016/0167-9473(90)90062-M}.
  \DOIprefix\doi{10.1016/0167-9473(90)90062-M}.
%Type = Manual
\bibitem[{{R Core Team}(2016)}]{Rcore}
\bibinfo{author}{{R Core Team}} (\bibinfo{year}{2016}).
\newblock {\it \bibinfo{title}{R: A Language and Environment for Statistical
  Computing}\/}.
\newblock \bibinfo{organization}{R Foundation for Statistical Computing}
  \bibinfo{address}{Vienna, Austria}.
\newblock \URLprefix \url{https://www.R-project.org/}.
%Type = Article
\bibitem[{Rahmani \& Atia(2017)}]{Rahmani2016}
\bibinfo{author}{Rahmani, M.}, \& \bibinfo{author}{Atia, G.~K.}
  (\bibinfo{year}{2017}).
\newblock \bibinfo{title}{Coherence pursuit: Fast, simple, and robust principal
  component analysis}.
\newblock {\it \bibinfo{journal}{IEEE Transactions on Signal Processing}\/},
  {\it \bibinfo{volume}{65}\/}, \bibinfo{pages}{6260--6275}.
%Type = Article
\bibitem[{Rousseeuw \& Croux(1993)}]{Rousseeuw1993}
\bibinfo{author}{Rousseeuw, P.~J.}, \& \bibinfo{author}{Croux, C.}
  (\bibinfo{year}{1993}).
\newblock \bibinfo{title}{{Alternatives to the Median Absolute Deviation}}.
\newblock {\it \bibinfo{journal}{Journal of the American Statistical
  Association}\/},  {\it \bibinfo{volume}{88}\/}, \bibinfo{pages}{1273--1283}.
  \URLprefix \url{http://www.jstor.org/stable/2291267}.
  \DOIprefix\doi{10.1080/01621459.1993.10476408}.
%Type = Article
\bibitem[{Rousseeuw \& Van~Driessen(1999)}]{Rousseeuw1999}
\bibinfo{author}{Rousseeuw, P.~J.}, \& \bibinfo{author}{Van~Driessen, K.}
  (\bibinfo{year}{1999}).
\newblock \bibinfo{title}{A fast algorithm for the minimum covariance
  determinant estimator}.
\newblock {\it \bibinfo{journal}{Technometrics}\/},  {\it
  \bibinfo{volume}{41}\/}, \bibinfo{pages}{212--223}. \URLprefix
  \url{http://dx.doi.org/10.2307/1270566}. \DOIprefix\doi{10.2307/1270566}.
%Type = Article
\bibitem[{Salibi\'{a}n-Barrera et~al.(2006)Salibi\'{a}n-Barrera, Aelst \&
  Willems}]{Salibian-Barrera2006a}
\bibinfo{author}{Salibi\'{a}n-Barrera, M.}, \bibinfo{author}{Aelst, S.~V.}, \&
  \bibinfo{author}{Willems, G.} (\bibinfo{year}{2006}).
\newblock \bibinfo{title}{Principal components analysis based on multivariate
  {MM} estimators with fast and robust bootstrap}.
\newblock {\it \bibinfo{journal}{Journal of the American Statistical
  Association}\/},  {\it \bibinfo{volume}{101}\/}, \bibinfo{pages}{1198--1211}.
  \URLprefix \url{http://dx.doi.org/10.1198/016214506000000096}.
  \DOIprefix\doi{10.1198/016214506000000096}.
  \href{http://arxiv.org/abs/http://dx.doi.org/10.1198/016214506000000096}{\tt
  arXiv:http://dx.doi.org/10.1198/016214506000000096}.
%Type = Article
\bibitem[{Salibi\'{a}n-Barrera \& Yohai(2006)}]{Salibian-Barrera2006}
\bibinfo{author}{Salibi\'{a}n-Barrera, M.}, \& \bibinfo{author}{Yohai, V.~J.}
  (\bibinfo{year}{2006}).
\newblock \bibinfo{title}{{A Fast Algorithm for S-Regression Estimates}}.
\newblock {\it \bibinfo{journal}{Journal of Computational and Graphical
  Statistics}\/},  {\it \bibinfo{volume}{15}\/}, \bibinfo{pages}{414--427}.
  \DOIprefix\doi{10.1198/106186006X113629}.
%Type = Article
\bibitem[{Serneels \& Verdonck(2008)}]{Serneels2008}
\bibinfo{author}{Serneels, S.}, \& \bibinfo{author}{Verdonck, T.}
  (\bibinfo{year}{2008}).
\newblock \bibinfo{title}{Principal component analysis for data containing
  outliers and missing elements}.
\newblock {\it \bibinfo{journal}{Computational Statistics \& Data Analysis}\/},
   {\it \bibinfo{volume}{52}\/}, \bibinfo{pages}{1712 -- 1727}. \URLprefix
  \url{http://www.sciencedirect.com/science/article/pii/S0167947307002241}.
  \DOIprefix\doi{https://doi.org/10.1016/j.csda.2007.05.024}.
%Type = Article
\bibitem[{{She} et~al.(2016){She}, {Li} \& {Wu}}]{She2016}
\bibinfo{author}{{She}, Y.}, \bibinfo{author}{{Li}, S.}, \&
  \bibinfo{author}{{Wu}, D.} (\bibinfo{year}{2016}).
\newblock \bibinfo{title}{{Robust Orthogonal Complement Principal Component
  Analysis}}.
\newblock {\it \bibinfo{journal}{Journal of the American Statistical
  Association}\/},  {\it \bibinfo{volume}{111}\/}, \bibinfo{pages}{763--771}.
  \URLprefix \url{http://dx.doi.org/10.1080/01621459.2015.1042107}.
  \DOIprefix\doi{10.1080/01621459.2015.1042107}.
%Type = Article
\bibitem[{Stewart(1980)}]{Stewart1980}
\bibinfo{author}{Stewart, G.~W.} (\bibinfo{year}{1980}).
\newblock \bibinfo{title}{The efficient generation of random orthogonal
  matrices with an application to condition estimators}.
\newblock {\it \bibinfo{journal}{SIAM Journal on Numerical Analysis}\/},  {\it
  \bibinfo{volume}{17}\/}, \bibinfo{pages}{403--409}.
%Type = Inproceedings
\bibitem[{Tharrault et~al.(2008)Tharrault, Mourot \& Ragot}]{Tharrault2008}
\bibinfo{author}{Tharrault, Y.}, \bibinfo{author}{Mourot, G.}, \&
  \bibinfo{author}{Ragot, J.} (\bibinfo{year}{2008}).
\newblock \bibinfo{title}{Fault detection and isolation with robust principal
  component analysis}.
\newblock In {\it \bibinfo{booktitle}{2008 16th Mediterranean Conference on
  Control and Automation}\/} (pp. \bibinfo{pages}{59--64}).
\newblock \DOIprefix\doi{10.1109/MED.2008.4602224}.
%Type = Article
\bibitem[{Todorov \& Filzmoser(2009)}]{rrcov}
\bibinfo{author}{Todorov, V.}, \& \bibinfo{author}{Filzmoser, P.}
  (\bibinfo{year}{2009}).
\newblock \bibinfo{title}{An object-oriented framework for robust multivariate
  analysis}.
\newblock {\it \bibinfo{journal}{Journal of Statistical Software}\/},  {\it
  \bibinfo{volume}{32}\/}, \bibinfo{pages}{1--47}. \URLprefix
  \url{http://www.jstatsoft.org/v32/i03/}.
%Type = Article
\bibitem[{Vardi \& Zhang(2000)}]{Vardi2000}
\bibinfo{author}{Vardi, Y.}, \& \bibinfo{author}{Zhang, C.-H.}
  (\bibinfo{year}{2000}).
\newblock \bibinfo{title}{The multivariate l1-median and associated data
  depth}.
\newblock {\it \bibinfo{journal}{Proceedings of the National Academy of
  Sciences}\/},  {\it \bibinfo{volume}{97}\/}, \bibinfo{pages}{1423--1426}.
  \URLprefix \url{http://www.pnas.org/content/97/4/1423.abstract}.
  \DOIprefix\doi{10.1073/pnas.97.4.1423}.
  \href{http://arxiv.org/abs/http://www.pnas.org/content/97/4/1423.full.pdf}{\tt
  arXiv:http://www.pnas.org/content/97/4/1423.full.pdf}.
%Type = Article
\bibitem[{Wilcox(2008)}]{Wilcox2008}
\bibinfo{author}{Wilcox, R.~R.} (\bibinfo{year}{2008}).
\newblock \bibinfo{title}{{Robust principal components: A generalized variance
  perspective}}.
\newblock {\it \bibinfo{journal}{Behavior Research Methods}\/},  {\it
  \bibinfo{volume}{40}\/}, \bibinfo{pages}{102--108}.
  \DOIprefix\doi{10.3758/BRM.40.1.102}.

\end{thebibliography}

\end{document}